\documentclass[aps,prL,twocolumn,superscriptaddress,showpacs]{revtex4-1}
\usepackage{graphicx}% Include figure files
\usepackage[dvipsnames,usenames]{color}
\usepackage{amsmath} %maths
\usepackage{amssymb}
\usepackage{bm}
\usepackage{multirow}
\usepackage{wasysym}
\usepackage{subfigure}
\usepackage{verbatim}
\usepackage{marvosym}
\usepackage{latexsym}
\usepackage[dvipsnames,usenames]{color}
\usepackage{float} % to use [H] and put the figure exactly where you want

\usepackage{xr}
\makeatletter
\newcommand*{\addFileDependency}[1]{
  \typeout{(#1)}
  \@addtofilelist{#1}
  \IfFileExists{#1}{}{\typeout{No file #1.}}
}
\makeatother

\newcommand*{\myexternaldocument}[1]{
    \externaldocument{#1}
    \addFileDependency{#1.tex}
    \addFileDependency{#1.aux}
}
\myexternaldocument{Scotti_hollow_in_hollow_overcrowded_SM}

\begin{document}
%%TC:ignore
\title{Do cavities matter? Hollow microgels resist crystallization}
%%TC:endignore

%%TC:ignore
\author{A.~Scotti}
\affiliation{Institute of Physical Chemistry, RWTH Aachen University, 52056 Aachen, Germany}
\author{A.~R.~Denton}
\affiliation{Department of Physics, North Dakota State University, Fargo, ND 58108-6050 USA}
\author{M.~Brugnoni}
\affiliation{Institute of Physical Chemistry, RWTH Aachen University, 52056 Aachen, Germany}
\author{R.~Schweins}
\affiliation{Institut Laue-Langevin ILL DS/LSS, 71 Avenue des Martyrs, F-38000 Grenoble, France}
\author{W.~Richtering}
\affiliation{Institute of Physical Chemistry, RWTH Aachen University, 52056 Aachen, Germany}
%%TC:endignore

%%TC:ignore
\date{\today}
\begin{abstract}
Solutions of microgels have been widely used as model systems to gain insight into atomic condensed matter and complex fluids. We explore the thermodynamic phase behavior of hollow microgels, which are distinguished from conventional colloids by possessing a central cavity.
Small-angle neutron and X-ray scattering are used to probe hollow microgels in crowded environments. These measurements reveal an interplay between interpenetration and deswelling, and an unusual absence of crystals. Monte Carlo simulations of hollow microgel solutions confirm that, due to the cavity, they form a supercooled liquid more stable than in the case of microgels with a crosslinked core.
\end{abstract}
%%TC:endignore

\maketitle 

Colloidal suspensions provide important insights into phase transitions in condensed matter, e.g., crystallization and glass formation, since they offer a versatile model for studying fundamental processes in atomic systems on length and time scales accessible with conventional techniques (confocal microscopy, dynamic light scattering) \cite{Pus86, Kim13, Ros96, Gas01}. A particular class of colloids is represented by solutions of soft, deformable microgels, polymeric crosslinked networks swollen in a good solvent that respond to external stimuli, e.g., changes in temperature \cite{Sti04} or pH \cite{Iye09}. Microgels have helped to address questions regarding the formation of soft and strong glasses \cite{Mat09,vdS17, Ma19} and the crystallization of spherical particles interacting \emph{via} soft pair potentials \cite{Sen99,Als05,Pel15,Sco17,Xia19}, in both two \cite{Han08, Rey16, Pen17} and three dimensions \cite{Zha09,Pal13}.

Microgels synthesized by precipitation polymerization typically comprise a dense polymeric core surrounded by a periphery (fuzzy shell) with lower polymer density (henceforth termed ``regular" microgels). Improved synthesis protocols allow creating microgels with internal structures substantially different from those of regular microgels, such as hollow microgels -- polymeric networks with central, solvent-filled cavities \cite{Nay05,Dub14}. Studies of microgels with diverse architectures have revealed that the particle ``softness" depends on more than the concentration of crosslinking agent used during synthesis \cite{Sco19a}. The central cavity allows a hollow microgel to respond to the compression of a matrix of regular microgels of comparable size and softness by rearranging its polymer chains into the cavity \cite{Sco18}. Hollow microgels, having no atomic counterpart, represent a new class of materials that show unique and unexplored properties. 

In this work, we investigate the phase behavior of solutions of hollow poly(\emph{N}-isopropylacrylamide)-based (pNIPAM) microgels as a function of concentration, crosslinker density, and cavity size. We address two main questions: (i)~How do hollow microgels respond to crowding? (ii)~Does the internal structure of single microgels affect colloidal phase behavior?

The thermodynamic quantity that determines the phase behavior of microgel solutions, at fixed temperature and pressure, is the generalized volume fraction, $\zeta$ \cite{Sti04, Sco17}, defined as the fraction of the sample volume occupied by the microgels in their fully swollen state. The capability of microgels to deform \cite{Agu17,Con17}, compress \cite{Iye09, Sco16}, or interpenetrate each other \cite{Moh17, Con19} is reflected by values of $\zeta$ ranging above 0.74, the close-packed limit for hard spheres \cite{Gas09}. For all microgels studied here, $\zeta=kc$ is linked to the weight concentration, $c$, by a conversion constant, $k$, obtained \emph{via} viscosimetry (Fig.~S1).

Our solutions of hollow microgels cover a range between $\zeta = 0.40 \pm 0.01$ and $\zeta = 1.05 \pm 0.02$, allowing the possibility of observing transitions from fluid to crystalline phases \cite{Deb03,Sti04,Sco17,Sco19}. We stress that a crystalline phase has been observed for regular microgels and core-shell particles with a rigid, incompressible core surrounded by either polymers \cite{Cra06,Cra08} or DNA \cite{Che10,Zha14}. Thus, the presence of a fuzzy shell \emph{per se} does not suppress crystallization. All of the above-mentioned colloids possess a hard-sphere-like dense core. Thermodynamic phase behavior of colloidal suspensions relies on the fact that the equilibrium state minimizes the Helmholtz free energy, $F = E - TS$, where $E$ and $S$ are the internal energy and entropy of the system, respectively. For a suspension of hard-sphere colloids, $E$ is independent of configuration and simply proportional to the temperature $T$ \cite{Sco18ILL_SM}. For certain concentrations, the entropy is maximized (i.e., $F$ is minimized) when hard spheres self-assemble on an ordered lattice. Indeed the loss of global configurational entropy due to collective ordering of particles is overcome by the gain in local configurational entropy deriving from the increased free volume around particles due to crystallization \cite{Bau87}. 

In contrast to colloids and microgels with a dense core, the hollow microgels studied here do not form crystals in the concentration range studied. Four different types of microgels were synthesized using either 5 or 2.5~mol\% of crosslinker (Table~S1). We prepared these solutions following standard procedures to favor crystal formation, e.g., heating/cooling cycles, and letting rest at constant $T$ for longer than one year (\emph{Supplemental Material}). 

The capability of microgels to spontaneously deswell \cite{Iye09, Fre09, Sco16} significantly decreases the size polydispersity $p$ of the solution with increasing $\zeta$ \cite{Den16, Sco17, Gas19}. Consequently, crystals can form well above the polydispersity limit known for hard spheres ($p=12$\%) \cite{Gas09}, as is the case for the regular microgels with outer radius $209\pm4$~nm  and $p=13.2\pm0.9$~\% (Fig.~S3). 

In contrast, none of the hollow microgel solutions studied here crystallized. As seen in Table~S2, at least two of the hollow microgels have $p\ll 10$\%, well below the value suppressing crystallization, not only for regular microgels, but also for hard spheres \cite{Sco17, Fre09}. Therefore, polydispersity \emph{per se} cannot explain the absence of crystals in the phase behavior of hollow microgels.

\begin{figure}[ht]
	\includegraphics[width=0.45\textwidth ]{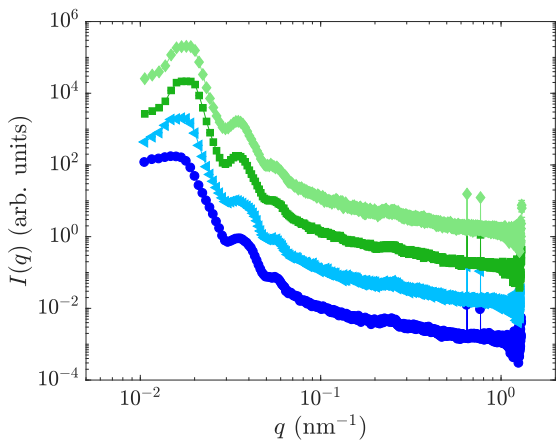}
	\caption{SAXS intensities, $I(q)$, vs. scattering vector, $q$, of solutions of hollow 5~mol\% crosslinked microgels with outer radius $210\pm8$~nm and internal cavity radius $91\pm4$~nm. Concentrations: $\zeta = 0.60\pm0.01$ (circles), $\zeta = 0.75\pm0.02$ (triangles), $\zeta = 0.85\pm0.02$ (squares), $\zeta = 0.90\pm0.02$ (diamonds). All measurements at 20~$^{\circ}$C.}
	\label{fig:SAXS_iqs}
\end{figure}

The SAXS intensities in Fig.~\ref{fig:SAXS_iqs} are relative to hollow 5~mol\% crosslinked microgels with outer radius $210\pm8$~nm and internal cavity radius $91\pm4$~nm. The absence of Bragg peaks indicates a disordered arrangement of microgels \cite{Gas13,Sco17}. The shift of the peaks to higher $q$-values with increasing $\zeta$ reflects a decrease of the microgel center-to-center separation.

To probe the response of these hollow 5~mol\% crosslinked microgels in crowded environments, we used SANS with contrast variation \cite{Gas14,Sco16,Moh17,Sco18}. Relatively few hollow microgels made of pNIPAM ([C$_6$H$_{11}$NO]$_n$) are dispersed in a matrix of 5~mol\% crosslinked D7-pNIPAM-based hollow microgels ([C$_6$D$_{7}$H$_{4}$NO]$_n$) of comparable size (Table~S2 and Fig.~S11). The solvent is a mixture of water and heavy water matching the neutron scattering length density of the deuterated microgels. Consequently, the form factor and, therefore, the shape and architecture of the protonated hollow microgels can be directly detected \cite{Sco18ILL_SM}. 

The generalized volume fraction of the protonated hollow microgels is kept constant at $\zeta_{H}=0.080\pm0.003$ in all SANS measurements. In contrast, for the matrix of hollow deuterated microgels, $\zeta_D$ ranges between 0 and $1.11\pm0.03$. Therefore, $\zeta = \zeta_{H} + \zeta_{D}$ spans minimum and maximum values of 0.08 and 1.19, respectively. 

\begin{figure}[ht]
	\subfigure{\includegraphics[width=0.49\textwidth ]{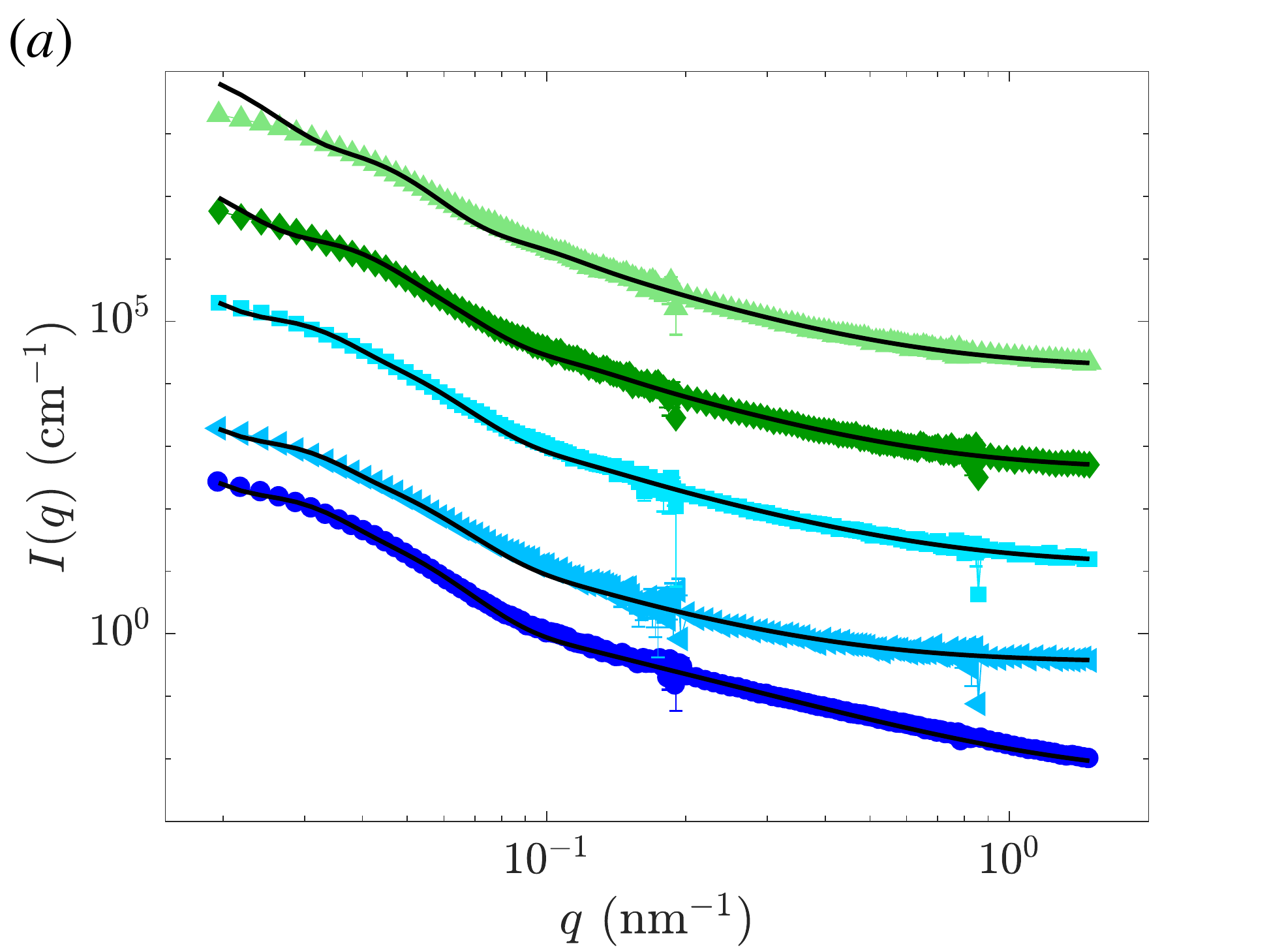}}
	\subfigure{\includegraphics[width=0.49\textwidth]{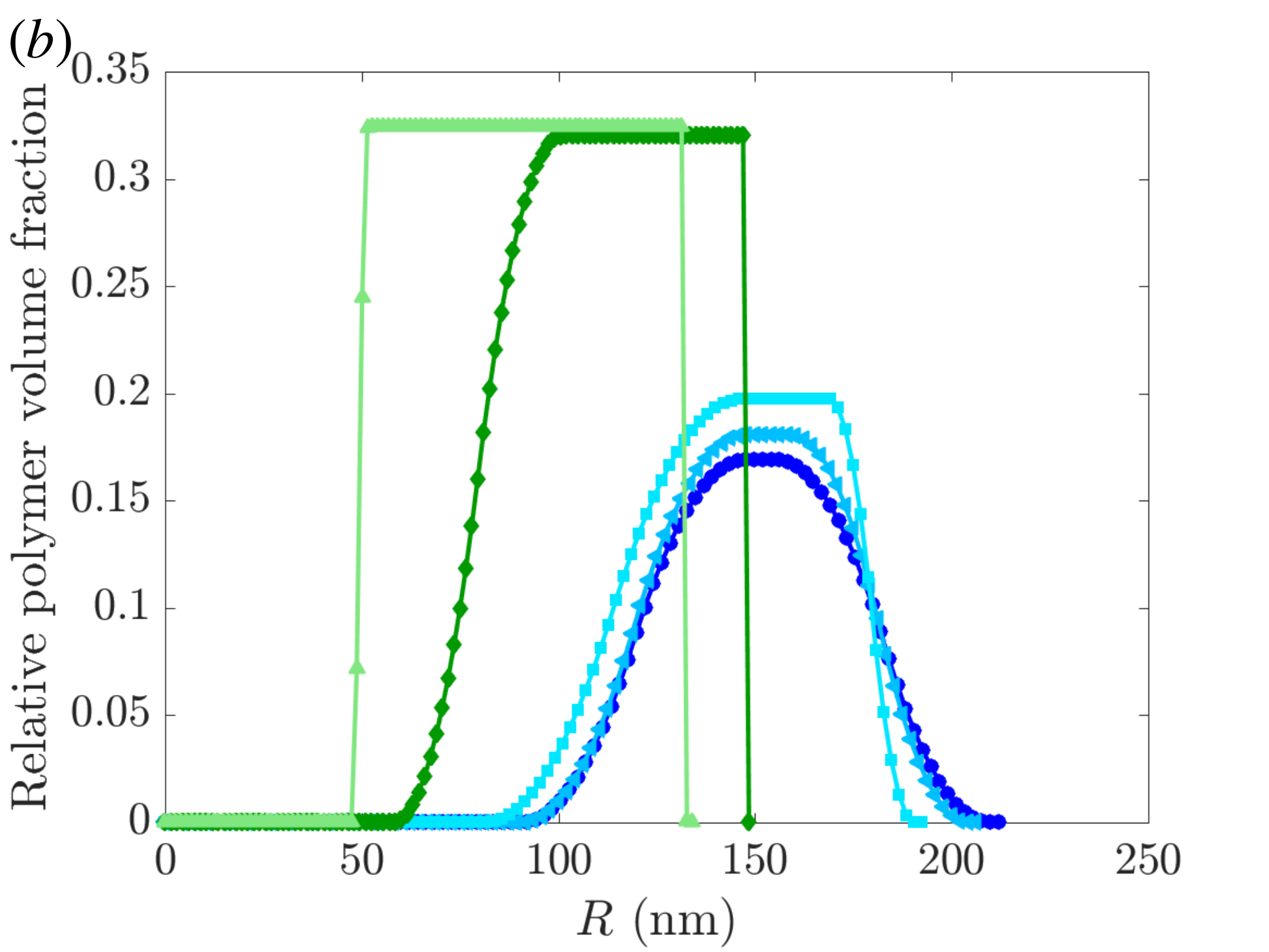}}
	\caption{(a) SANS intensities, $I(q)$, vs. scattering vector, $q$, of hollow protonated 5~mol\% crosslinked microgels. (b) Radial distribution of relative polymer volume S8fraction obtained by fitting curves in (a) using model of Ref.~\cite{Dub15}. Concentrations: $\zeta = 0.080 \pm 0.003$ (circles); $0.30 \pm 0.01$ (left-side-triangles); $0.65 \pm 0.01$ (squares); $0.87 \pm 0.02$ (diamonds); $1.19 \pm 0.03$ (upside-triangles). All measurements at 20~$^{\circ}$C. Colors, symbols, and concentrations in panel (b) correspond to those in panel (a).}
	\label{fig:HS_SANS}
\end{figure}

Figure~\ref{fig:HS_SANS}(a) shows the measured form factors of the protonated hollow 5~mol\% crosslinked microgels at different $\zeta$ at 20~$^\circ$C. The data are fitted (solid lines) with a core-fuzzy-shell model (\emph{Supplemental Material} and Fig.~S7~\cite{Sco18ILL_SM}) previously used to fit SANS data obtained from similar systems \cite{Dub14,Sch16,Sco18}. From the fits, the radial profiles of the relative polymer volume fraction within the microgels are obtained and plotted in Figs.~\ref{fig:HS_SANS}(b) and S8.

Below the volume phase transition temperature (VPTT), the radial profile of the hollow protonated microgels (circles) shows an internal cavity of radius $91\pm4$~nm. The hollow microgels display fuzziness toward both the cavity and the outer periphery. Between these two regions, we observe a region of constant polymer concentration. The outer radius of the microgels is $R_{SANS} = 210\pm8$~nm with $p = 14\pm1$~\%. The fit of static light scattering intensity confirms the radial profile obtained from the SANS data (Fig.~S9) \cite{Sco18ILL_SM}. The cavity is maintained even above the VPTT (Fig.~S8), as expected \cite{Dub14,Dub15,Sch16,Bru18,Sco18}.   

Figure~\ref{fig:HS_SANS}(b) shows that at $\zeta = 0.30\pm0.01$ the radial profile of the hollow protonated microgels (left-side-triangles) virtually coincides with the radial profile at $\zeta = 0.080 \pm 0.003$ (circles). The outer radius is $204\pm5$~nm, $p=13.7\pm0.9$~\%, and the internal cavity radius is $91\pm2$~nm. 

At $\zeta = 0.65\pm0.01$, the radial profile of the hollow protonated microgels shows a decrease of the external fuzziness (squares) and partial occupation of the cavity by the polymer. From the fit, we obtain $R_{SANS} = 190\pm8$~nm, $p = 14\pm1$~\%, and a cavity radius of $80\pm2$~nm. 

In Fig.~\ref{fig:HS_SANS}(b), the diamonds correspond to $\zeta = 0.86\pm0.02$, well above the limit where hard spheres make contact in a random close-packed state ($\zeta_\text{rcp} = 0.64$). A significant compression is observed: $R_{SANS} = 147\pm6$~nm. The external fuzziness disappears and the internal cavity radius decreases to $58\pm3$~nm. At the highest concentration measured, $\zeta=1.19\pm0.03$, the outer radius further decreases to $R_{SANS} = 133\pm5$ and the cavity radius equals $47\pm2$~nm. %upside-triangles in Fig.~\ref{fig:HS_SANS}(b).
Notably, the cavity is preserved throughout the entire $\zeta$-range probed (Fig.~S10). 

The size polydispersity increases with increasing $\zeta$ and reaches $p=19\pm2$~\% for $\zeta = 1.19 \pm 0.03$ (Fig.~S10).
This trend can be explained by the faceting observed for microgels with increasing $\zeta$ \cite{Sco18, Con19, vdS17}. Indeed, in our simple form factor model of spherical microgels, the only way to account for deformation or faceting results in an increase of the polydispersity fitting parameter, now seen as an apparent polydispersity, which reflects deformations. An increase of the apparent polydispersity has been observed also for nanoemulsion droplets \cite{Sch09}. Together with faceting, the presence of anisotropic deformation seems to play a role for very high packing fractions (see \emph{Supplemental Material}, Fig.~S12).

Figure~\ref{fig:R_vs_VF} summarizes how the outer diameter of the hollow protonated microgels, $2R_{SANS}$, varies with $\zeta$ (empty blue circles). Four regimes can be identified: (i) for $\zeta < 0.50\pm0.01$, the hollow protonated microgels maintain their size; (ii) for $0.50\pm0.01 < \zeta < 0.76\pm0.01$, the size first decreases; (iii) for $0.76\pm0.02<\zeta<1.10\pm0.3$, the size further steeply decreases; and (iv) for $\zeta>1.10\pm0.02$, the size remains constant.

\begin{figure}[ht!]
	\centering
	\includegraphics[width=.49\textwidth]{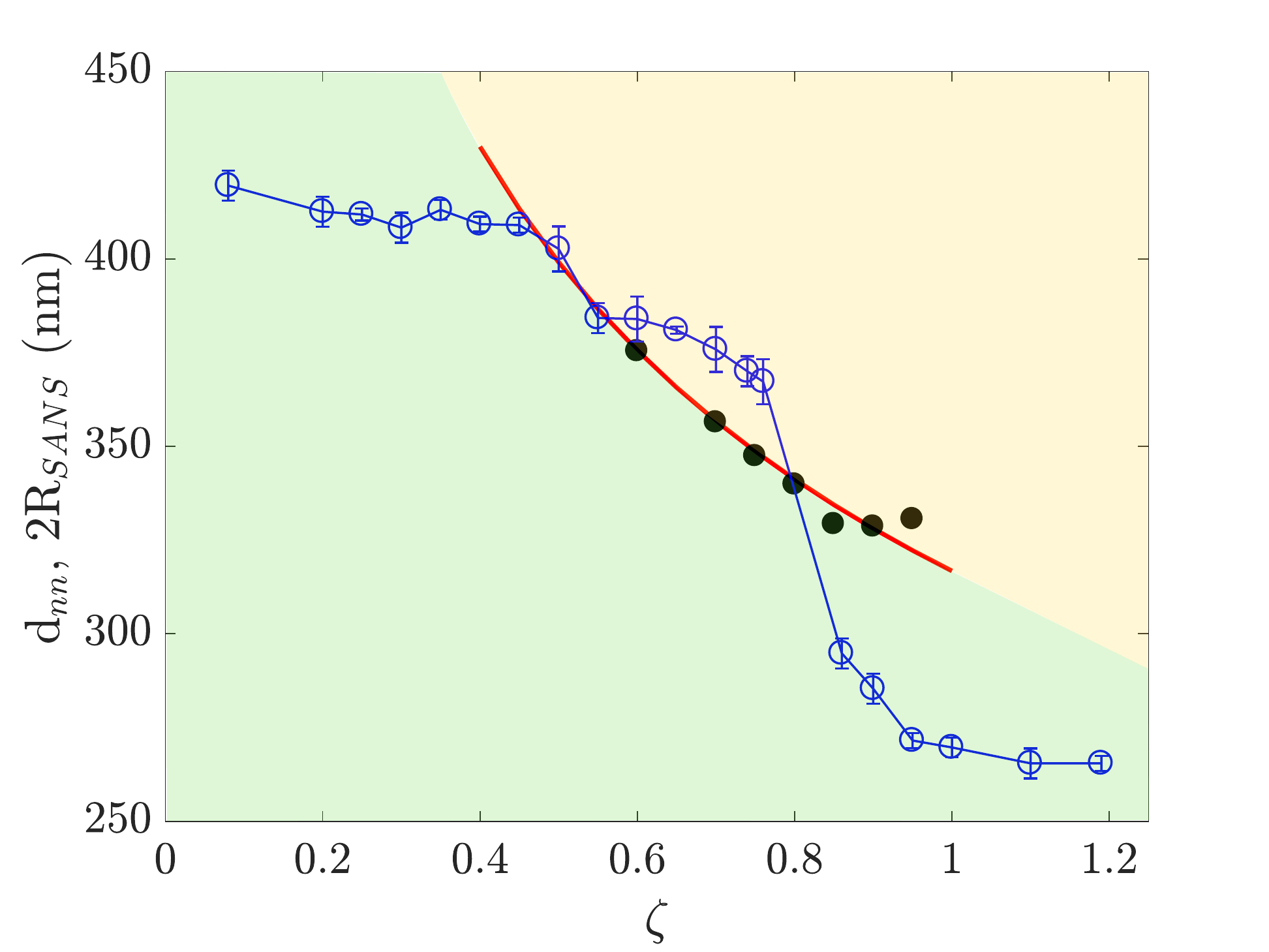}
	\caption{Particle diameter $2R_{SANS}$ (measured with SANS, blue empty circles) and nearest-neighbor distance $d_{nn}$ (measured with SAXS, black solid circles) vs. generalized volume fraction $\zeta$ for hollow microgel solutions. Red curve is a fit of the $d_{nn}$ data to the function $c\zeta^{-1/3}$ with fit parameter $c$.
	}
	\label{fig:R_vs_VF}
\end{figure}

The onset of deswelling in regime (ii) occurs at concentrations below that at which the microgels make contact. This phenomenon has been reported also for regular microgels \cite{Iye09,Fre09,Sco17,Sco18} and attributed to overlapping of counterion clouds surrounding the microgels, significantly increasing the osmotic pressure of the solution~\cite{Sco16, Gas19}. In contrast, the second stage of deswelling in regime (iii) is due to a collapse of the external fuzzy shell and rearrangement of polymer within the cavity.

To understand whether microgel deswelling is sufficient to forestall interpenetration, the diameter of the hollow microgels obtained by SANS is compared to the nearest-neighbor distance, $d_{nn}$ (black circles in Fig.~\ref{fig:R_vs_VF}). The latter is obtained from the structure factor (Fig.~S13) of the solutions shown in Fig.~S2(a), measured with SAXS. The red curve in Fig.~\ref{fig:R_vs_VF} represents a fit of the data with the power law $d_{nn} = a\zeta^{-1/3}$, $a$ being the value of $d_{nn}$ at $\zeta = 1$. This curve divides the graphs in two areas, indicated in yellow and in green in the figure. If a the microgel diameter (blue circles) lies above the red curve, in the yellow area, its dimension is larger than the microgel-to-microgel distance and, therefore, interpenetration is dominant. In contrast, if the diameter of a microgel (blue circles) lies below the red curve, in the green area, it has been deformed/deswollen and its dimension is smaller than the microgel-to-microgel distance and, therefore, interpenetration is negligible. For $0.5\lesssim\zeta\lesssim0.8$, the blue circles lies in the yellow area, indicating interpenetration of the hollow microgels with their neighbors. In contrast, for values of $\zeta\gtrsim0.8$, a stronger deswelling/deformation is observed and interpenetration is negligible, as indicated by $2R_{SANS}$ being consistently smaller than $d_{nn}$.

This deswelling behavior can be explained by the absence of a polymeric core, making the hollow microgels more compressible and interpenetrable than regular microgels with a dense core surrounded by a fuzzy shell. The hollow protonated microgels embedded in a matrix of hollow deuterated microgels collapse, at $\zeta = 1.19\pm0.03$, to a size that is $\approx 63$~\% of their fully swollen size (Fig.~S6). In contrast, regular microgels embedded in the very same matrix of hollow deuterated microgels deswell to a size that is, at $\zeta = 1.17\pm0.03$, only 90~\% of the fully swollen size (Figs.~S5 and S6), implying that regular microgels are much less compressible than hollow microgels. The presence of a dense core limits the rearrangement of polymeric chains inside the microgels, while hindering interpenetration. 

\begin{figure}[h!]
\begin{center}
\includegraphics[width=0.49\textwidth]{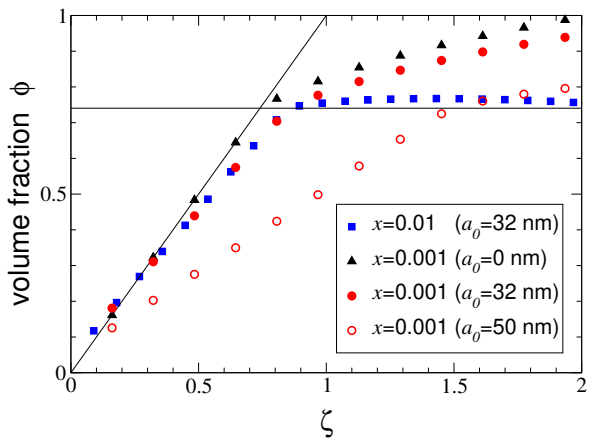}
\vspace*{-0.5cm}
\end{center}
\caption{Simulation data for actual volume fraction $\phi$ vs. generalized volume fraction $\zeta$ in solutions of hollow microgels with collapsed outer radius $b_0=63$ nm, collapsed inner radius $a_0=32$ or 50 nm, Flory solvency parameter $\chi=0.2$, and crosslink fraction $x=0.01$ or 0.001. The straight line, $\phi=\zeta$, represents incompressible microgels and the horizontal line, $\phi=0.7405$, the close-packed limit for hard spheres. Data are shown also for softer, regular microgels with no cavity ($a_0=0$). Statistical error bars are smaller than symbols.
}\label{fig:phi-vs-zeta}
\end{figure}

To complement our experiments, we performed Monte Carlo (MC) simulations of microgel solutions modeled by the Flory-Rehner single-particle free energy and the Hertz pair potential (Eqs.~S10 and S11 in \emph{Supplemental Material}). For solutions of $N=256$ particles with typical system parameters (Fig.~\ref{fig:phi-vs-zeta}), we performed runs over a range of concentrations. Each MC step comprises a trial displacement and {\it independent} trial changes in size and internal structure -- outer and inner (cavity) radii -- of every microgel. After equilibrating for $5\times 10^4$ steps, we collected statistics for single-particle and collective properties at intervals of 100 steps for $10^6$ steps.

First, we map $\zeta$ onto the actual volume fraction $\phi$, which represents the real fraction of the volume occupied by the collapsed/deformed/interpenetrated microgels. Figure~\ref{fig:phi-vs-zeta} presents our simulation data for $\phi$ vs.~$\zeta$. At low concentrations ($\zeta\ll 1$), where interparticle interactions are weak, the microgels are barely compressed and $\phi\simeq\zeta$. 

With increasing $\zeta$, however, the data progressively deviate from the line $\phi=\zeta$, demonstrating both compression and faceting of the particles. With increasing $\zeta$, the volume fraction of stiffer microgels ($x=0.01$) plateaus just above the hard-sphere close-packing limit of $\phi\simeq 0.74$. In contrast, softer microgels ($x=0.001$) both deswell and facet, as reflected by steady increase of $\phi$. We verified that this trend continues as $x$ is further lowered. These data show that the higher the crosslink density, the stronger the tendency for microgels to deswell rather than facet in response to mutual crowding. Furthermore, enlarging the cavity from a collapsed inner radius of $a_0=32$ nm to $a_0=50$ nm results in a much more gradual increase of $\phi$ with increasing $\zeta$. 

\begin{figure}[h!]
\begin{center}
\includegraphics[width=0.49\textwidth]{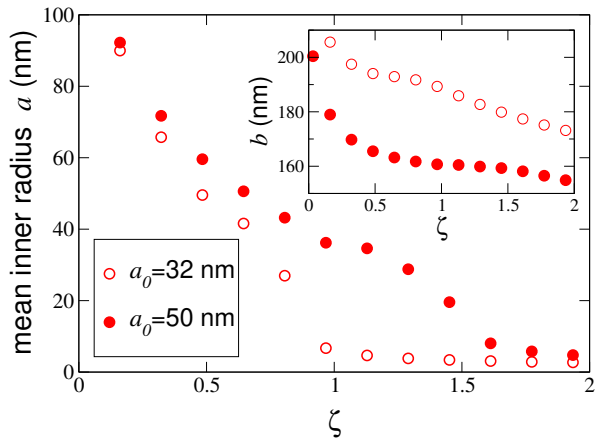}
\vspace*{-0.5cm}
\end{center}
\caption{Simulation data for mean swollen inner radius $a$ and outer radius $b$ (inset) vs.~generalized volume fraction $\zeta$ for systems of Fig.~\ref{fig:phi-vs-zeta}, with collapsed inner radius $a_0=32$ nm (empty circles) or $50$ nm (solid circles), and crosslink fraction $x=0.001$. Statistical error bars are smaller than symbols.
}\label{fig:radii-vs-zeta}
\end{figure}

The tendency for the volume fraction of hollow microgels with larger cavities to rise more gradually with increasing concentration is correlated with a weaker contraction of the cavity and stronger compression of the particles. These correlations are illustrated in Fig.~\ref{fig:radii-vs-zeta}, which shows that the mean outer radius progressively decreases with increasing $\zeta$ and $a_0$, while the mean inner radius levels off when $\phi$ reaches close packing ($\zeta\simeq 1$ for $a_0=32$ nm and $\zeta\simeq 1.6$ for $a_0=50$ nm).

Beyond single-particle structure, we also computed collective thermodynamic and structural properties. As detailed in the \emph{Supplemental Material}, our analyses of the bulk osmotic pressure, mean pair interaction energy, pair distribution function, and static structure factor consistently indicate that a microgel solution freezes into either a close-packed crystal or an amorphous solid when its volume fraction exceeds $\phi\simeq 0.5$, i.e., roughly the freezing density of a hard-sphere fluid. The key new insight from our modeling is that hollow microgels, because of their additional internal degrees of freedom, can delay the increase of $\phi$ with $\zeta$ by collapsing their cavities, and thus remain stable in the fluid phase up to considerably higher concentrations, more so the larger the cavity, consistent with experimental observations.

To summarize, we presented the phase behavior of solutions of hollow microgels, soft colloidal particles composed of crosslinked polymeric networks swollen in a good solvent and distinguished by central solvent-filled cavities. Crystals were not observed, regardless of cavity size and amount of crosslinker agent used during synthesis. From SANS and SAXS experiments, we observed that at intermediate concentrations interpenetration between microgels is dominant. In contrast, at higher concentrations, in overcrowded conditions, the data indicate both faceting and anisotropic deformation of the microgels, in agreement with recent studies \cite{Sco18, Con19, vdS17}. Monte Carlo simulations support our experimental findings on deswelling of hollow microgels and the inferences that cavities enhance stability of the fluid phase and are progressively occupied by polymer with increasing concentration. For the same concentration, a supercooled fluid is more stable for hollow microgels than for regular microgels. 

The absence of a dense core must be considered to rationalize the unique phase behavior of hollow microgels compared with that of both hard and soft spheres \cite{Aue01, Pam09}. With increasing concentration, conventional colloids form crystals, the only way to increase their configurational entropy \cite{Fre94,Fre99}, as observed not only for hard-sphere-like systems (e.g., pMMA particles \cite{Pus87, Pus86}), but also for regular microgels, core-polymer, and DNA-coated particles, which resemble hard spheres once the external soft shell is compressed and the dense, rigid core is reached \cite{Cra06,Cra08,Che10,Zha14, Sco17, Sco19}. 

In contrast, hollow microgels can deswell/deform with lower energetic cost due to fewer internal constraints. Furthermore, the polymer chains can rearrange within the empty cores. These two mechanisms offer an alternative pathway to increase configurational entropy without the need to crystallize. For sufficiently high concentrations, however, the collapsed chains may completely occupy the cavity. In this case, if the microgels are not jammed, crystals might form. 

The absence of a crystalline phase for solutions of hollow microgels allows studying glass formers or complex fluids without significant size disparities, making clear whether the unique properties of glasses, e.g., storage and loss moduli, depend on colloidal size disparities. Furthermore, it has been reported that relatively few hollow microgels embedded in a matrix of regular microgels of like size do not hinder crystallization \cite{Sco19a}. Therefore, the role of hollow microgels as defects that suppress crystallization can be studied in binary mixtures of hollow and regular microgels. This study would open a path to a new kind of defect that suppresses crystallization because of its internal structure, i.e., the absence of the core, and not because of a size mismatch, as is usually observed \cite{Pus86, Aue01}. Pure solutions of only hollow microgels, or binary mixtures in which hollow microgels act as defects, offer the possibility to study the glass and jamming transitions without introducing large size mismatches to hinder crystallization \cite{Mat09}. Finally, solutions of hollow microgels can be used to test some recent findings of simulations on the roles of deformation, interpenetration, and deswelling in the dynamics of soft colloids \cite{Gna19}. 

%%TC:ignore
\begin{acknowledgments}
Financial support from the Alexander von Humboldt Foundation, SFB 985 - Functional Microgels and Microgel Systems and from the International Helmholtz Research School of Biophysics and Soft Matter (IHRS BioSoft) is gratefully acknowledged. All SANS experiments were performed on D11 at the Institut Laue-Langevin (ILL), Grenoble, France.\\
\end{acknowledgments}
%%TC:endignore

%%TC:ignore
\noindent \textbf{Data availability} 
\noindent All the data used for this study are available under request at: https://hdl.handle.net/21.11102/74326e3e-1b59-11ea-9a63-e41f1366df48. SANS data are available at  DOI: 10.5291/ILL-DATA.9-11-1855.
%%TC:endignore

\bibliographystyle{apsrev4-1}
\bibliography{Scotti_hollowINhollow}

\end{document}

%\title{Phase behavior of hollow microgel solutions: Cavities suppress crystallization}
%Other possibilities:
%\title{\emph{Supporting Information}: Do cavities matter? Suppression of crystallization in hollow microgel solutions}
\title{\emph{Supplemental Material}: Do cavities matter? Hollow microgels resist crystallization}
%\title{\emph{Supporting Information}: Do cavities matter? Hollow microgels successfully avoid crystallization}
%\title{Cavities suppress crystallization in solutions of hollow microgels}
%\title{Phase behavior of solutions of compressible hollow microgels: does the presence of a cavity affect the crystallization?}

\author{A.~Scotti}
\affiliation{Institute of Physical Chemistry, RWTH Aachen University, 52056 Aachen, Germany}
\author{A.~R.~Denton}
\affiliation{Department of Physics, North Dakota State University, Fargo, ND 58108-6050 USA}
\author{M.~Brugnoni}
\affiliation{Institute of Physical Chemistry, RWTH Aachen University, 52056 Aachen, Germany}
\author{R.~Schweins}
\affiliation{Institut Laue-Langevin ILL DS/LSS, 71 Avenue des Martyrs, F-38000 Grenoble, France}
\author{W.~Richtering}
\affiliation{Institute of Physical Chemistry, RWTH Aachen University, 52056 Aachen, Germany}

\date{\today}

\maketitle 

\section{Synthesis}

The exact synthesis of the hollow 2.5\,mol\% and 5\,mol\% crosslinked microgels with a sacrificial core of 60 nm has been described previously \cite{Sco18}.
5\,mol\% crosslinked hollow microgels, deuterated and protonated, are produced by core dissolution of silica-core pNIPAM-shell microgels \cite{Dub15}. The silica cores were obtained by the St\"ober synthesis \cite{Sto68} with surface modification \cite{Zha02}. Briefly, 80\,mL of ammonia solution was added to 700\,mL of preheated ethanol. After equilibration, 24\,mL of tetra\-ethyl ortho\-sili\-cate~(TEOS) was added to start the reaction. Centrifugation was applied for purification and the solvent was evaporated for storage.

Core-shell microgels are produced similarly to regular microgels, except for the presence of silica seeds during the polymerization. The monomer solution consists of NIPAM or D7-NIPAM, \textit{N,N'}-methyl\-ene\-bis\-acryl\-am\-ide~(BIS), and sodium do\-de\-cyl sulfate~(SDS) in filtered (0.2~$\mu$m RC membrane filter) double-distilled water (see Table \ref{tab:Synthesis} for exact amounts). To generate core-shell microgels, a solution of 1.4\,g of the sacrificial silica cores in 3.8\,mL of ethanol was added. The reaction solution was purged with nitrogen under stirring at 200~rpm and heated to 60~$^\circ$C. At once, a solution of potassium per\-oxy\-di\-sul\-fate~(KPS) in 5~mL of water was degassed. The KPS solution was transferred into the monomer solution to initiate the reaction. The polymerization was left to proceed for 4~hours at constant stirring and 60~$^\circ$C. The microgels were purified by threefold centrifugation at 157\,000~g and redispersion in fresh water. Lyophilization was performed for storage.

To generate the hollow microgels, the silica cores of the core-shell microgels were dissolved by means of a sodium hydroxide (NaOH) solution \cite{Dub15}. Finally, the resulting hollow microgels were centrifuged at 360\,000~g for purification and lyophilization was carried out for storage.

\begin{table*}

\caption{Composition of reaction solutions during synthesis of regular and core-shell microgels and hydrodynamic radii at 20\,$^\circ$C and 50\,$^\circ$C of corresponding regular and hollow microgels. For the synthesis of the deuterated hollow microgels, D7-NIPAM was used as monomer; for the protonated microgels, NIPAM was used.}
\label{tab:Synthesis}
\begin{tabular}{ccccccccc}
\hline\hline
	\textbf{Sample name}& \textbf{Category}	& \textbf{n(Monomer)} & \textbf{n(BIS)} & \textbf{n(SDS)} & \textbf{n(KPS)} &$ \textbf{V}_{total}$ &\textbf{R$_h^{20 ^\circ \textup{C}}$} & \textbf{R$_h^{50 ^\circ \textup{C}}$}\ \\
\hline 
MB-HS-105-5-D7PNIPAM & Hollow	& 9.5\,mmol 	& 0.5\,mmol 	& 0.22\,mmol 	& 0.31\,mmol 	& 200\,mL 	& (232$\pm$6)\,nm		& (85$\pm$1)\,nm\\ 
{MB-HS-105-5-PNIPAM} 		& Hollow	& 9.5\,mmol 	& 0.5\,mmol 	& 0.22\,mmol 	& 0.31\,mmol 	& 200\,mL 	&(230$\pm$4)\,nm 		& (106$\pm$1)\,nm\\
{MB-HS-60-5-PNIPAM} 		& Hollow	& 10.0\,mmol 	& 0.53\,mmol 	& 0.5\,mmol 	& 0.39\,mmol 	& 250\,mL 	&(117$\pm$2)\,nm 		& (56$\pm$1)\,nm\\
{MB-HS-60-2.5-PNIPAM} 		& Hollow	& 10.2\,mmol 	& 0.26\,mmol 	& 0.5\,mmol 	& 0.39\,mmol 	& 250\,mL 	&(152$\pm$3)\,nm 		& (55$\pm$1)\,nm\\
{MB-PNIPAM-5-225} 	& Regular	& 45.0\,mmol 	& 2.37\,mmol 	& 0.04\,mmol 	& 0.47\,mmol 	& 300\,mL 	& (223$\pm$2)\,nm 		& (122$\pm$1)\,nm \\ 
\hline\hline

\end{tabular}
\end{table*}

\section{Viscosimetry}
The generalized volume fraction, $\zeta$, is related to the weight fraction of polymer in the solution, $c$, by a conversion constant $k$: $\zeta = kc$. Viscosimetry is the standard method to experimentally determine this constant since the Einstein-Batchelor equation links the relative viscosity, $\eta_r$, to the generalized volume fraction:
\begin{equation}
    \eta_r = 1 + 2.5\zeta + 5.9\zeta^2 =  1 + 2.5kc + 5.9(kc)^2\,. \label{eq:EB_visc}
\end{equation}

\begin{figure}[ht]
	\subfigure{\includegraphics[width=0.4\textwidth ]{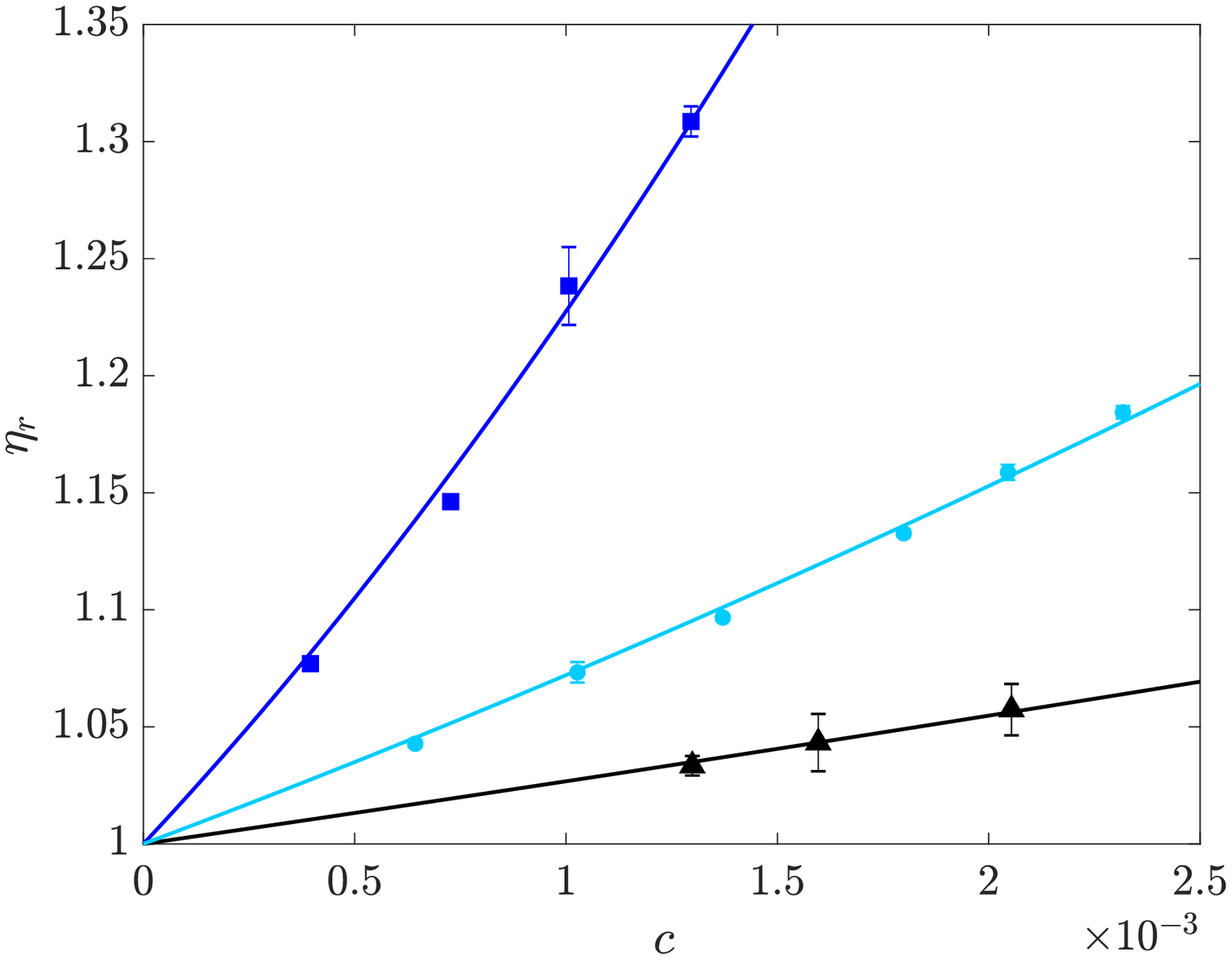}}
	\caption{Relative viscosity versus polymer concentration in solutions for: pNIPAM-based hollow 5~mol\% crosslinked microgels in Fig.~\ref{fig:PB}(a), light blue circles; D7-pNIPAM-based hollow 5~mol\% crosslinked microgels in Fig.~\ref{fig:PB}(b), light blue circles; regular 5~mol\% crosslinked microgels in Fig.~\ref{fig:PB_reg}, black triangles. Solid lines are fit of the data according to \ref{eq:EB_visc}.
	}
	\label{fig:visc}
\end{figure}

We prepared solutions of microgels with weight fractions in the range $5\cdot10^{-4}<c<3\cdot10^{-3}$. For every solution, the time of fall, $t$, of a fixed volume of the solutions through a thin capillary of an Ubbelohde viscosimeter immersed in a water bath at $20.0 \pm 0.1\,^{\circ}$C was measured. The time of fall is linked to the kinematic viscosity, $\nu$, by an instrumental constant, $K$, given by the producer: $\nu = Kt$. For the instrument used for these experiments, $K = 3.156\cdot10^{-9}\,{\rm m}^2 {\rm s}^{-2}$. After computing the $\nu$-values for the different samples, the values of the viscosity were obtained as $\eta = \nu\rho_\text{H$_2$O}$ with $\rho_\text{H$_2$O}$ being the density of the water at $20\,^\circ$C. This relation is valid under the assumption that the solution mass density is approximately equal to that of water due to the low concentration of microgels. Figure~\ref{fig:visc} shows data for the relative viscosity, $\eta_r = \eta/\eta_\text{H$_2$O}$, with  $\eta_\text{H$_2$O}$ the viscosity of water at $20\,^\circ$C. The conversion constant for the smaller 5~mol\% crosslinked hollow microgels mentioned in the text and of the 2.5~mol\% crosslinked hollow microgels shown in Fig.~\ref{fig:PB}(c) have been experimentally determined in Ref.~\cite{Sco19a}.

\section{Partition Function for Hard-Sphere-Like Systems}\label{sec:Z_HS}

As mentioned in the main text, hard-sphere-like systems crystallize because their positional ordering increases the configurational entropy, since the particles have more free space to jiggle around their equilibrium positions than in a disordered arrangement. Examples of hard-sphere-like particles suspended in a solvent include pMMA hard colloidal microspheres and colloids consisting of a more rigid core surrounded by a softer shell, such as microgels or DNA-coated silica particles. To demonstrate that the decrease in the Helmholtz free energy ($F = E-TS$) is purely driven by an increase in the entropy ($S$), and does not depend on a decrease of the internal energy ($E$), the partition function $Z$ of the system must be considered:
\begin{equation}\label{eq:Z}
    Z = \frac{1}{h^{3N}N!}\int\dots\int dp^N dq^N \exp{[-\beta H(p^N,q^N)]}\,,
\end{equation}
where $\beta = (k_BT)^{-1}$ and $H(p^N,q^N)$ is the Hamiltonian of the system, which is a function of the momenta $p^N$ and of the coordinates $q^N$ of the $N$ hard spheres. The Hamiltonian can be decomposed into the sum of the kinetic energy of the system, $K(p^N)$, and the potential energy, $U(q^N)$. Colloidal dispersions of hard spheres (or hard-sphere-like particles) can be treated neglecting quantum mechanical effects and, therefore, the integration of Eq.~(\ref{eq:Z}) can be performed over the momenta, leading to a term depending on $T$ only: $(3/2)Nk_BT$. In this way, we can decouple the kinetic part of the Hamiltonian from the configurational part, $Q$:
\begin{equation}
    Q =  \frac{1}{N!}\int\dots\int dq^N \exp{[-\beta U(q^N)]}\,.
\end{equation}
$Q$ is in general a function of $N$, $T$, and $V$, the total volume of the system. Now assuming that the particles in our system interact with a hard-sphere pair potential, i.e., zero for distances larger than the radius of the hard-spheres and infinity elsewhere, we can write the average potential energy as:
\begin{equation}
    \langle U \rangle = -\frac{\partial\ln Q}{\partial \beta} = 0\,.
\end{equation}
This means that the average energy for a system of hard-sphere-like particles equals the average kinetic energy only. The latter depends on $T$ only and, therefore, once $T$ is fixed, $E$ is constant. The only means of minimizing $F = E -TS$ is then to maximize the entropy $S$. This is achieved by a hard-sphere-like system forming an ordered lattice: the loss in configurational energy is overcome by the gain in entropy because of the increased space the hard spheres have to jiggle around their equilibrium positions \cite{Bau87, Low00}. The hollow microgels studied here, which are more easily deformable and possess an empty cavity in which the polymeric chains can rearrange upon increasing the generalized volume fraction of the solution, have an alternative way to minimize the Helmholtz free energy -- one that avoids long-range positional ordering.

\section{Preparation of the solutions of hollow microgels}

For solution of colloids at high concentrations there is the risk of having supercooled liquid depending on the preparation protocols. In the literature, it has been reported that high levels of size-polydispersity, as well as the presence of microgels with significantly different sizes (e.g., binary mixture), dramatically slow down crystallization kinetics \cite{Sco16, Sco17}. Nevertheless, within a few months from their synthesis, solutions of microgels, both binary mixtures and highly polydisperse (18.5\%), reach their thermodynamic equilibrium.

All the solutions of hollow microgels presented in this study have been equilibrated for more than one and a half years. Therefore, the time allowed for the system to evolve toward equilibrium is considerably longer than in previous studies. All the samples are hermetically sealed to avoid problems with evaporation of the solvent. The vials containing the samples are regularly weighed to verify that solvent is not evaporated and that their packing fraction is constant. The samples are stored in a room at constant temperature $T = 20.0\pm0.5$~$^\circ$C.

Microgel assemblies in a supercooled liquid-like state can be released from their kinetically trapped, high-energy states by thermal annealing via heating, followed by slow cooling to form well-ordered crystals. These quasi-static cooling processes, as well as gentle shearing of the samples in an oscillatory sweep, have been reported to promote crystallization of microgel solutions as reported in Ref.~10 or in the studies of Brijitta et al.~\cite{Bri09} or of Lyon and co-workers \cite{StJ07,Lyo04,Deb02}. 

To lead our solutions of hollow microgels to equilibrium, we have performed heating/cooling cycles with quasi-static temperature changes ($\pm0.2~^\circ$C). The samples were annealed by raising the sample temperature to 38~$^\circ$C (well above the VPTT) and kept at that temperature for different amounts of time (from 1 hr to 1 day). Then the samples were slowly cooled down to 20~$^\circ$C with a cooling rate of 0.2~$^\circ$C/hour.

\noindent The absence of crystals after all these ``standard'' procedures ensures us that our solutions of hollow microgels do not crystallize within a time scale accessible in experiments.
\section{Phase Behavior of Hollow Microgels}

\begin{figure}[ht]
	\includegraphics[width=0.35\textwidth ]{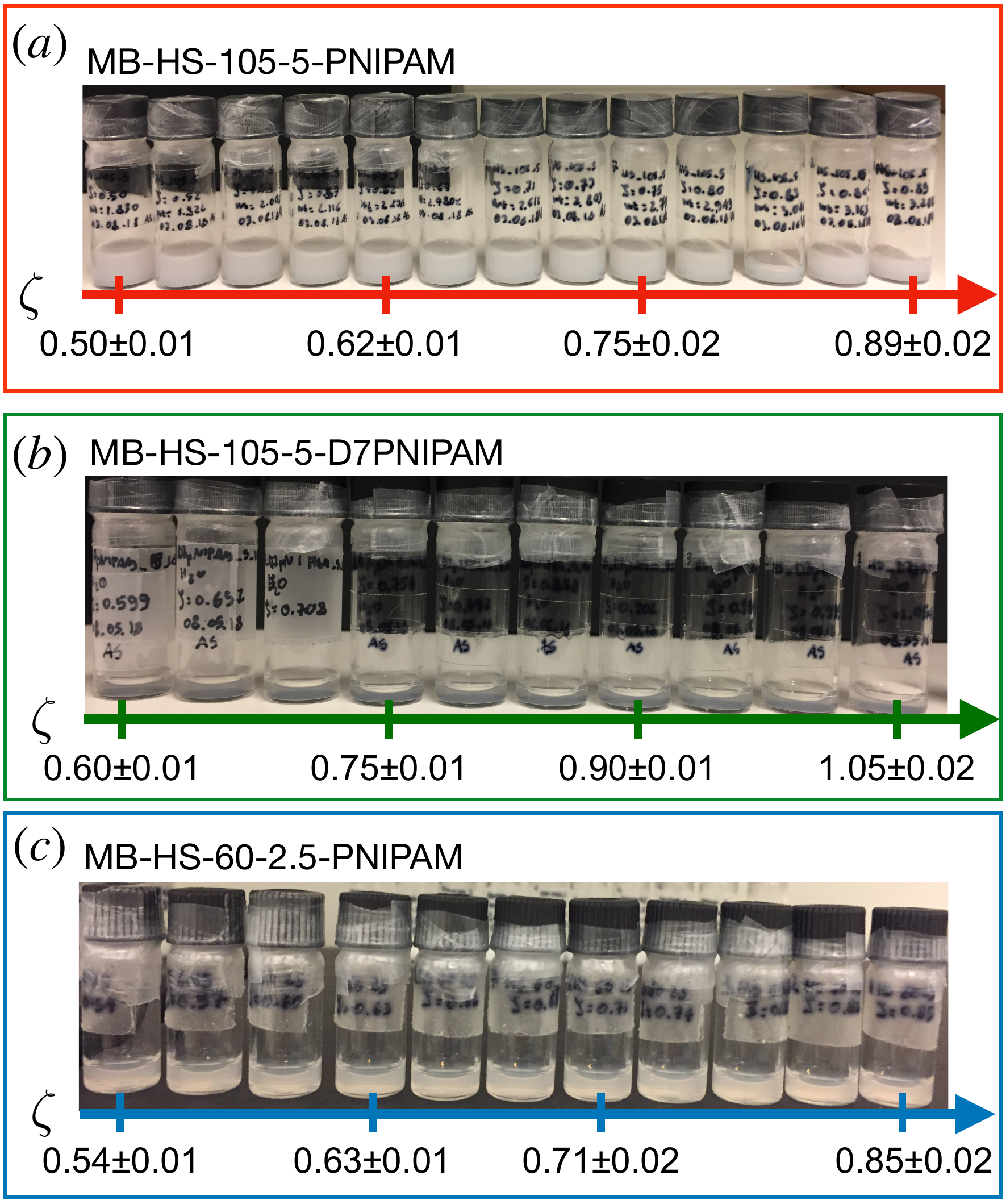}
	\caption{Photographs of samples at different concentrations: (a) MB-HS-105-5-PNIPAM; (b) MB-HS-105-5-D7PNIPAM; (c)MB-HS-60-2.5-PNIPAM. The data relative to the synthesis and characterization of these samples are reported in Tables~\ref{tab:Synthesis} and \ref{tab:samples}. All samples were stored at $20.0\pm0.5~^{\circ}$C.}
	\label{fig:PB}
\end{figure}

As illustrated in Fig.~\ref{fig:PB}, none of the samples exhibits a stable crystalline structure in the studied $\zeta$-range. We observed crystals neither in solutions of hollow 5~mol\% crosslinked microgels with an internal cavity of radius $91\pm4$~nm (Figs.~\ref{fig:PB}(a) and (b)) nor in solutions of hollow 5~mol\% crosslinked microgels with a much smaller cavity of radius $25.5\pm0.8$~nm (not shown). Finally, Fig.~\ref{fig:PB}(c) shows that crystallization is as well suppressed in a solution of less crosslinked (2.5~mol\%) hollow microgels with a cavity of radius $24\pm2$~nm. 

%The hollow microgels of Fig.~\ref{fig:PB}(a) have a size polydispersity $p=14\pm$1~\%, virtually the same as the value $p=13.2\pm0.9$~\% obtained for regular microgels of comparable radius $209\pm4$~nm (Fig.~\ref{fig:SANS_reg}), which show crystals (Fig.~\ref{fig:PB_reg}). It has been demonstrated that the capability of microgels to spontaneously deswell in response to increasing $\zeta$ \cite{Iye09, Sco16} significantly decreases the size polydispersity of the microgel solution \cite{Den16, Sco17, Gas19}. Consequently, crystals can form well above the limit known for hard spheres (12\%) \cite{Gas09}. Solutions of regular microgels crystallize for size polydispersity as high as 18\% \cite{Sco17}.

\begin{figure}[ht]
	\subfigure{\includegraphics[width=0.35\textwidth ]{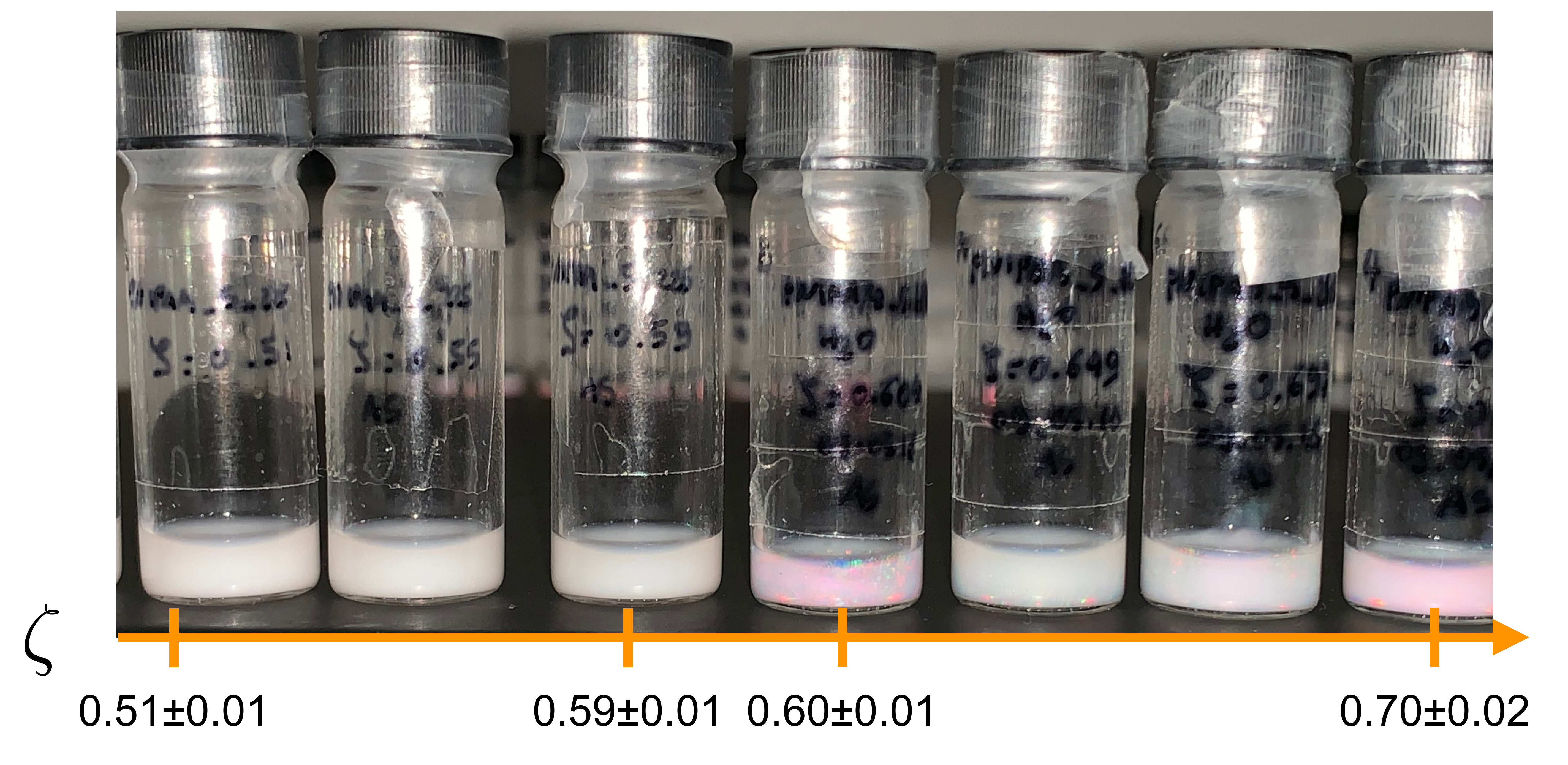}}
	\caption{Solutions of regular microgels at different generalized volume fractions $\zeta$.
	}
	\label{fig:PB_reg}
\end{figure}

%A similar behavior can be expected for solutions of hollow microgels as well. However, 
Crystals are absent also for the hollow microgels with polydispersities lower than 12~\%, i.e., MB-HS-60-5-PNIPAM and MB-HS-60-2.5-PNIPAM (Fig.~\ref{fig:PB}(c)). Both of these hollow microgels have a cavity with a radius $\approx25$~nm (Table~\ref{tab:samples}). Crystals are observed in concentrated solutions of both hard spheres and regular microgels with such low polydispersities \cite{Pus87, Sco17}. Furthermore, regular microgels with a comparable size ($209\pm4$~nm) and polydispersity ($13.2\pm0.9$~\%) exhibit crystals (see Fig.~\ref{fig:PB_reg}). These observations preclude polydispersity as the limiting factor for the crystallization of hollow microgels and point to the lack of a dense core as the key factor.

\section{Dynamic Light Scattering}

\begin{figure}[ht!]
	\subfigure{\includegraphics[width=0.35\textwidth ]{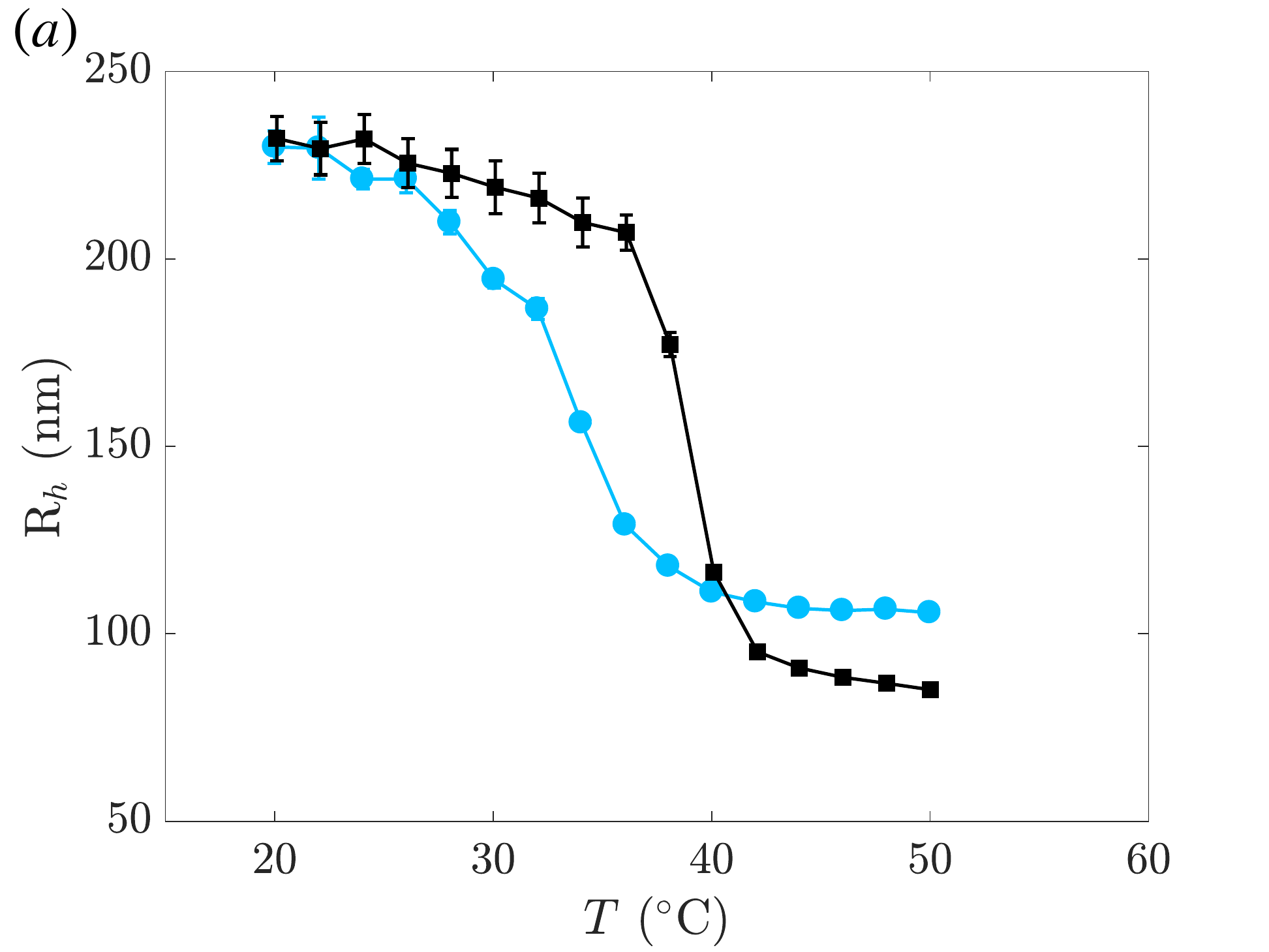}}
	\subfigure{\includegraphics[width=0.35\textwidth]{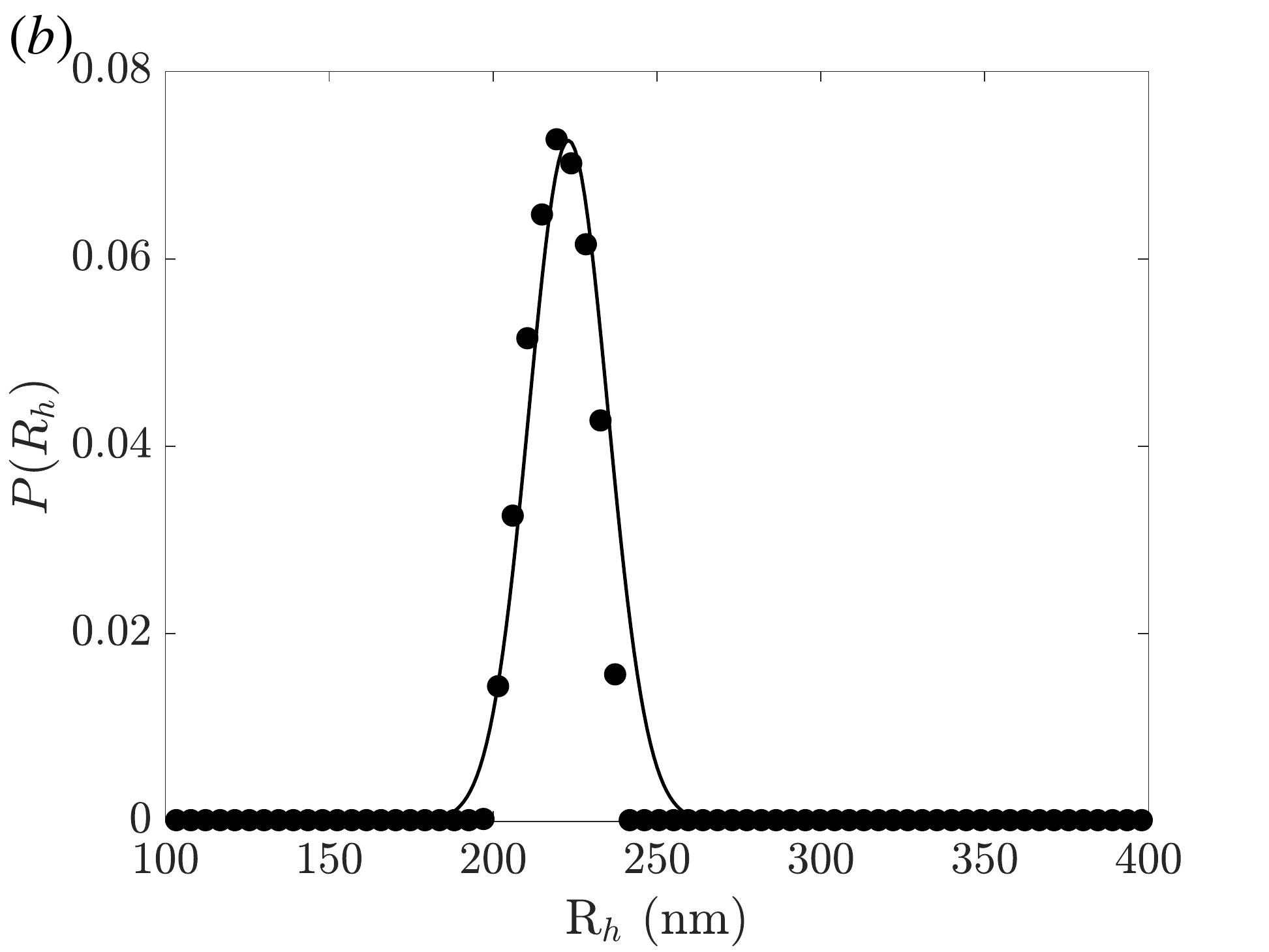}}
	\caption{(a) Hydrodynamic radius, $R_h$, versus temperature, $T$ for protonated and deuterated 5~mol\% crosslinked hollow microgels, light blue circles and black squares, respectively. (b) Size distribution, $P(R)$, of the hollow protonated microgels at 20~$^\circ$C and $\theta = 100^\circ$. The curve is a Gaussian fit to the data.
	}
	\label{fig:DLS}
\end{figure}

The dynamic light scattering (DLS) instruments used a mounted laser with vacuum wavelength $\lambda_0 = 633$~nm. We studied dilute solutions of the different microgels in water, with refractive index $n({\lambda_0}) = 1.33$. In the dilute regime, microgel-microgel interactions were negligible. The temperature was controlled using a thermal bath filled with toluene to match the refractive index of the glass. The scattering vector $q = (4\pi n/\lambda_0)\sin(\theta/2)$ was changed by varying the scattering angle, $\theta$, between 30 and 130 degrees, in steps of 5 degrees. 

Every intensity autocorrelation function acquired was analyzed with the second cumulant analysis \cite{Kop72}. After computing all the average decay rates, $\Gamma$, the relation, $\Gamma = D_0q^2$, linking $\Gamma$ to the scattering vector, $q$, was used to obtain the average diffusion coefficient, $D_0$, by fitting the data of $\Gamma$ versus $q^2$ with linear regression \cite{Bur89}. Finally, the Stokes-Einstein relation, $R_h =k_BT/(6\pi\eta_{H_2O} D_0)$, was used to obtain the hydrodynamic radii of the microgels in solution at different temperatures. The values of $R_h$ for the hollow protonated and hollow deuterated microgels in Fig.~\ref{fig:PB}(a)-(b) are plotted in Fig.~\ref{fig:DLS}(a) as a function of temperature (light blue circles and black squares, respectively).

The size distributions of the different microgel solutions were obtained by analyzing the intensity autocorrelation functions measured at different scattering angles using the Contin \cite{Pro82} algorithm modified according to Ref.~\cite{Sco15} to have a more robust capability to choose the regularizor parameter. The values for the radii and for the size polydispersity were obtained as the mean and the standard deviation (errors) of ten measurements taken at scattering angles between 30 and 130 degrees. An example of the outcome of the analysis for the hollow microgels in Fig.~\ref{fig:PB}(a) is shown in Fig.~\ref{fig:DLS}(b).

\begin{table*}[ht]
\caption{Characteristic lengths for the hollow and the regular microgels as obtained from the fit of the data with the core-fuzzy-shell model (section~\ref{subsec:model_FCS})\cite{Dub14} for the hollow microgels and with the fuzzy-shell model \cite{Sti04FF} for the regular microgels. All the measurements were conducted in pure D$_2$O at 20~$^\circ$C. All the samples have been measured by small-angle neutron scattering. The meanings of the different characteristic lengths correspond to the sketches in Fig.~\ref{fig:HS_sketch}. For the regular microgels, we identify $w_s$ with the extension of the core region and 2$\sigma_{ext}$ with the external fuzzy shell. $R_{SANS}$ is $w_\text{core}+2\sigma_\text{int}+ w_\text{s}+2\sigma_\text{ext}$ for the hollow microgels ($w_\text{s}+2\sigma_\text{ext}$ for the regular microgels). If a parameter obtained from the fit had an unrealistically low value (i.e., $\leq 2$~nm), it was considered zero. The data and the fit relative to MB-HS-60-5-PNIPAM and MB-HS-60-2.5-PNIPAM are reported in Ref.~\cite{Sco19a}.\\}
\label{tab:samples}
\begin{tabular}{ccccccc}
\hline\hline
\textbf{Sample name}            	&   {$R_{SANS}$}    			&   {$p$}           			&   {$w_{core}$}    	&   {$2\sigma_{int}$}   	&   {$w_s$}         		&   {2$\sigma_{ext}$}  	 	\\
\hline 
{MB-HS-105-5-D7PNIPAM}    	&   $212\pm7$~nm			& $15\pm3$~\%       		&   $89\pm2$~nm     &   $49\pm2$~nm	        &   $23\pm2$~nm 		&   $51\pm1$~nm        	 	\\
{MB-HS-105-5-PNIPAM} 	        &   $210\pm8$~nm    		& $14\pm1$~\% 		&   $91\pm4$~nm 	&   $57\pm2$~nm 	    	&   $10\pm1$~nm 		&   $52\pm1$~nm	       	 	\\
{MB-HS-60-5-PNIPAM} 	        &   $115\pm3$~nm 			&  $8.9\pm0.7$~\%   	&  $26\pm1$~nm     	&       - 	            		&  $14.8\pm0.8$~nm 	&    $74.4\pm0.9$~nm     		\\
{MB-HS-60-2.5-PNIPAM} 	        &   $147\pm4$~nm			& $7\pm1$~\% 	    		&   $24\pm2$~nm 	&   $34.6\pm0.5$~nm 	&   $18.5\pm0.3$~nm	&   $69.9\pm0.9$~nm     		\\
{MB-PNIPAM-5-225} 	        &   $209\pm4$~nm    		& $13.2\pm0.9$~\% 		&       - 	        		&       -    	        			&   $62\pm1$~nm	    	&   $147\pm3$~nm     		\\ 
\hline\hline
\end{tabular}
\end{table*}

\section{Small-Angle Neutron Scattering}

The small-angle neutron scattering (SANS) measurements were performed on D11 small-angle neutron scattering instrument at the Institut Laue-Langevin (ILL), Grenoble, France. Three configurations were used to cover the $q$-range of interest: $d_\text{SD} = 34\,$m with $\lambda =0.6\,$nm;  $d_\text{SD} = 8\,$m with $\lambda =0.6\,$nm; and $d_\text{SD} = 2\,$m with $\lambda =0.6\,$nm. Data were corrected by subtracting the dark count and background and by considering the instrument resolution due to the velocity selector, $\Delta\lambda/\lambda=9$~\%, and the fact that the instrument is equipped with a $^3$He detector with a pixel size $=7.5$~mm.

The solvent to realize the solution for SANS experiments with contrast matching of the D7-pNIPAM polymer was composed of 90~wt\% heavy water in a water/heavy water mixture. This solvent composition was already used to contrast match deuterated microgels realized with the very same monomer and was determined experimentally~\cite{Sco18, Sco19a}.

As a comparison to the deswelling of hollow protonated 5~mol\% crosslinked microgels embedded in a matrix of hollow deuterated 5~mol\% crosslinked microgels, we used small-angle neutron scattering (SANS) to measure the response of a few regular protonated 5~mol\% crosslinked microgels embedded in the very same matrix of hollow deuterated 5~mol\% crosslinked microgels as before. All the microgels used have comparable swollen sizes ($\approx 210$~nm), as listed in Table~\ref{tab:Synthesis}. Furthermore, the same amount of crosslinking agent was used to synthesize all the microgels used for the SANS experiments, i.e., all the polymeric networks have a comparable number of crosslinks and, therefore, softness.

\begin{figure}[h!]
	\subfigure{\includegraphics[width=0.37\textwidth ]{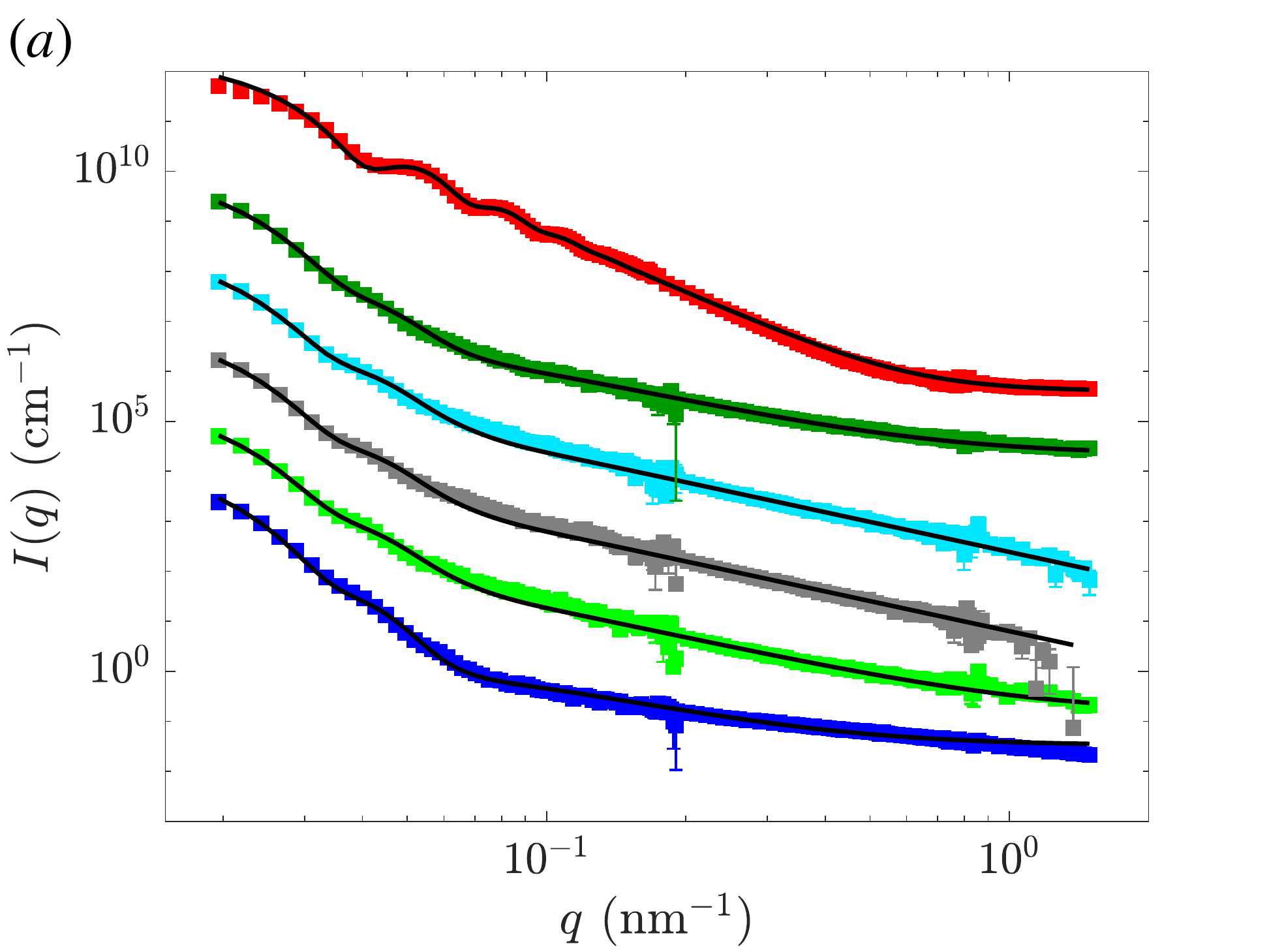}}
	\subfigure{\includegraphics[width=0.37\textwidth]{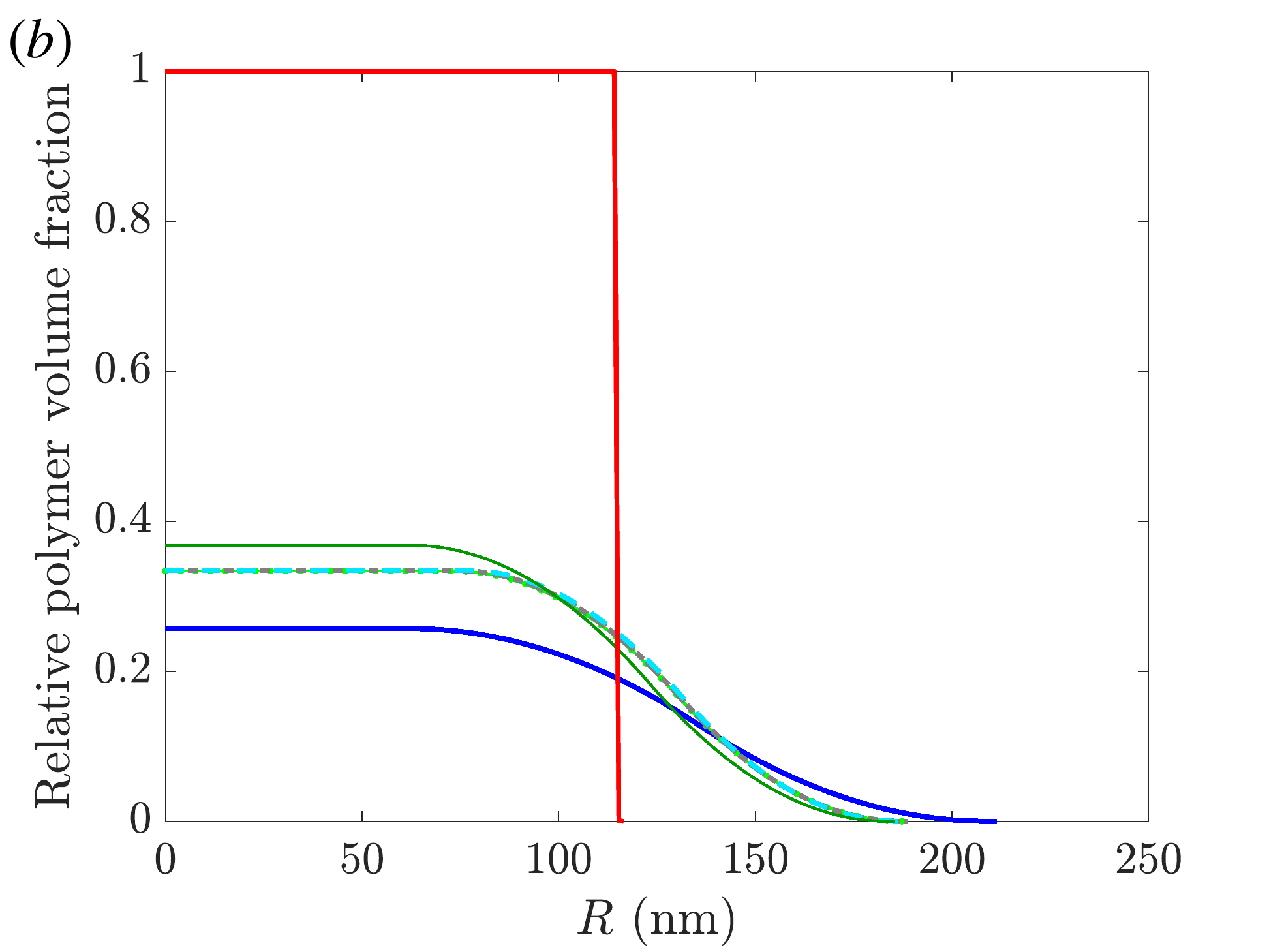}}
	\caption{(a) SANS intensities, $I(q)$, vs. scattering vector $q$, of the regular protonated 5~mol\% crosslinked microgels. (b) Radial distribution of the relative polymer volume fraction as obtained by the fits of the curves in (a) using the model of Ref.~\cite{Dub15}. In (a), the concentrations from bottom to top are: $\zeta = 0.080 \pm 0.003$; $0.30 \pm 0.01$; $0.65 \pm 0.02$; $0.87 \pm 0.02$; $1.19 \pm 0.03$; and $\zeta < 0.080 \pm 0.003$. The top curve is at $T = 40.0 \pm 0.1$~$^\circ$C, while all the other measurements are at $T = 20.0 \pm 0.1$~$^\circ$C. In (b) the colors and concentrations correspond to the colors in (a).
	}
	\label{fig:SANS_reg}
\end{figure}

\begin{figure*}[t]
	\subfigure{\includegraphics[width=0.32\textwidth ]{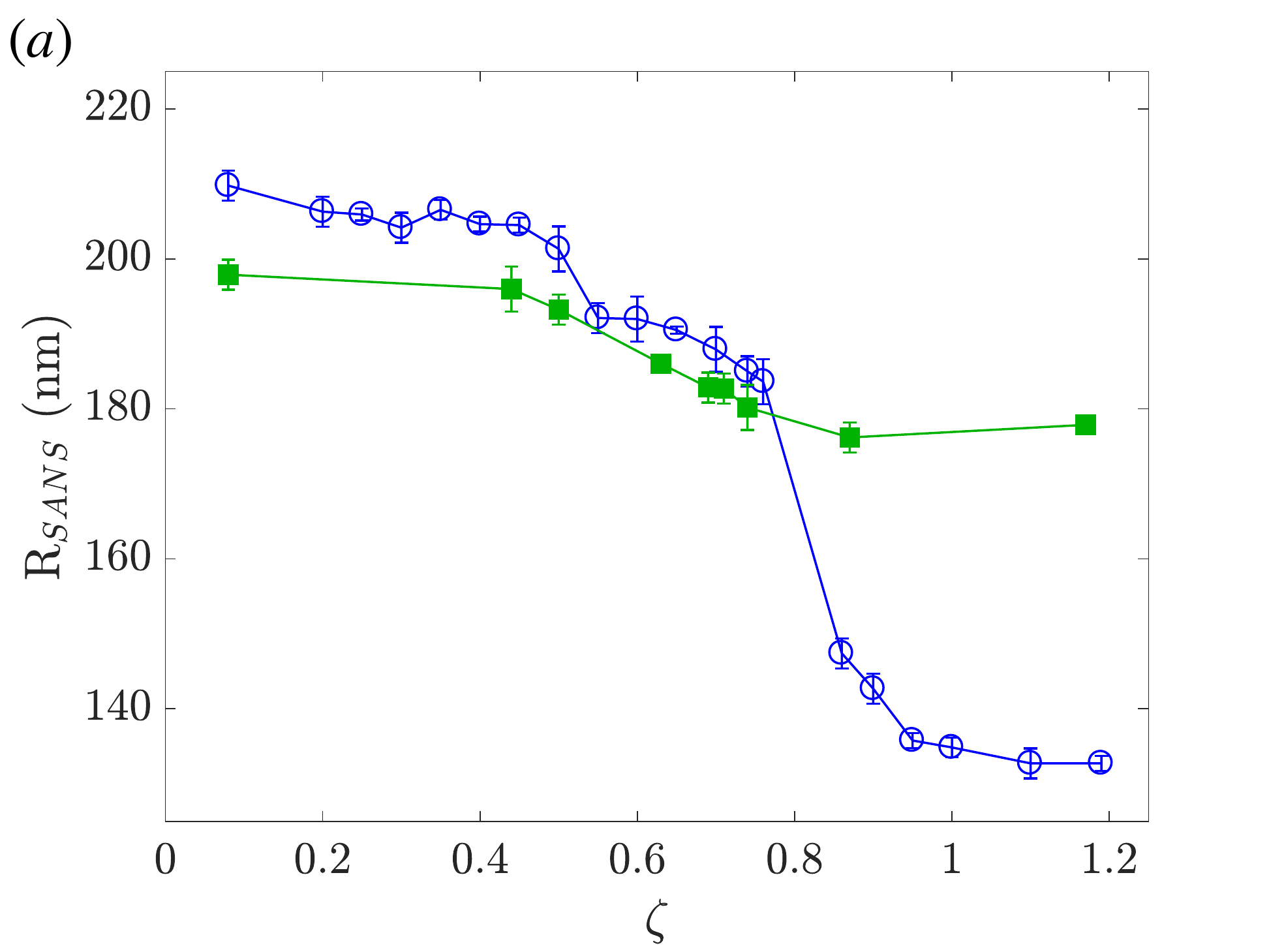}}
	\subfigure{\includegraphics[width=0.32\textwidth ]{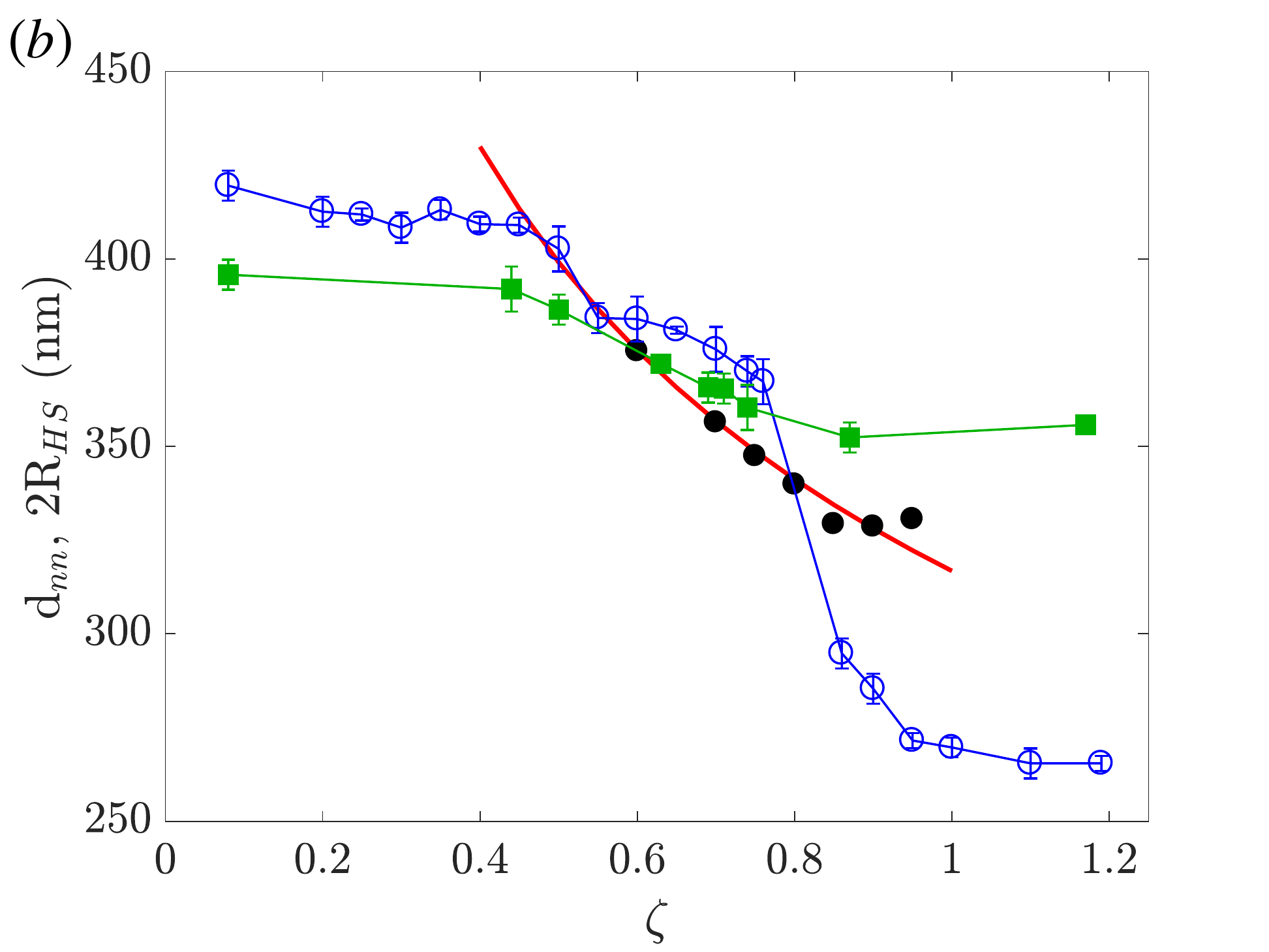}}
	\subfigure{\includegraphics[width=0.32\textwidth ]{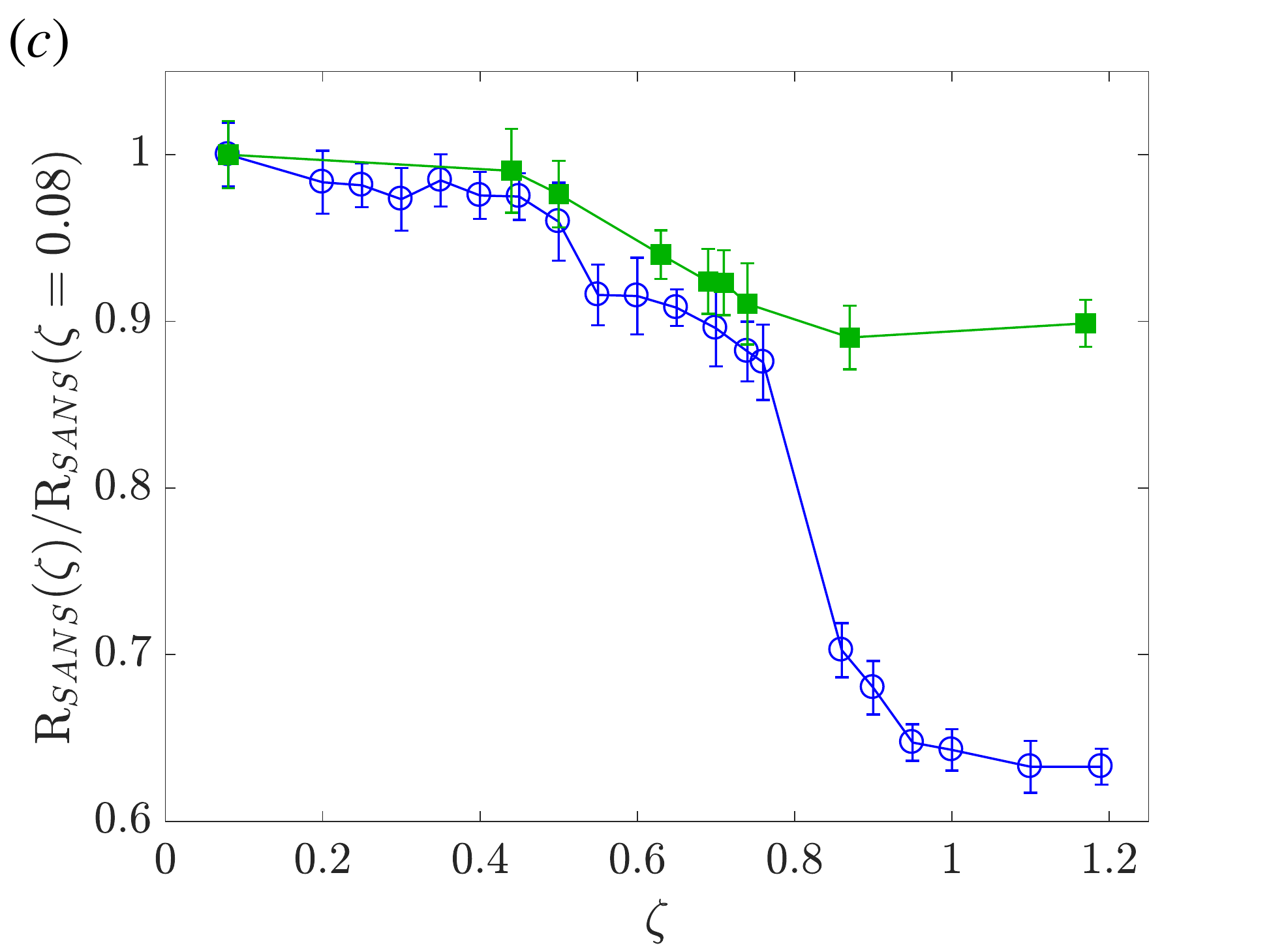}}
	\caption{Variation of the radius (a), and the diameter (b) as determined by SANS and the variation of the radius measured with SANS divided by the radius at $\zeta = 0.080 \pm 0.003$ and $T = 20.0 \pm 0.01\,^{\circ}$C (c) as a function of the total generalized volume fraction, $\zeta$, for: hollow protonated 5~mol\% crosslinked microgels (blue empty circles) and regular 5~mol\% crosslinked microgels with an almost homogeneous core surrounded by a fuzzy shell (green solid squares). In (b) the black solid circles represent the values of the nearest-neighbor distances as determined by SAXS. The red curve is a fit of the  nearest-neighbor distance with a power law of $a\zeta^{-1/3}$.
	}
	\label{fig:R_vs_Z_SI}
\end{figure*}

The results of these experiments, SANS scattered intensities vs.~scattering vector, are plotted in Fig.~\ref{fig:SANS_reg}(a). As seen in Fig.~\ref{fig:SANS_reg}(b), the radial distributions of the protonated 5~mol\% crosslinked microgels with an almost homogeneous core surrounded by a fuzzy shell change only slightly with increasing $\zeta$. In Fig.~\ref{fig:R_vs_Z_SI}, the results for the total size, $R_{SANS}$, are summarized and compared to the corresponding values obtained for the hollow protonated 5~mol\% crosslinked microgels, blue empty circles and green solid squares, respectively.

%\newpage

\subsection{Form Factor Model: Core-Fuzzy-Shell Model}\label{subsec:model_FCS}

\begin{figure}[h!]
	\subfigure{\includegraphics[width=.5\textwidth ]{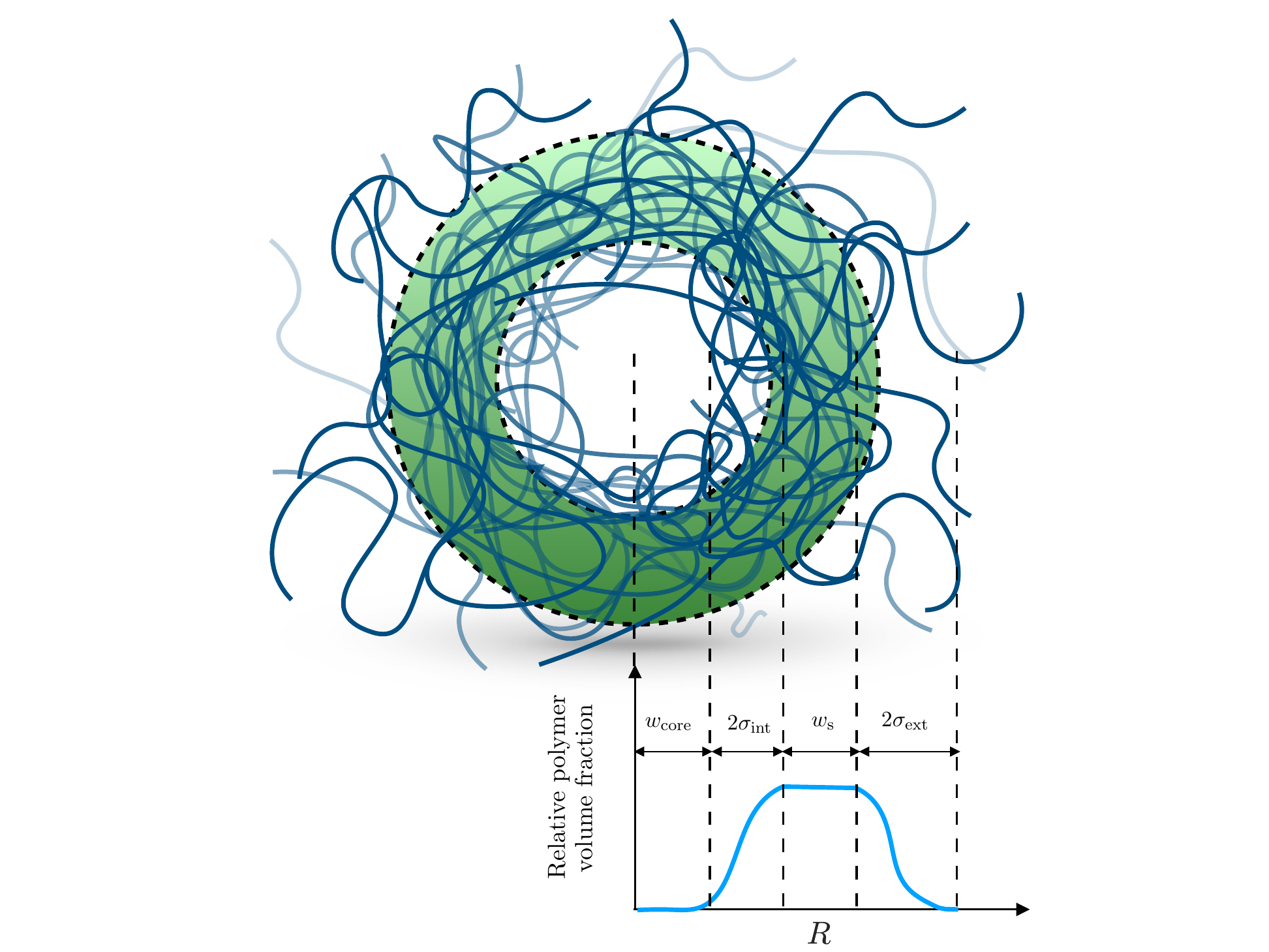}}
	\caption{Sketch of a hollow microgel and of the relative polymer volume fraction density profile with the corresponding characteristic length scales.}
	\label{fig:HS_sketch}
\end{figure}

The core-fuzzy-shell model was used to fit the scattering data of the hollow microgels \cite{Ber06}. In the following, the characteristic length scales for the model can be seen in the sketch of Fig.~\ref{fig:HS_sketch}. The model describes a fuzzy core-shell structure with an interpenetrating layer of core and shell of the length $2\sigma_\text{int}$ and a fuzzy outer surface with an extension $\sigma_\text{ext}$. The widths of the core and shell boxes are labeled as $w_\text{core}$ and $w_\text{s}$. The scattering amplitude of a core-shell particle $A(q)$ is expressed as
	\begin{align}
		A(q)=&\ \Delta\rho_\text{s}V_\text{s}\mathnormal{\mathnormal{\Phi}}_{s}(q,R_\text{ext},\sigma_\text{ext})\nonumber
		\\&+(\Delta\rho_\text{core}-\Delta\rho_\text{s})V_\text{core}\mathnormal{\mathnormal{\Phi}}_\text{core}(q,R_\text{int},\sigma_\text{int})\,,
		\label{Aq}
	\end{align}
where $\Delta\rho$ is the difference between the scattering length density of the solvent and the core (or the shell) and $V_\text{core}$ and $V_\text{s}$ are the volumes of the core and the shell, respectively. The radii are defined as $R_\text{int} = w_\text{core}+\sigma_\text{int}$ and $R_\text{ext}~=~w_\text{core}+2\sigma_\text{int}+ w_\text{s}+\sigma_\text{ext}$, while
$\mathnormal{\Phi}(q,R,\sigma)$ represents the normalized Fourier transform of the radial density profile
	\begin{align}
	\mathnormal{\Phi}(q,R,\sigma) =\ & \frac{1}{V_n} \Big[ \left( \frac{R}{\sigma^2}+\frac{1}{\sigma} \right)\frac{\cos[q(R+\sigma)]}{q^4}\nonumber\\
& + \left( \frac{R}{\sigma^2}-\frac{1}{\sigma} \right) \frac{\cos[q(R-\sigma)]}{q^4} -\frac{3\sin[q(R-\sigma)]}{q^5\sigma^2}\nonumber\\\
& -\frac{2R\cos(qR)}{q^4\sigma^2}+\frac{6\sin(qR)}{q^5\sigma^2}\Big]\,,  
\label{phi}
 	\end{align}
with $V_n=R^3/3+R\sigma^2/6$ and $q$ being the scattering vector \cite{Ber06, Bru18}. Finally, the form factor, $P(q)$, is proportional to the scattering length amplitude squared:
	\begin{align}
		P(q)\,\alpha\,A^2(q)\,.
	\end{align}

In the literature, it is common to express $R_{SANS} = w_\text{core}+2\sigma_\text{int}+ w_\text{s}+2\sigma_\text{ext}$ \cite{Ber06, Deb03, Sti04FF, Sco16}. The size polydispersity is accounted for by convolution of the form factor with a Gaussian of width $2\sigma_{poly}$. The size polydispersity is usually expressed as $p = 2\sigma_{poly}/R_{SANS}$. The scattering at high $q$-values resulting from inhomogeneities in the polymer network due to the crosslinking are accounted by a Lorentzian term, which is added to the fitting models: $I_\text{chain}(0)/(1+q^2\xi^2)$ with $I_\text{chain}(0)$ being the value of the scattered intensity due to the chain at $q=0$ and and $\xi$ the average mesh size of the polymeric network \cite{Sti04, Fer02}. A constant background is added to account for the incoherent scattering.
Finally, the model is convoluted with a resolution function to account for the smearing due to the instrument \cite{Ped90}.\\

The value of $2\sigma_{int}$ determined from the fit was below the resolution of SANS for the sample MB-HS-60-5-PNIPAM, therefore, we decided to set it to zero.

The reason why sample MB-HS-60-5-PNIPAM do not show internal fuzziness is that the internal interface of the cavity is, before the dissolution of the silica, connected to the rigid core. This can lead to different density within the polymeric network and consequent higher crosslinking of the microgels in this region. Once the silica core is dissolved, the polymeric network maintains a different density and/or crosslinking and this can hinder the internal fuzziness leading to the observed value  $2\sigma_{int} = 0$. A similar behavior has been reported in the literature for similar hollow-shell and hollow-double-shell microgels \cite{Dub14, Sch16}.

\subsection{Data Fits}

\begin{figure}[h!]
	\subfigure{\includegraphics[width=0.35\textwidth ]{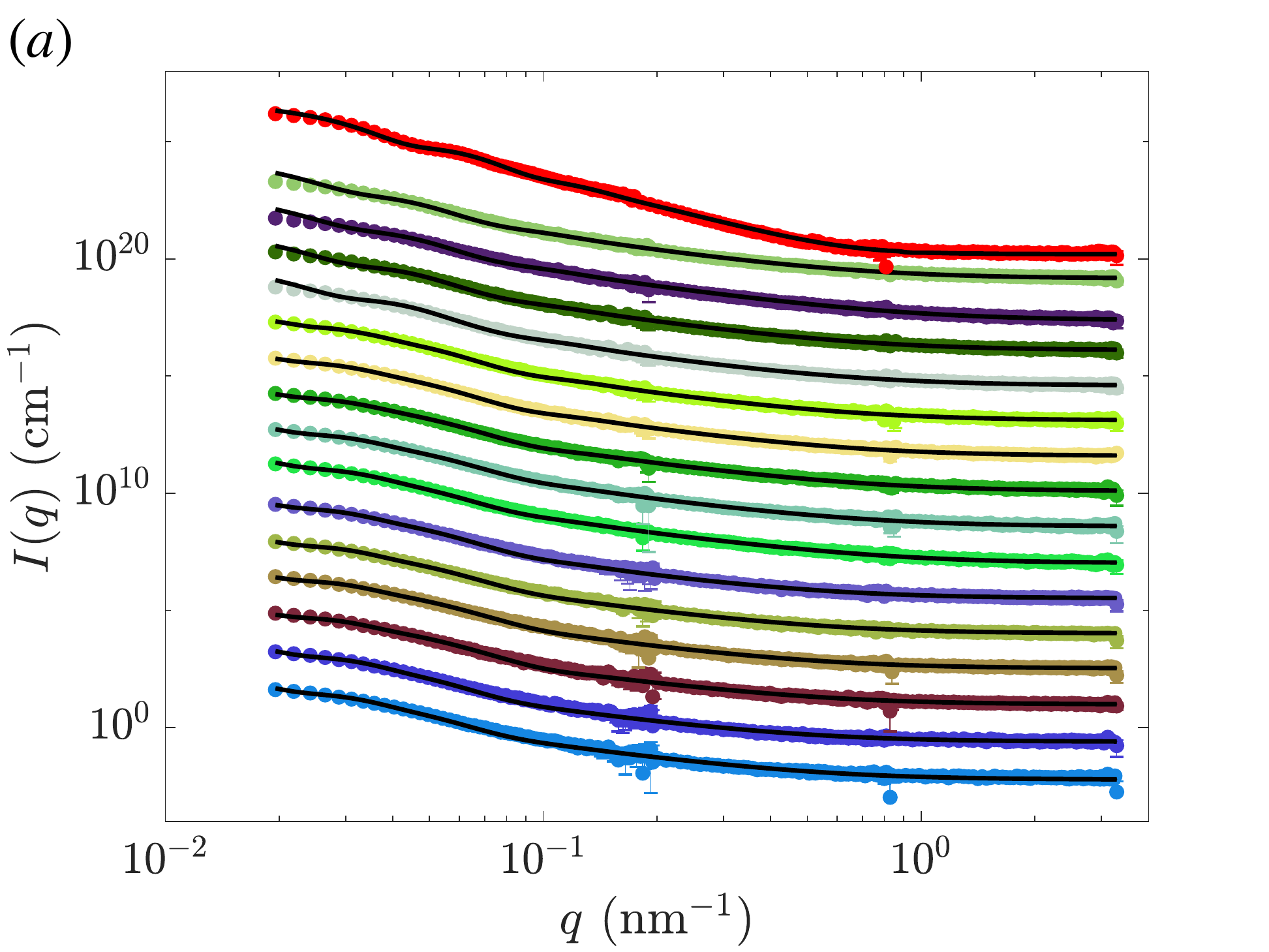}}
	\subfigure{\includegraphics[width=0.35\textwidth]{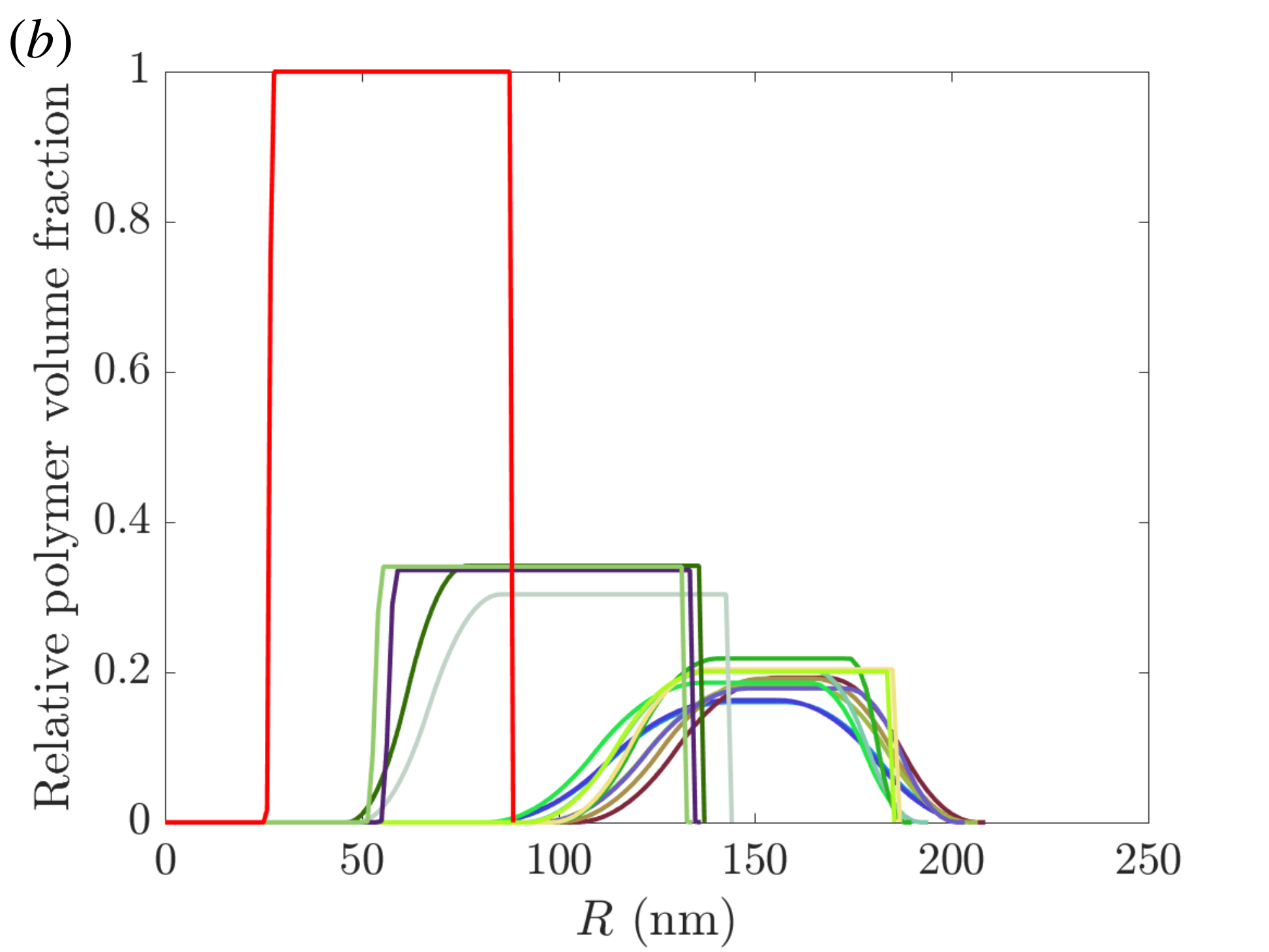}}
	\caption{(a) SANS intensities, $I(q)$, vs. scattering vector $q$, of the hollow protonated 5~mol\% crosslinked microgels. (b) Radial distribution of the relative polymer volume fraction as obtained by fitting the curves in (a) using the model of Ref.~\cite{Dub15}. In (a), the concentrations from bottom to top are: $\zeta = 0.20 \pm 0.01$; $0.25 \pm 0.01$; $0.35 \pm 0.01$; $0.40 \pm 0.01$; $0.45 \pm 0.01$; $0.50 \pm 0.01$; $0.55 \pm 0.01$; $0.60 \pm 0.01$; $0.70 \pm 0.01$; $0.74 \pm 0.01$; $0.76 \pm 0.02$; $0.90 \pm 0.02$; $0.95 \pm 0.02$; $1.00 \pm 0.02$; $1.10 \pm 0.02$; and $\zeta < 0.080 \pm 0.003$. The top curve is at $T = 40.0 \pm 0.1$~$^\circ$C, while all the other measurements are at $T = 20.0 \pm 0.1$~$^\circ$C. In (b) the colors and, therefore, the concentrations correspond to the colors in (a).}
	\label{fig:HS_SANS_SI}
\end{figure}

%The small-angle neutron scattering (SANS) measurements were performed on D11 small-angle neutron scattering instrument at the Institut Laue-Langevin (ILL), Grenoble, France. Three configurations were used to cover the $q$-range of interest: $d_\text{SD} = 34\,$m with $\lambda =0.6\,$nm;  $d_\text{SD} = 8\,$m with $\lambda =0.6\,$nm; and $d_\text{SD} = 2\,$m with $\lambda =0.6\,$nm. Data were corrected subtracting the dark count, the background and considering the instrument resolution due to the velocity selector, $\Delta\lambda/\lambda=9$~\%, and the fact that the instrument is equipped with a $^3$He detector with a pixel size $=7.5$~mm.

%The solvent to realize the solution for SANS experiments with contrast matching of the pD7-pNIPAM polymer was composed of 90~wt\% heavy water in a water/heavy water mixture. This solvent composition was already used to contrast match deuterated microgels made from the D7-NIPAM monomer realized with the very same monomer and was determined experimentally \cite{Sco18, Sco19a}.

Above the VPTT, the hollow protonated microgels collapse to $88\pm2$~nm and no fuzziness is detected (red curve in Fig.~\ref{fig:HS_SANS_SI}). The internal cavity shrinks with respect to the swollen state to a radius of $25.9 \pm 0.8$~nm, as already reported both from SANS experiments and molecular dynamics simulations \cite{Dub14,Dub15,Sch16,Bru18,Sco18}.

The other curves in Fig.~\ref{fig:HS_SANS_SI} are the SANS form factors of the hollow 5 mol~\% crosslinked protonated microgels measured with the tracing method at different concentrations (a) and the radial distributions of the relative polymer volume fraction within the microgels as obtained by fitting the data with the core-fuzzy-shell model (b). All the curves in Fig.~\ref{fig:HS_SANS_SI} are measured at 20~$^\circ$C, except for the red curves, which correspond to 40~$^\circ$C. 

We fixed the value of the volume fraction of $w_s$ (Fig.~\ref{fig:HS_sketch}) to 1 during the fits. The data were then normalized to the area under the radial distribution of the collapsed state since: (i) above the VPTT the microgels have a more defined profile with higher contrast due to the collapsed polymer and (ii) the mass of the polymer is conserved. The graphs in Fig.~\ref{fig:HS_SANS_SI}(b) show that in the swollen state the polymer occupies $\approx 40$ and $20~\%$ of the volume occupied in the collapsed state. As reported by Dubbert et al.~\cite{Dub14}, as well as more recently by Schmid et al.\cite{Sch16}, SANS shows that the polymer volume fraction within a microgel well above the VPTT is $\approx 75\%$. This property can be considered valid also for our microgels, since they have been synthesised with similar protocols and are based on pNIPAM as well. Thus, in the swollen state the volume fraction of polymer in the core is $\approx 20\%$. Also, this value agrees with the values reported in the literature \cite{Dub14, Sch16}.

\begin{figure}[h!]
	\subfigure{\includegraphics[width=0.35\textwidth ]{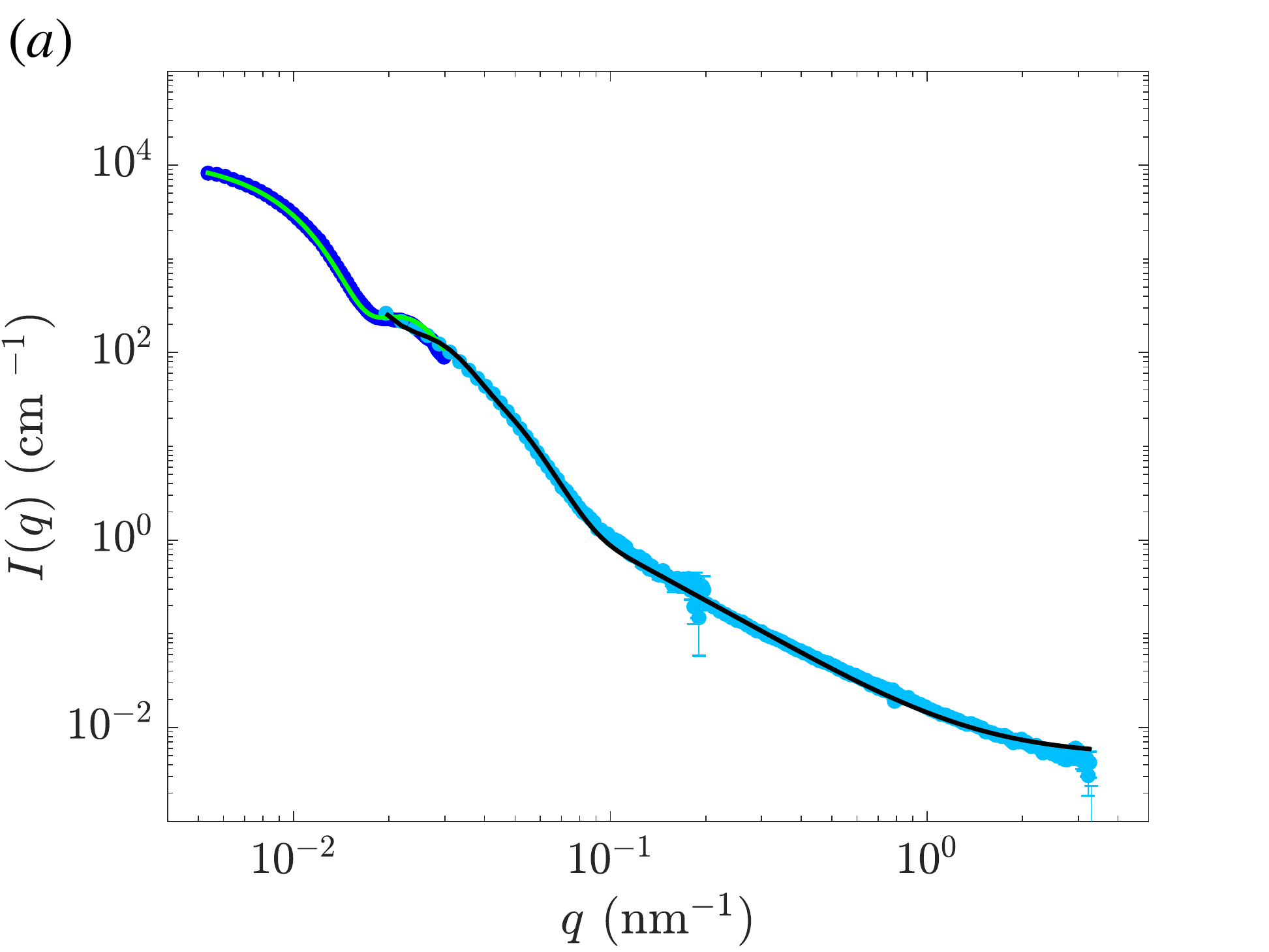}}
	\subfigure{\includegraphics[width=0.35\textwidth]{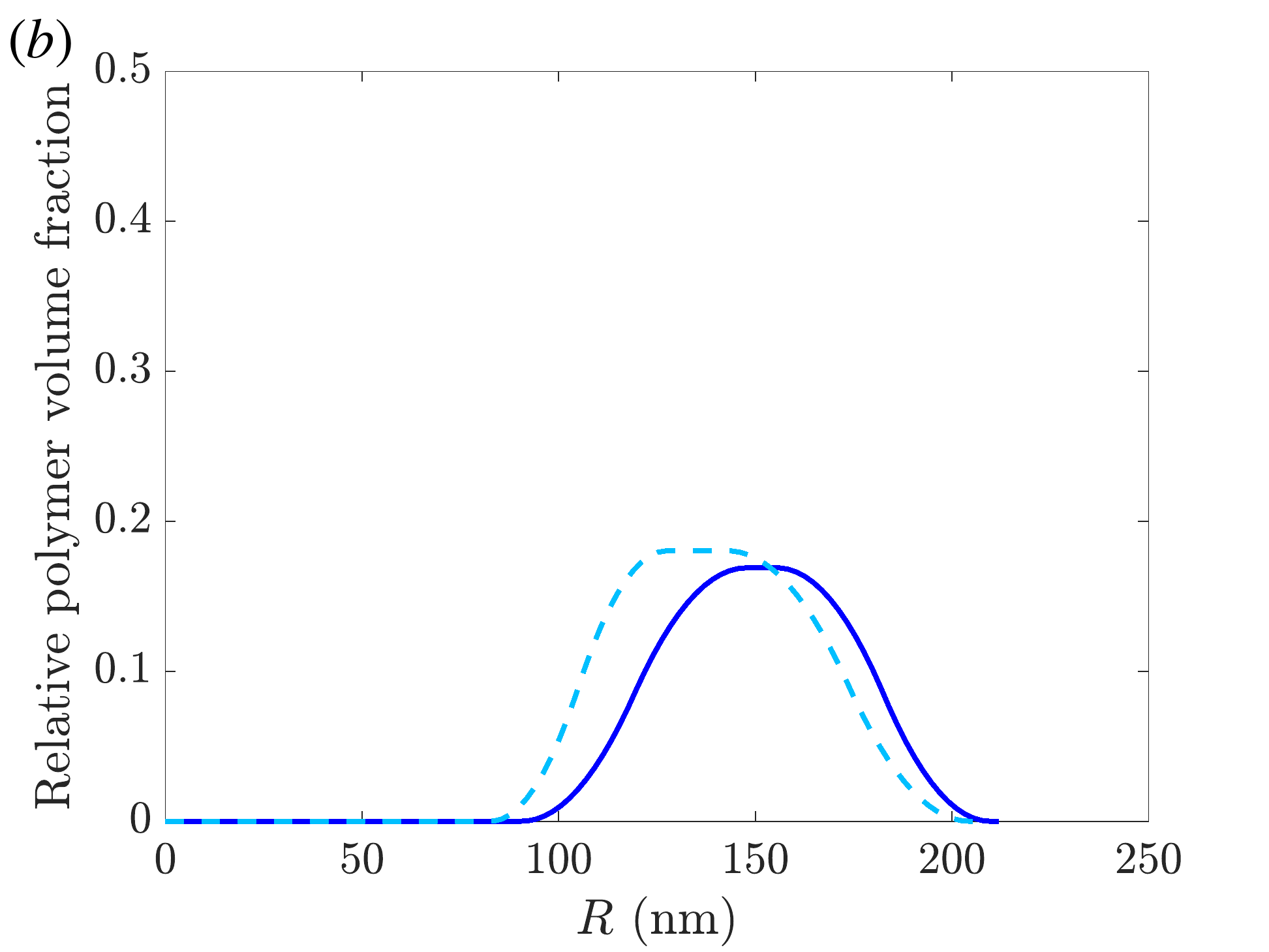}}
	\caption{(a) SANS and SLS scattered intensities of the hollow protonated 5~mol\% crosslinked microgels, light blue and blue circles, respectively. The solid curves represent the data fits using the model in Ref.\ \cite{Dub15}. The sample had a $\zeta = 0.080 \pm 0.003$ and a temperature $T = (20.0\pm0.01)\,^{\circ}C$. (b) Radial distribution of the relative polymer volume fraction within the microgel for: $\zeta = 0.080 \pm 0.003$ and $T = 20.0 \pm 0.01\,^{\circ}C$ as obtained from the SANS and SLS data fit, solid and dashed lines, respectively.
	}
	\label{fig:SLS}
\end{figure}

Figure \ref{fig:SLS}(a) shows the scattered intensities obtained from SANS and SLS of the hollow protonated 5~mol\% crosslinked microgels at $\zeta = 0.08\pm0.01$ combined together, light blue and blue circles, respectively. The radial distribution of the relative polymer volume fractions within the microgels are reported in Fig.~\ref{fig:SLS}(b).

\begin{figure}[h!]
	\subfigure{\includegraphics[width=0.35\textwidth ]{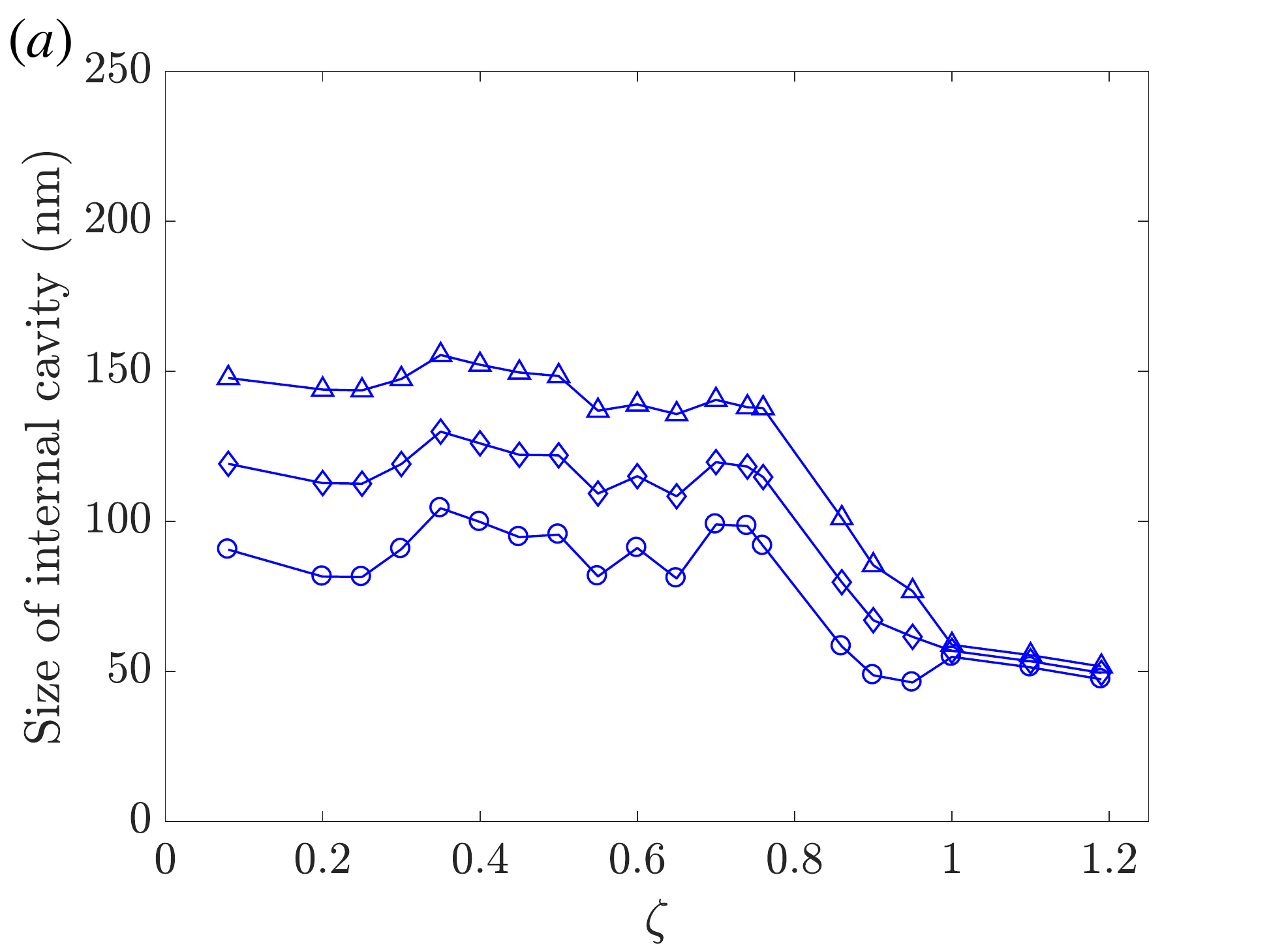}}
	\subfigure{\includegraphics[width=0.35\textwidth]{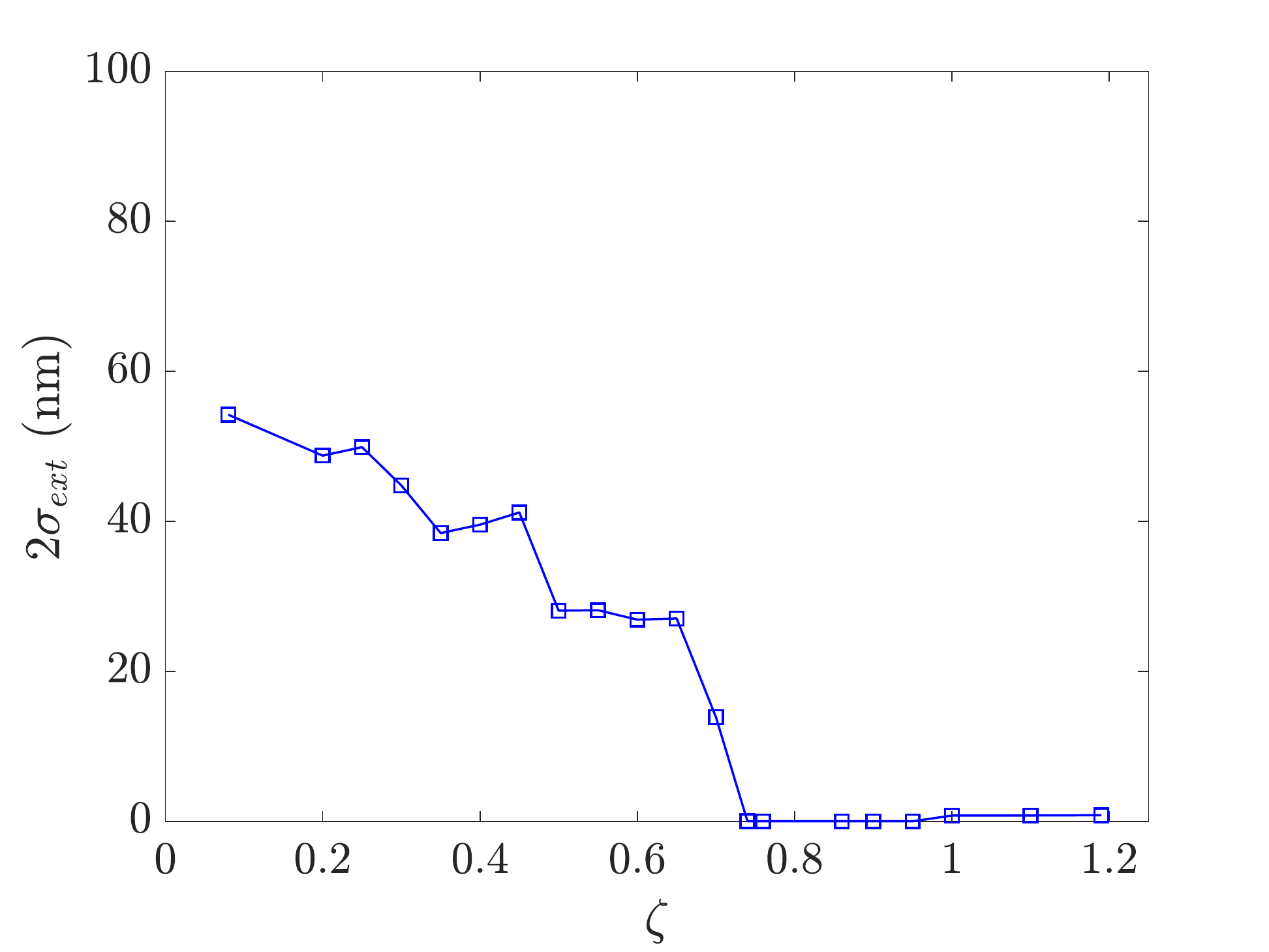}}
	\subfigure{\includegraphics[width=0.35\textwidth]{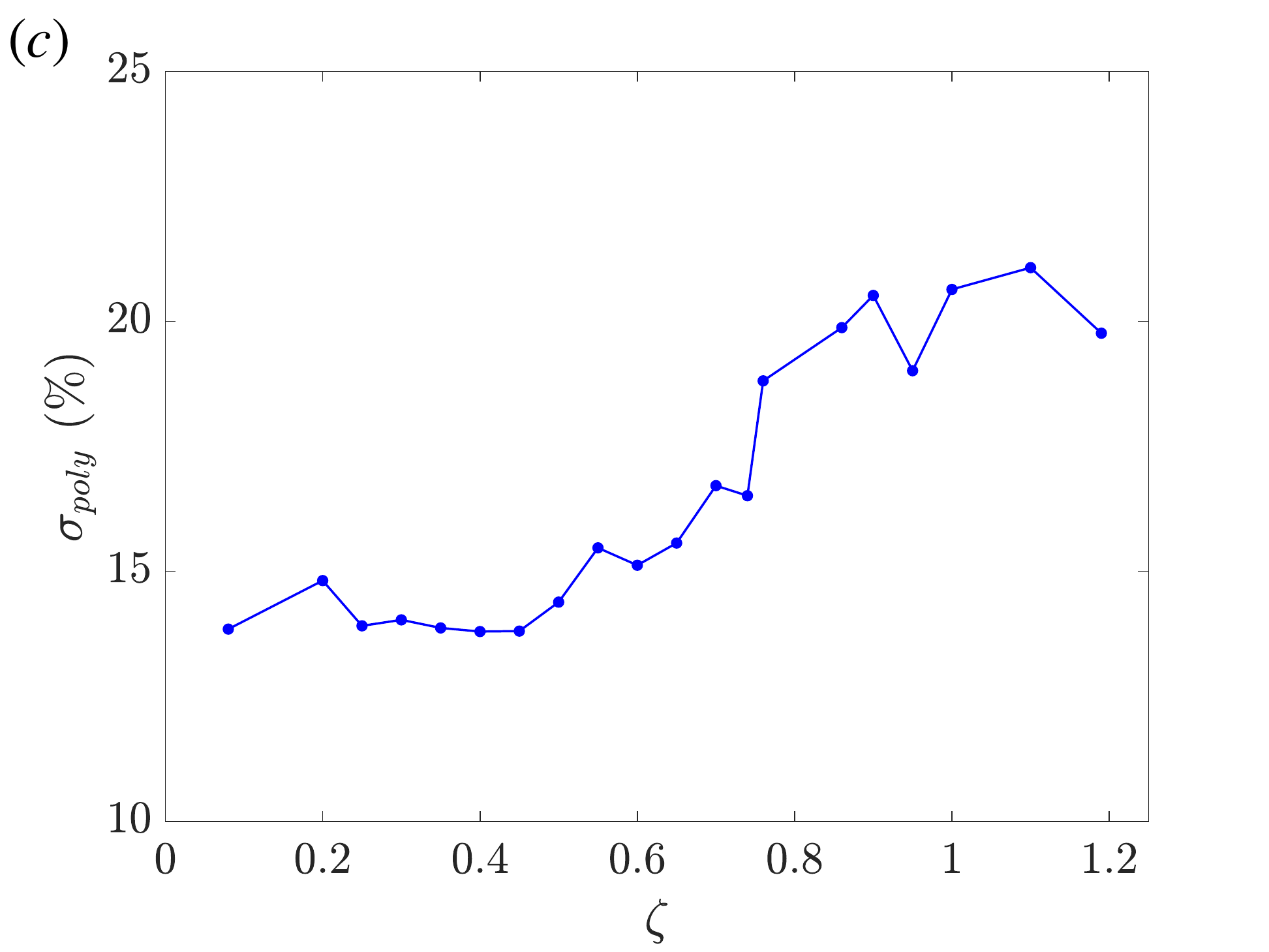}}
	\caption{(a) Extension of the cavity versus $\zeta$. The width of the cavity is computed as  $w_{core}$ (empty circles), $w_{core}+\sigma_{int}$ (empty diamonds), $w_{core} + 2\sigma_{int}$ (empty triangles). (b) Fuzzy shell width, $2\sigma_{ext}$ versus $\zeta$. (c) Size polydispersity, $\sigma_{poly}$, versus $\zeta$. 
	}
	\label{fig:parameters}
\end{figure}

The variation of the characteristic length scales of a hollow microgel (corresponding to the sketch in Fig.~\ref{fig:HS_sketch}) as obtained by fitting the data in Figs.~2 and \ref{fig:HS_SANS_SI} with the model described in \ref{subsec:model_FCS} are reported in Fig.~\ref{fig:parameters}. Since it is challenging to define the extension of the internal cavity, due to the fuzzy surface of the microgels, three different way are proposed in Fig.~\ref{fig:parameters}(a): $w_{core}$ (empty circles), $w_{core}+\sigma_{int}$ (empty diamonds), $w_{core} + 2\sigma_{int}$ (empty triangles). In the main text, we used as radius of the cavity the values of $w_{core}$.  

We synthesized four different types of hollow microgels using either 5 or 2.5~mol\% of crosslinker to study the phase behavior of hollow microgels. As can be seen in Table~\ref{tab:samples}, at least two of the hollow microgels have a size polydispersity $\ll 10$\%, MB-HS-60-5-PNIPAM and MB-HS-60-2.5-PNIPAM, respectively. These values are well below the value of size polydispersity suppressing crystallization not only for regular microgels but also for hard sphere. Crystallization is hindered even for these hollow microgels with low polydispersity, which, akin to regular microgels, are soft deformable spheres, for which crystals are expected for $\zeta$ between 0.51 and 0.61 since their size polydispersity is $\ll 10$\%.

Then, to investigate the response of hollow microgels to crowding, deuterated hollow microgels have been synthesized to be used in SANS with contrast variation. The synthesis of hollow microgels is a multi-step procedure, which requires considerable effort, therefore, once we have characterized the deuterated hollow microgels, MB-HS-105-5-D7PNIPAM, we have chosen the sample with the closest internal structure and total size,  MB-HS-105-5-PNIPAM, to be studied further with SANS. The sample MB-HS-105-5-PNIPAM has a relatively high polydispersity that increases with $\zeta$, i.e., with crowding. 

Nevertheless, in Figure~\ref{fig:parameters}(c), the value of the size polydispersity is $\lesssim 16.5$\% for $\zeta\lesssim 0.75$, a value well above the freezing and melting point of both hard spheres ($\phi_f = 0.49$ and $\phi_m = 0.52$) and regular microgels ($\zeta_f = 0.56$ and $\zeta_m \lesssim 0.65$). Hollow microgels should crystallize in a similar range of concentration since in this region, for $\zeta<0.75$, their nominal polydispersity is smaller than the limit reported in Ref.~\cite{Sco17}. Above $\zeta = 0.75$, the value of the polydispersity is so high that crystallization is suppressed even for regular microgels. We think that, in any case, at such high packing fractions the microgels are jammed and that rearrangement is very difficult independent of the polydispersity.
%In Figure~\ref{fig:parameters}(c), the value of the size polydispersity is $\lesssim 16.5$\% for $\zeta\lesssim 0.75$, a value well above the freezing and melting point of both hard spheres ($\phi_f = 0.49$ and $\phi_m = 0.52$) and regular microgels ($\zeta_f = 0.56$ and $\zeta_m \lesssim 0.65$). Hollow microgels should crystallize in a similar range of concentration since in this region, for $\zeta<0.75$, their nominal polydispersity is smaller than the limit reported by Scotti et al.~\cite{Sco17}. Above $\zeta = 0.75$, the value of the polydispersity is so high that crystallization is suppressed even for regular microgels. We think that, in any case, at such high packing fractions the microgels are jammed and rearrangement is very difficult independent of the polydispersity.\\ 

%\noindent Also, the course of the parameter describing the polydispersity has not been studied by Scotti et al.~in the above mentioned work \cite{Sco17}, and similarly in the study of Scotti et al.\cite{Sco16}. Therefore, it is not possible to state what is the value of the polydispersity of binary or polydisperse solutions when these crystallize and it is not clear whether the trend is different with respect to our study.

\begin{figure}[h!]
	\subfigure{\includegraphics[width=0.35\textwidth ]{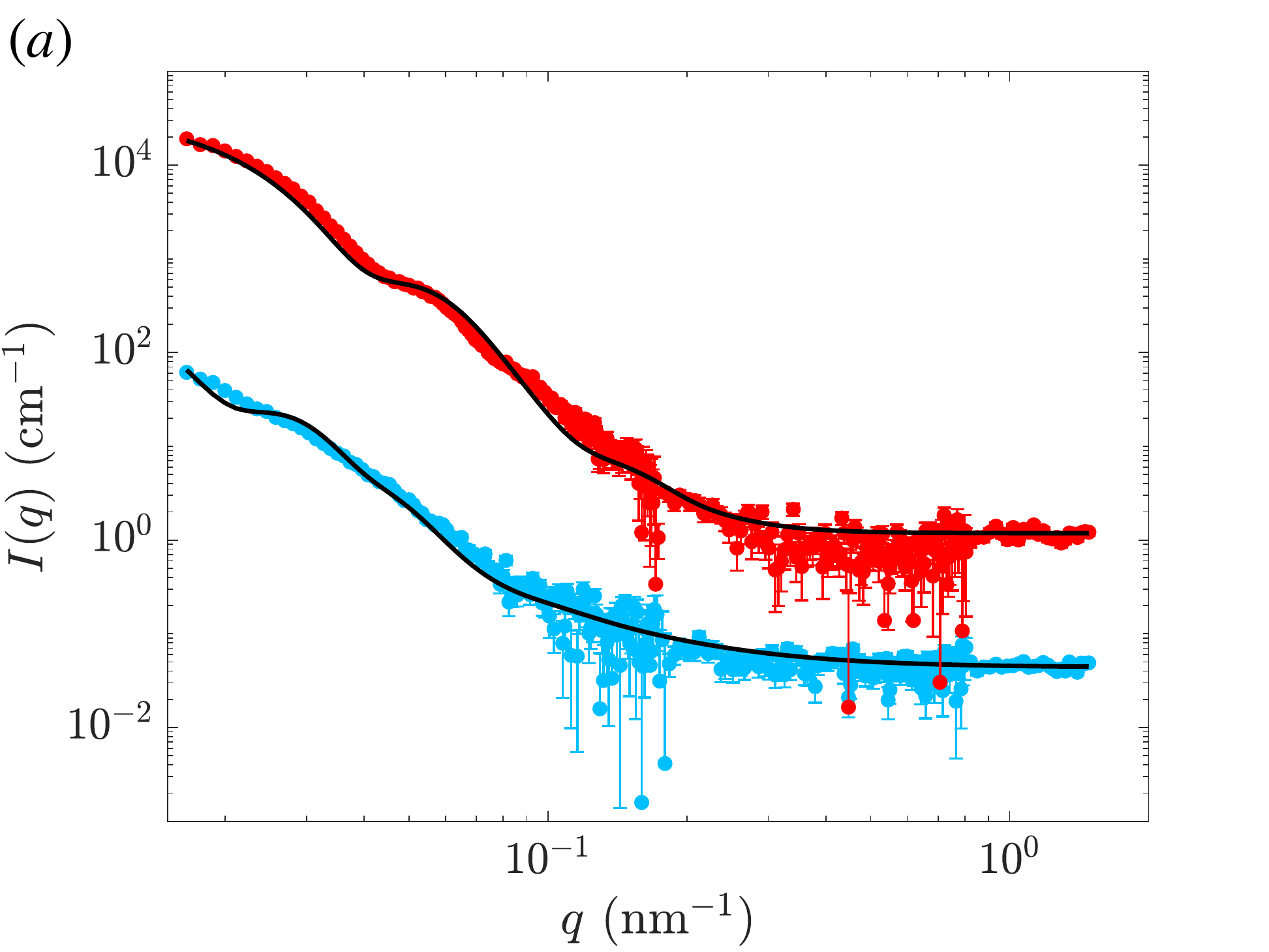}}
	\subfigure{\includegraphics[width=0.35\textwidth ]{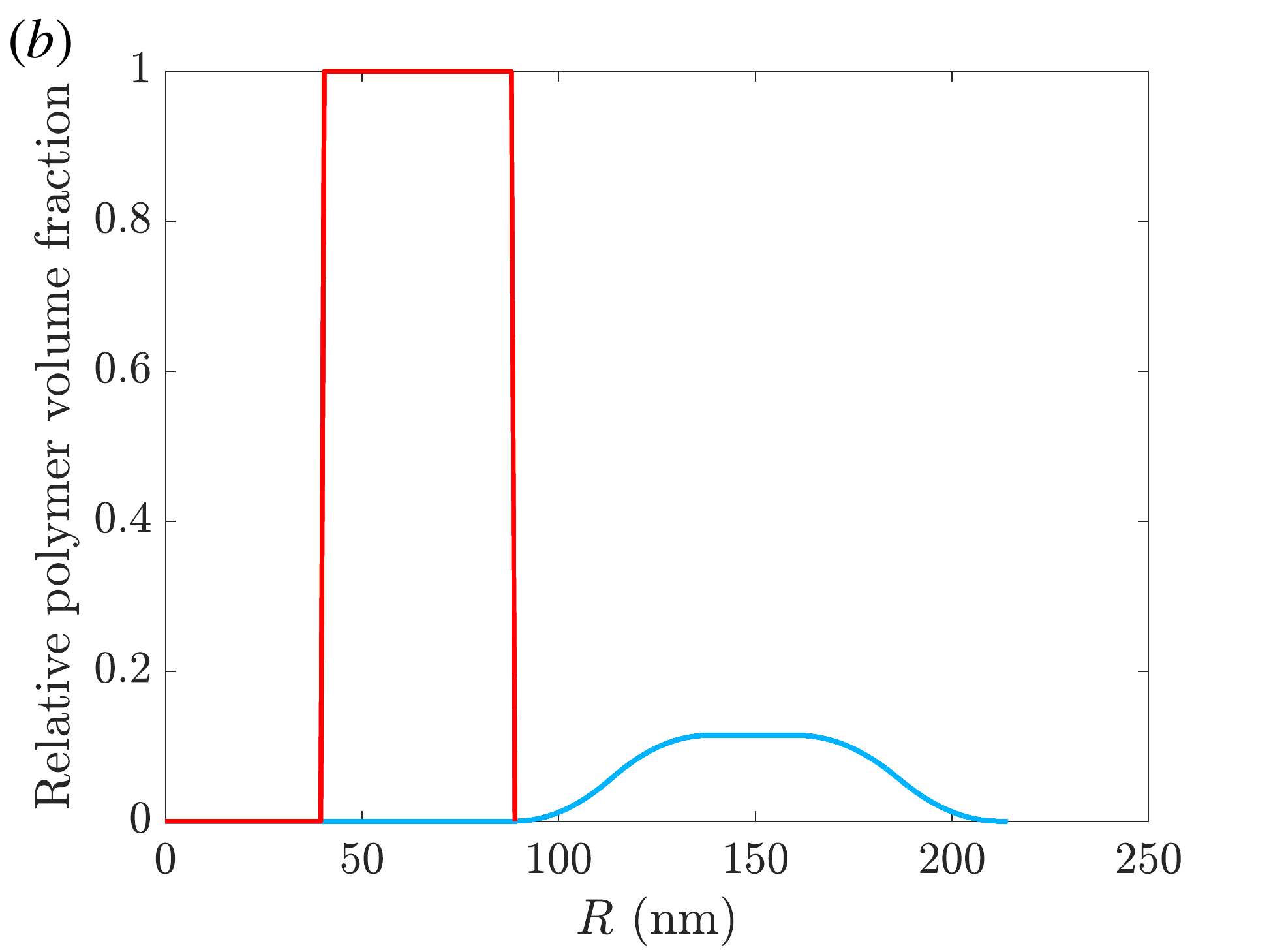}}
	\caption{(a) SANS intensities, $I(q)$, vs. scattering vector $q$, for the deuterated 5 mol~\% crosslinked hollow microgels below (blue circles) and above (red circles) the VPTT, at 20 and 40~$^\circ$C, respectively. The solid black curves represent the fit of the data with the model in Ref.~\cite{Dub15}. (b) Radial distributions of the relative polymer volume fraction as obtained by the fits of the curves in (a) using the model of Ref.~\cite{Dub15}. The sample concentration is $\zeta = 0.081 \pm 0.005$.}
	\label{fig:d7HS_SLS}
\end{figure}

Figure~\ref{fig:d7HS_SLS} shows the form factors (a) and the resulting radial distributions of the relative polymer volume fraction within the microgel as obtained by the fits of the curves in (a) using the model of Ref.~\cite{Dub15} (b). As can be seen, the deuterated hollow microgels maintain their cavity both below and above the VPTT, at 20 and 40~$^\circ$C, respectively. The total radius, $212 \pm 7$~nm, at  20~$^\circ$C. Above the VPTT, the radius is $88\pm5$~nm.

\subsection{Anisotropic deformation of the hollow microgels}

To check if the deformation of the hollow microgels, especially at high packing fractions, is anisotropic we tried to fit the data using a model for a hollow spheroid (i.e.~ an ellipsoid with $R_x = R_y \neq R_z$). We define the radius in the  $x$-$y$-plane (equatorial lane) $R_x = R_y = R_{tot, eq}$ and the radius in the  $z$-direction (polar plane) $R_z = R_{tot, pol}$.

This model for an empty ellipsoidal shell has as parameters the radius of the cavity of the ellipsoid in the $x$-$y$-plane, $R_{eq}$, and the length of the shell (without fuzziness) in the $x$-$y$-plane, $D_{eq}$. The total principal axis of the ellipsoid in the $x$-$y$-plane is $R_{tot, eq} = R_{eq} + D_{eq}$. Additional parameters of the fit are the ratios  $a = R_{pol}/R_{eq}$ and $b = D_{pol}/D_{eq}$, where $R_{pol}$ and $D_{eq}$ are the radius of the cavity and the length of the shell (without fuzziness) in the $z$-plane, respectively. From the values of these fitting parameters, the polar radius of the cavity of the ellipsoid, $R_{pol} = aR_{eq}$, and the length of the shell (without fuzziness) in the $z$-plane, $bD_{eq} = D_{pol}$, can be computed. The total polar principal axis of the ellipsoid is $R_{tot, pol} = R_{pol} + D_{pol}$. The model used is implemented in the free SasView software (version 4.2.2, http://www.sasview.org/) \cite{Ped97}. As for the spherical models, a Lorenzian term, to account for the inhomogenities of the polymeric network, and a constant background have been added to the ellipsoidal model to correctly reproduce the data at high-$q$.

\begin{figure}[ht]
    \centering
    \includegraphics[width=.5\textwidth]{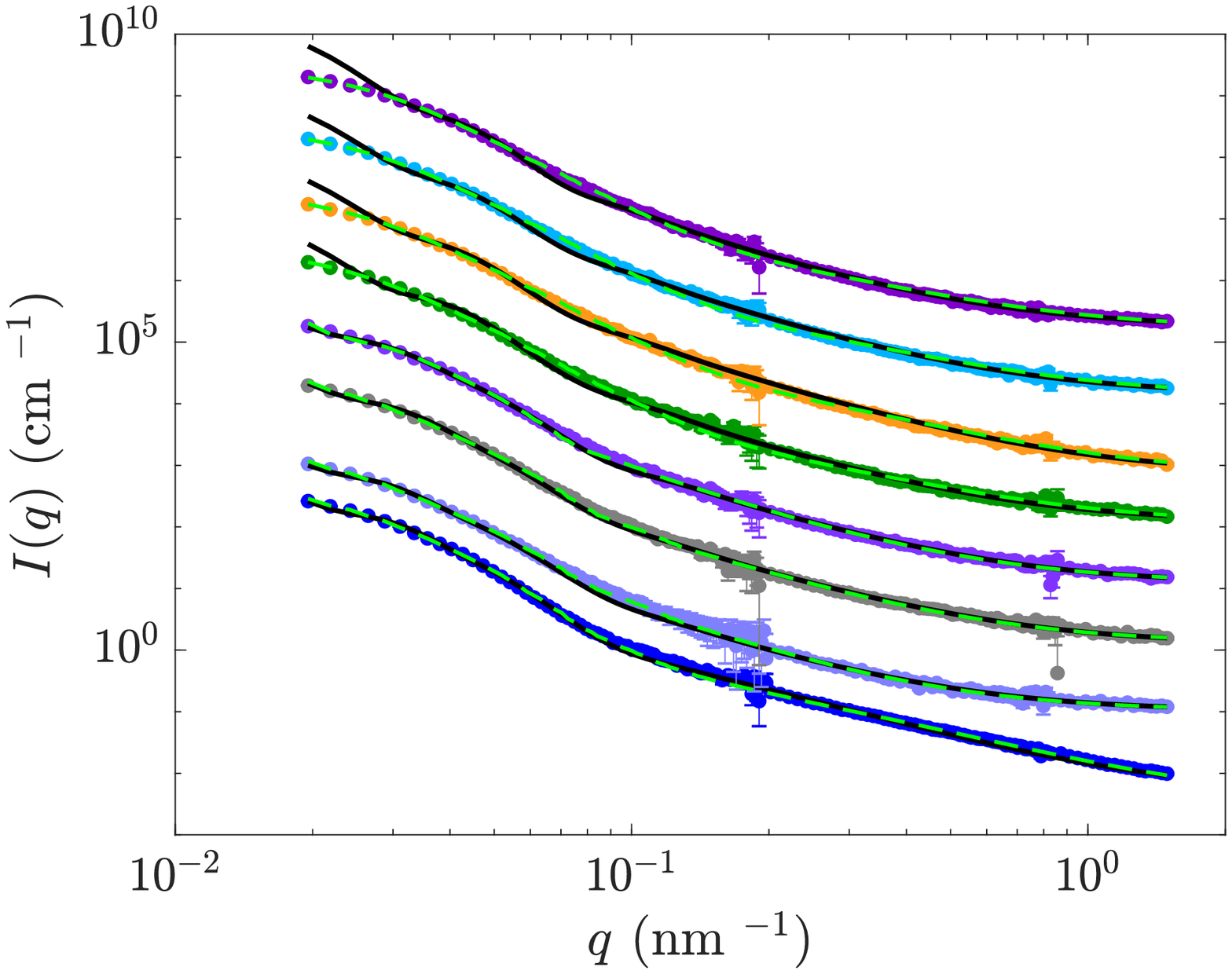}
	\caption{SANS intensities, $I(q)$, vs. scattering vector $q$, of hydrogenated microgels with $\zeta_{H}=0.080\pm0.003$ measured with contrast variation for concentrations, from bottom to top, $\zeta = 0.080 \pm 0.003$; $0.50 \pm 0.01$; $0.65 \pm 0.01$; $0.74 \pm 0.01$; $0.90 \pm 0.02$; $1.00 \pm 0.02$; $1.10 \pm 0.02$; $1.19 \pm 0.03$. The solid black lines represent fits with the hollow-fuzzy-shell model, while the dashed green lines represent fits of the data with the ellipsoidal model.}
	\label{fig:sphere_vs_ellipsoid}
\end{figure}

Figure.~\ref{fig:sphere_vs_ellipsoid} shows as an example the SANS intensities, $I(q)$, of hydrogenated microgels with $\zeta_{H}=0.080\pm0.003$ measured with contrast variation for concentrations, from bottom to top, $\zeta = 0.080 \pm 0.003$; $0.50 \pm 0.01$; $0.65 \pm 0.01$; $0.74 \pm 0.01$; $0.90 \pm 0.02$; $1.00 \pm 0.02$; $1.10 \pm 0.02$; $1.19 \pm 0.03$. The solid black lines represent the fit of the data using the hollow-fuzzy-sphere model reported in the first version of the manuscript, while the dashed green curves are fits of the same data using the ellipsoidal model.

\begin{table*}[tbh]
\caption{Generalized packing fraction, $\zeta$, and characteristic lengths for the spherical and ellipsoidal fits of the SANS data in Figure~\ref{fig:sphere_vs_ellipsoid}.}
\begin{center}
\label{tab:sphere_ellipsoid}
\begin{tabular}{c c c c c c c c c}
\hline
\hline
\multirow{2}{*}{\textbf{$\zeta$}}  	&\multicolumn{2}{c}{\textbf{Sphere}}		&\multicolumn{6}{c}{\textbf{Ellipsoid}}	\\
							        & $R_{SANS}$~(nm)   & $w_{core}$~(nm)  		        &$R_{eq}$~(nm)   &$a$ &$D_{eq}$~(nm) & $b$  &$R_{pol}$~(nm)  &$D_{pol}$~(nm)\\ 
							     \hline
							     
$0.080 \pm 0.003$   & $210\pm8$ &  $91\pm4$ & $135\pm10$ & $1.00\pm0.01$ & $90\pm3$ & $1.00\pm0.01$ & $135\pm11$ &$90\pm4$\\
$0.50 \pm 0.01$   & $201\pm7$ &  $96\pm3$ & $129\pm8$ & $1.01\pm0.01$ & $88\pm3$ & $1.00\pm0.01$ & $130\pm9$ &$87\pm4$\\
$0.65 \pm 0.01$   & $190\pm8$ &  $81\pm5$ & $120\pm8$ & $1.05\pm0.01$ & $88\pm2$ & $1.02\pm0.01$ & $126\pm10$ &$90\pm3$\\
$0.74 \pm 0.01$   & $185\pm7$ &  $99\pm3$ & $105\pm7$ & $1.02\pm0.03$ & $89\pm4$ & $1.04\pm0.03$ & $107\pm10$ &$93\pm7$\\
$0.90 \pm 0.02$   & $143\pm5$ &  $49\pm3$ & $30\pm4$ & $0.79\pm0.02$ & $87\pm3$ & $0.35\pm0.02$ & $24\pm4$ &$30\pm3$\\
$1.00 \pm 0.02$   & $135\pm6$ &  $55\pm3$ & $31\pm4$ & $0.91\pm0.03$ & $76\pm2$ & $0.35\pm0.03$ & $29\pm5$ &$26\pm3$\\
$1.10 \pm 0.02$   & $133\pm7$ &  $51\pm4$ & $29\pm3$ & $0.98\pm0.03$ & $77\pm5$ & $0.35\pm0.01$ & $28\pm4$ &$27\pm3$\\
$1.19 \pm 0.03$   & $133\pm5$ &  $47\pm2$ & $30\pm4$ & $0.90\pm0.05$ & $69\pm4$ & $0.32\pm0.03$ & $26\pm5$ &$22\pm3$\\
\hline
\hline
\end{tabular}
\end{center}
\end{table*}

As can be seen, the fits with the two models virtually coincide for concentrations $\zeta <0.8$. For these samples the fitting parameters that describe the anisotropy of the microgels, $a$ and $b$, are $\approx 1$, indicating that there is not a significant shape deformation (Table~\ref{tab:sphere_ellipsoid}). We have $R_{eq} = R_{pol} \approx R_{SANS}$ and also the size of the internal cavity are consistent with the values determined by SANS. The differences in the characteristic lengths of the microgels determined by the two models can be understood since, for simplicity and to not increase too much the number of fitting parameters, the ellipsoidal model does not account for internal or external fuzziness of the microgels. This might lead to the difference in the values of the length of the shell or to the differences in the cavity size. Nevertheless, for $\zeta\lesssim 0.75$ no significant anisotropy can be detected and, therefore, the increase of the value of the fitting parameter describing the apparent polydispersity is mainly due to the faceting of the microgels that cannot be resolved within the SANS resolution. We note that the fact that microgels in crowded environments first facet and then isotropically collapse is in agreement with direct imaging reported by de Aguiar et al.~(confocal) \cite{Agu17} and by Conley et al.~(super resolution) \cite{Con19}.

For concentrations $\zeta \geq 0.90 \pm 0.02$, the green dashed curves describe more accurately the SANS intensities, especially for very low values of $q$. For these fits the parameters reflecting the anisotropy of the microgels are smaller than one: $0.8\lesssim a \lesssim 0.95$ and $0.3\lesssim b \lesssim 0.35$ (Table~\ref{tab:sphere_ellipsoid}). This indicate that, well above the random close packing, the microgel become oblate spheroids ($R_{eq} > R_{pol}$). Furthermore, it seems that the cavity maintains a more spherical shape, $a \simeq 1$, while the shell becomes more anisotropic $b\ll 1$. 

Even in this case, we have to be careful and not over-interpret the data. Since the simple model we are using for the ellipsoidal particles does not account for both fuzziness and faceting, it might be that such a strong decrease of the parameter $b$ is a consequence of faceting. Acknowledging this possibility, we can still see that, at very high packing fractions, the microgels are better described by an anisotropic object and, therefore, it is reasonable to consider this effect being present, most likely at the same time with faceting.

\section{Static Light Scattering}

To avoid the influence of microgel-microgel interactions on the form factor measurements in our static light scattering (SLS) experiments, we worked with highly diluted samples, for which the structure factor, $S(q)$, can be approximated to unity and $I(q)\approx P(q)$, with $I(q)$ and $P(q)$ being the total scattered intensity and the form factor, respectively. A closed goniometer from \textit{SLS-Systemtechnik GmbH}, mounting a laser with a wavelength in vacuum of $\lambda_0 = 640$~nm, was used for all measurements. The instrument covers a $q$-range lower than that of small-angle neutron and X-ray scattering and, therefore, complements these methods. An index-matched toluene bath was used to keep the temperature constant during the measurements. The scattering angle, $\theta$, was varied between 15$^\circ$ and~150$^\circ$, thereby varying the wavevector $q = (4\pi n/\lambda_0)\sin{\theta/2}$. All measurements were corrected by subtracting the dark count and the scattered intensity of the solvent and the cell.

\begin{figure*}[ht]
	\subfigure{\includegraphics[width=0.25\textwidth ]{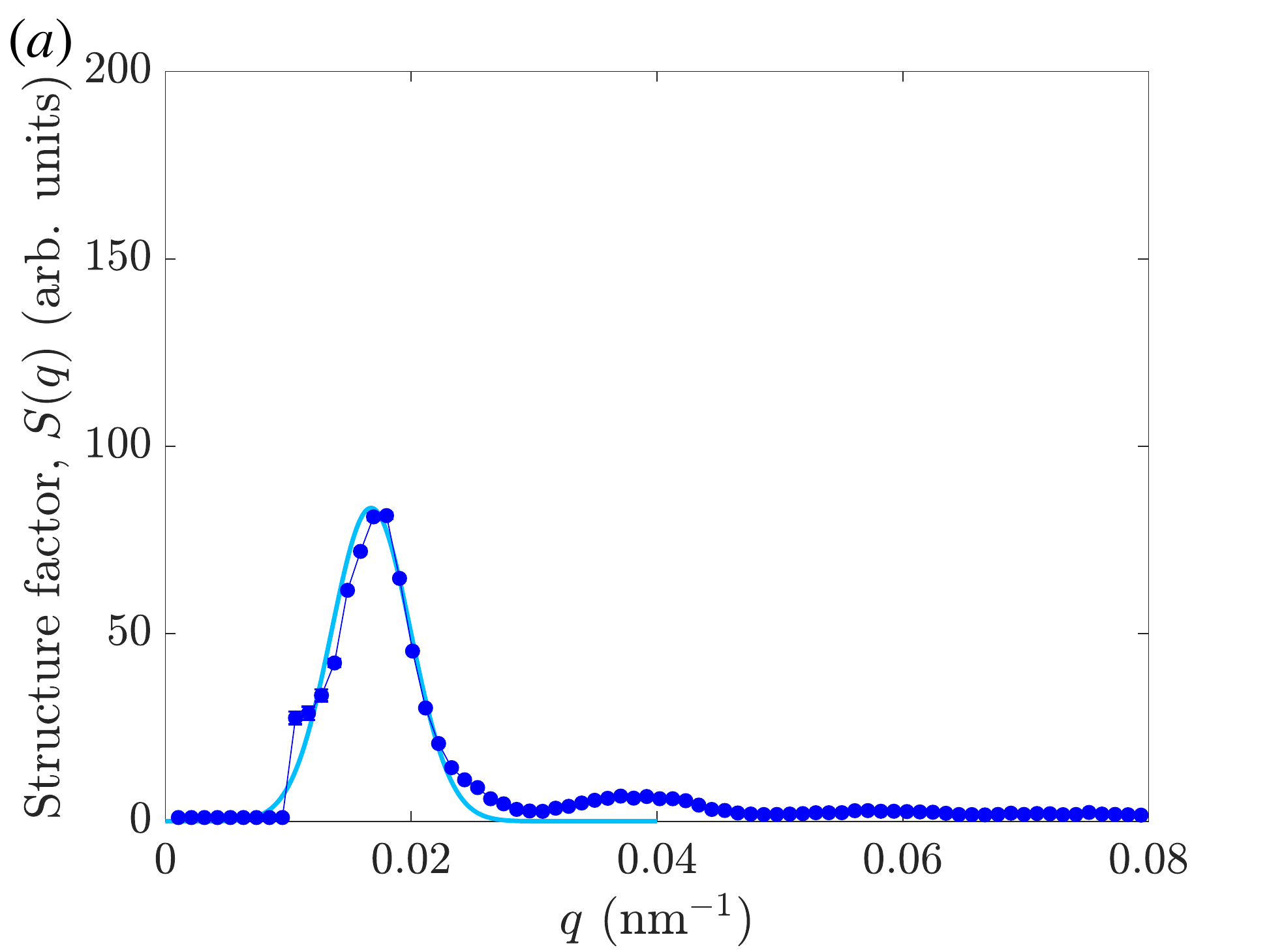}}
	\subfigure{\includegraphics[width=0.25\textwidth ]{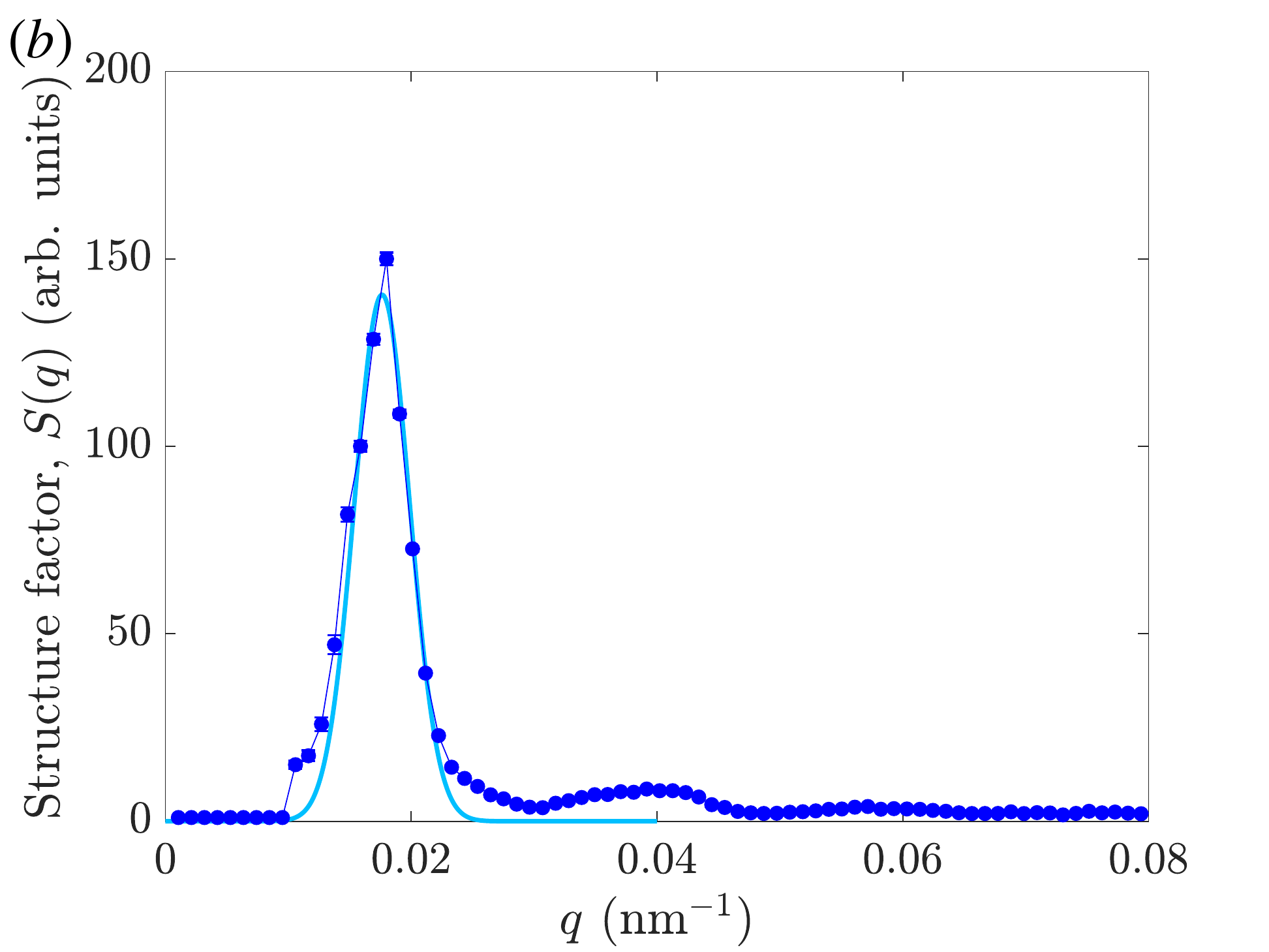}}
	\subfigure{\includegraphics[width=0.25\textwidth ]{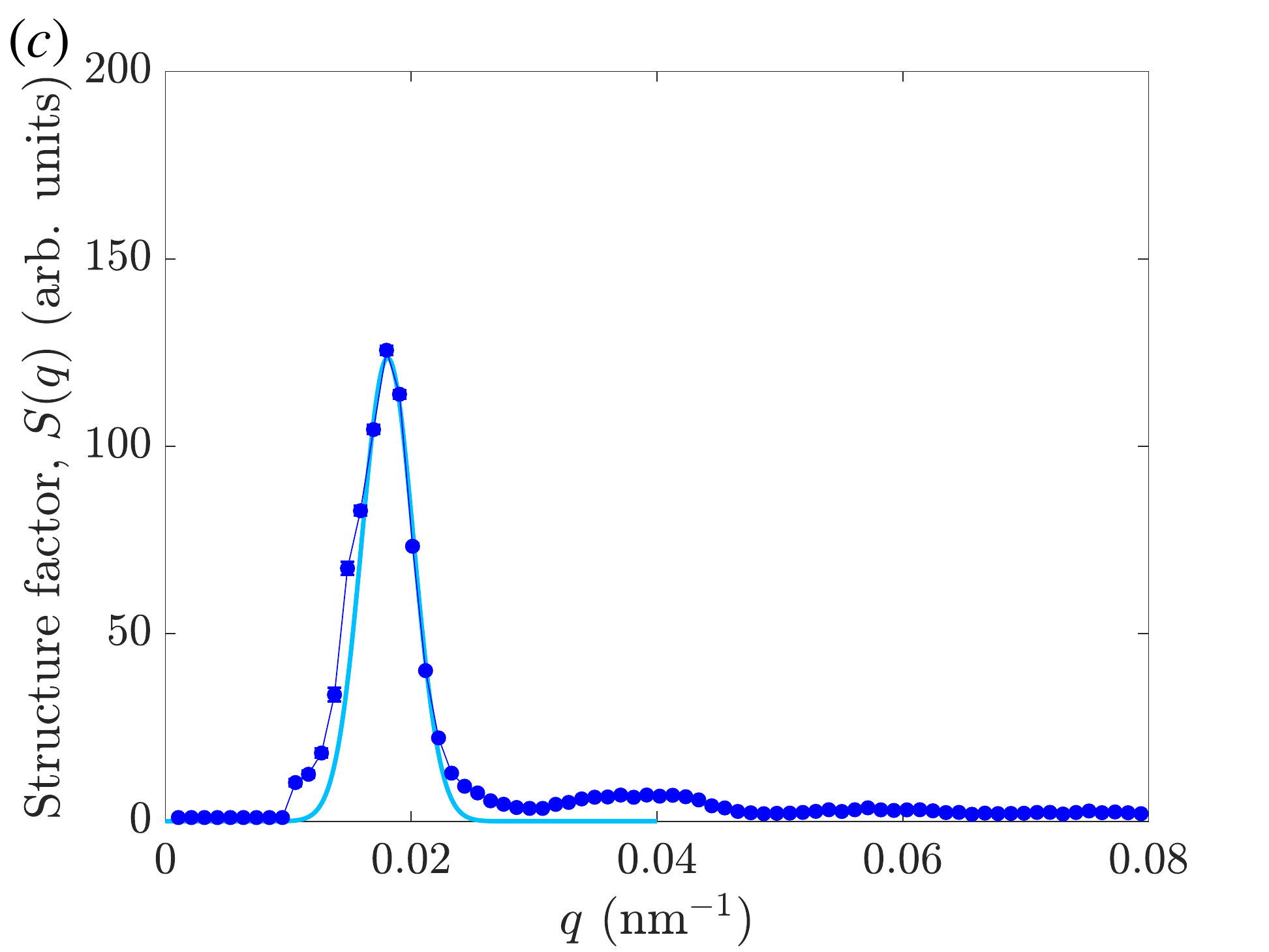}}
    \subfigure{\includegraphics[width=0.25\textwidth ]{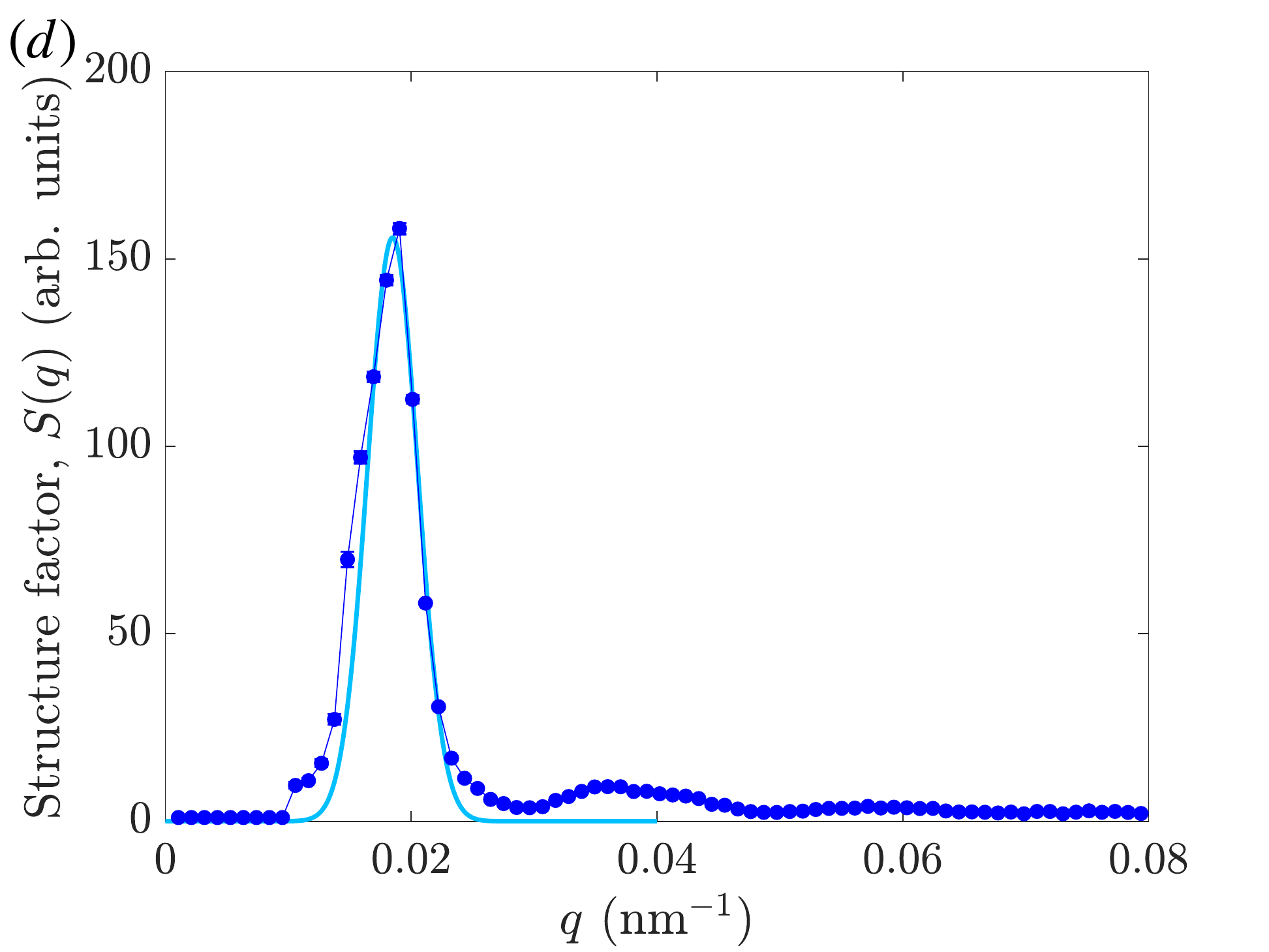}}
	\subfigure{\includegraphics[width=0.25\textwidth ]{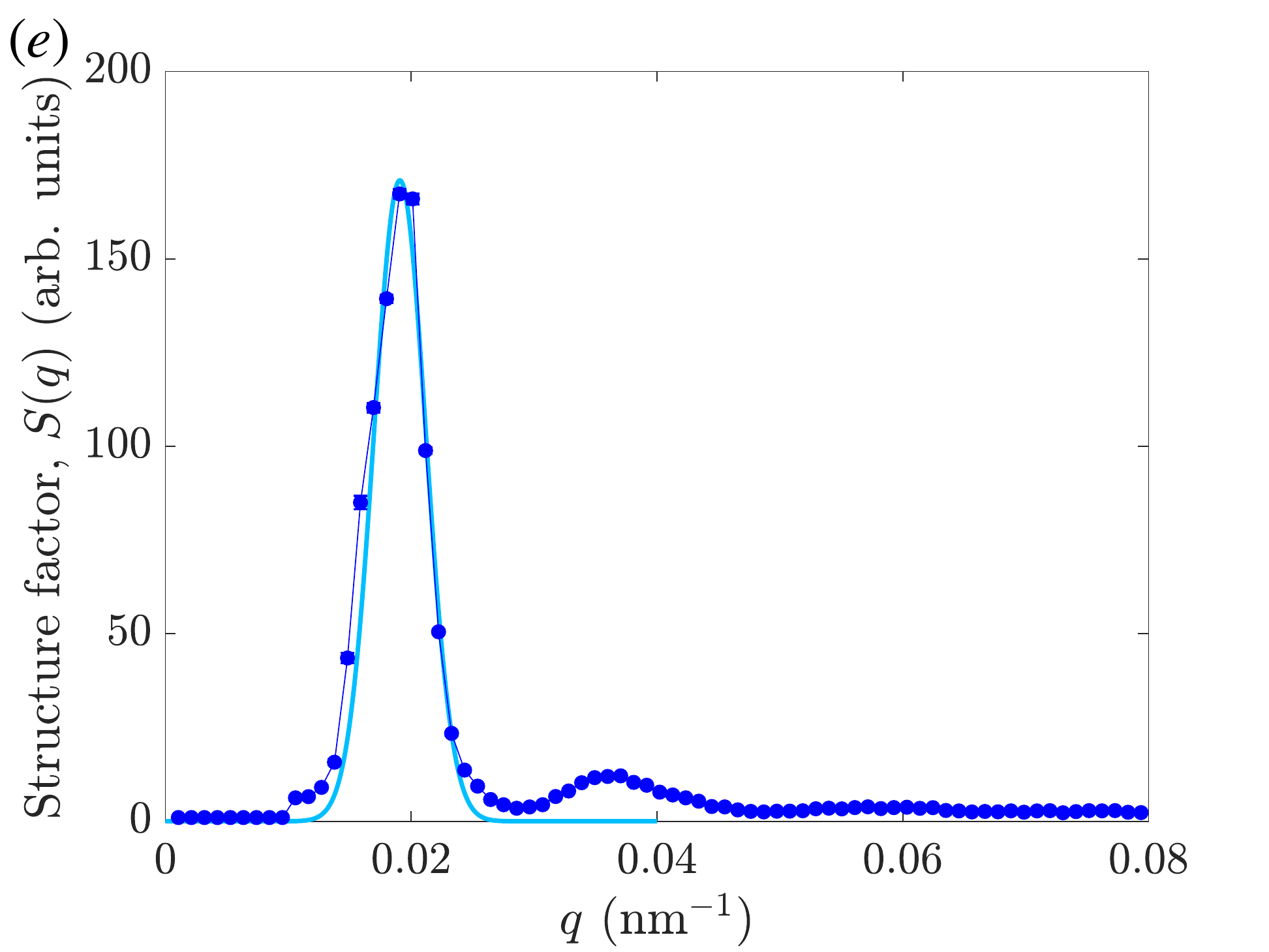}}
	\subfigure{\includegraphics[width=0.25\textwidth ]{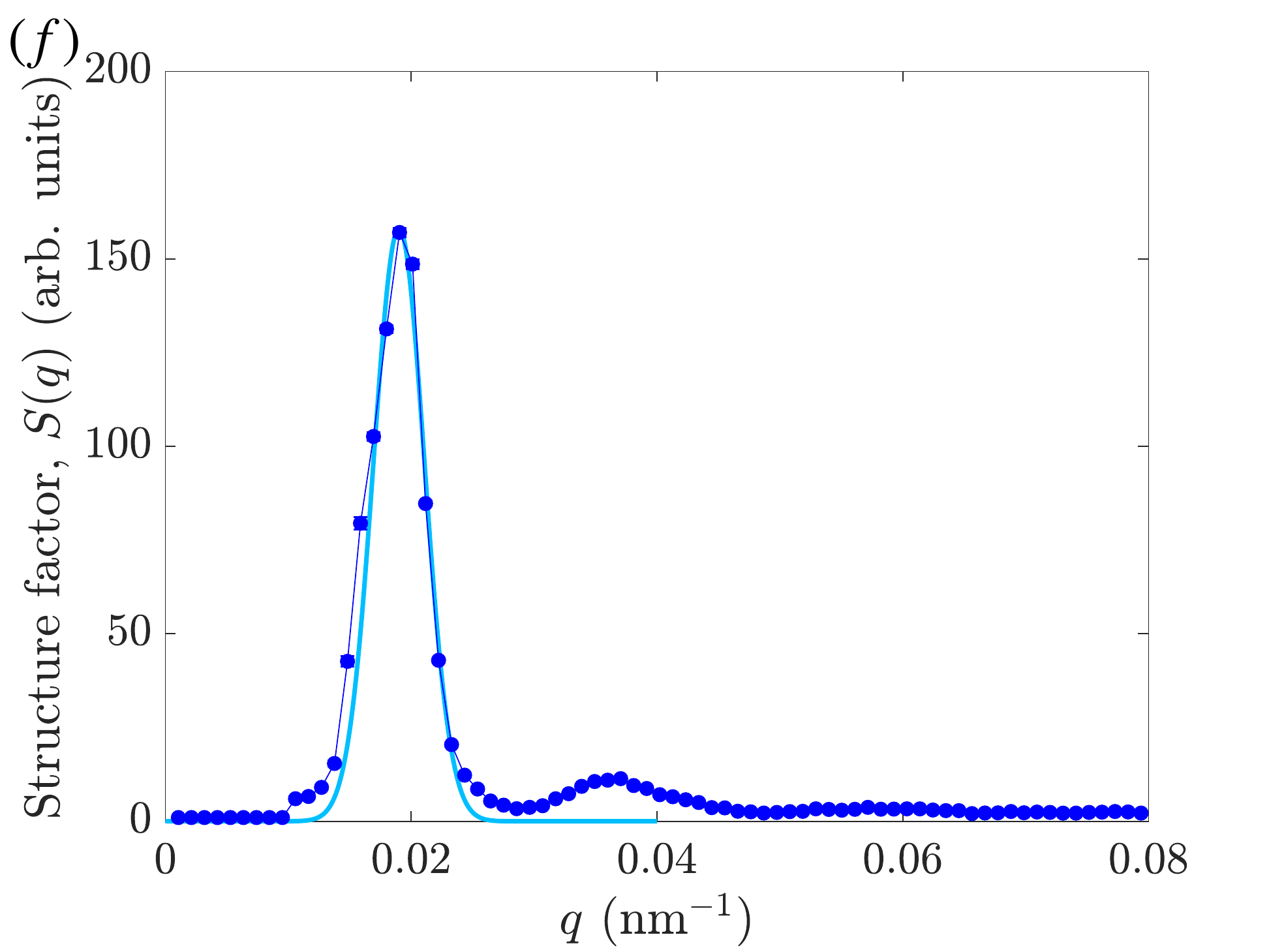}}
	\caption{Structure factors, $S(q) = I(q)/I(q, \zeta = 0.08)$, versus scattering vector, $q$, of solutions of the hollow 5~mol\% crosslinked microgels in Fig.~\ref{fig:PB}(a) at different concentrations (blue circles): $\zeta = 0.60 \pm 0.01$ (a); $0.70 \pm 0.01$ (b); $0.75 \pm 0.02$ (c); $0.80 \pm 0.02$ (d); $0.85 \pm 0.02$ (e); and $0.95 \pm 0.02$ (e). Light blue solid lines: Gaussian fits of the first peak of the structure factors.
	}
	\label{fig:SAXS_Sq_HS}
\end{figure*}

%\newpage
\section{Small-Angle X-Ray Scattering}

At the Swiss Light Source, Paul Scherrer Institut (Villigen, Switzerland), the cSAXS instrument was used to perform the small-angle X-ray scattering (SAXS) experiments. The X-ray wavelength was set to $\lambda=0.143$ nm with an error of 0.02~\% over $\lambda$ resolution. The detector was positioned at a distance of 7.12~m. The collimated beam illuminated an area of about $200~\mu$m $\times~200~\mu$m. The instrument mounts a 2D detector with a pixel size of 172~$\mu$m and 1475$\times$1679 pixels.\\

The structure factors, $S(q)$, shown in figure \ref{fig:SAXS_Sq_HS}, were obtained by dividing the scattered intensity, $I(q)$, by the form factor measured in a diluted solution, $\zeta = 0.08 \pm 0.01$. In this way we neglected the variation of the form factor of the microgels due to the increase of $\zeta$. Nevertheless, it has been shown that this approximation leads to an error on the position of the first peak in $S(q)$ of $\approx 2$~\% \cite{Sco17}.

\section{Flory-Hertz Model of Microgels} We consider $N$ hollow microgels dispersed in a solvent. Each microgel, consisting of $N_{{\rm mon}}$ monomers and $N_{{\rm ch}}$ crosslinked chains, has in its collapsed state an inner (cavity) radius $a_0$ and outer radius $b_0$, and in its swollen state an inner radius $a$ and outer radius $b$. For simplicity, we assume the crosslinks to be uniformly distributed throughout the shell volume. Under the assumption that monomers in the collapsed state are randomly close-packed with volume fraction 0.63, the monomer numbers are related to the collapsed radii according to
\begin{equation}
N_{\rm mon}=0.63~\frac{b_0^3-a_0^3}{r_{\rm mon}^3},
\label{Nmon}
\end{equation}
where $r_{\rm mon}\simeq 0.3$ nm is a typical monomer radius. To allow for independent variations of the inner and outer radii, we define the particle swelling ratio via the volume ratio of the swollen and collapsed particles:
\begin{equation}
\alpha=\left(\frac{v}{v_0}\right)^{1/3}
%=\left(\frac{b^3-a^3}{b_0^3-a_0^3}\right)^{1/3}
=\left(\frac{\alpha_{\rm out}^3-\gamma^3\alpha_{\rm in}^3}{1-\gamma^3}\right)^{1/3},
\label{alpha}
\end{equation}
where $\alpha_{\rm in}\equiv a/a_0$ and $\alpha_{\rm out}\equiv b/b_0$ are the inner and outer swelling ratios, respectively, and $\gamma\equiv a_0/b_0$ is the ratio of the collapsed inner and outer radii.

In a coarse-grained model, we describe swelling of microgels via the Flory-Rehner theory of polymer networks~\cite{flory1953,flory-rehner1943-I,flory-rehner1943-II}. Combining mixing entropy, polymer-solvent interactions, and elastic free energy, the free energy of a microgel of swelling ratio $\alpha$ is
\begin{eqnarray}
\beta F(\alpha)&=&N_m\left[(\alpha^3-1)\ln\left(1-\alpha^{-3}\right) 
+\chi\left(1-\alpha^{-3}\right)\right] 
\nonumber\\[1ex]
&+&\frac{3}{2}N_{\rm ch}\left(\alpha^2-\ln\alpha-1\right)~,
\label{FloryF}
\end{eqnarray}
where $\beta\equiv 1/(k_BT)$ at temperature $T$, $N_{\rm mon}$ and $N_{\rm ch}$ are the monomer and chain numbers of the particle, and $\chi$ is the Flory solvency parameter, associated with polymer-solvent interactions.  When applying the Flory-Rehner theory, choosing the reference particle radius to be the collapsed radius is equivalent to choosing the reference polymer volume fraction to be the random-close-packed volume fraction of monomers in the collapsed state [Eq.~(\ref{Nmon})]. Our choice is consistent with the experimental synthesis, in which crosslinking occurs at temperatures sufficiently high that the particles when unstrained are in their collapsed states. Previous studies~\cite{fernandez-nieves-pre2002,lopez-richtering2017} that have successfully fit the Flory-Rehner theory to light scattering data for swelling of thermoresponsive microgels have determined that the volume phase transition can be accurately predicted only if the Flory solvency parameter is treated as a function of temperature and the polymer volume fraction $\phi_p$, following the form
\begin{equation}
\chi(T,\phi_p)=\frac{1}{2}-A\left(1-\frac{\theta}{T}\right)+C\phi_p+D\phi_p^2,
\label{chi}
\end{equation}
where $\theta$ is the theta temperature and $A$, $C$, and $D$ are fit parameters. pNIPAM microgels in room-temperature water are typically characterized by $\chi<0.5$~\cite{fernandez-nieves-pre2002,lopez-richtering2017}.

Interactions between a pair of microgels of respective outer radii $b_i$ and $b_j$ are modeled via a Hertzian effective pair potential~\cite{landau-lifshitz1986}
\begin{equation}
v_H(r_{ij})=\left\{ \begin{array} 
{l@{\quad}l}
\epsilon_{ij}\left(1-\frac{\displaystyle r_{ij}}{\displaystyle b_i+b_j}\right)^{5/2},
& r_{ij}<b_i+b_j \\[1ex]
0~, 
& r_{ij}\ge b_i+b_j, \end{array} \right.
\label{Hertz}
\end{equation}
where $r_{ij}$ is the center-to-center separation of particles $i$ and $j$. Overlapping configurations ($r_{ij}<b_i+b_j$) can be physically associated with faceting of otherwise spherical microgels. Neglecting any influence of the cavity on the interaction strength, the amplitude of the pair potential is given by~\cite{landau-lifshitz1986}
\begin{equation}
\epsilon_{ij}=\frac{8}{15}\left(\frac{1-\nu_i^2}{Y_i}+\frac{1-\nu_j^2}{Y_j}\right)^{-1}
(b_i+b_j)^2(b_ib_j)^{1/2},
%\frac{4Y v}{5\pi(1-\nu^2)}~,
\label{Bij1}
\end{equation}
depending on the gel elastic properties through Young's moduli $Y_i$ and the Poisson ratios $\nu_i$. Assuming equal Poisson ratios $\nu$ of all particles, we have
\begin{equation}
\epsilon_{ij}=\frac{8}{15}\frac{Y_iY_j}{Y_i+Y_j}
\frac{(b_i+b_j)^2}{1-\nu^2}(b_ib_j)^{1/2}.
\label{Bij2}
\end{equation}
Scaling theory of polymer gels in good solvents~\cite{deGennes1979} predicts that the bulk modulus scales linearly with temperature and crosslinking density: $Y\sim k_BTN_{\rm ch}/(b^3-a^3)$. Thus,
\begin{equation}
\beta\epsilon_{ij}=C\frac{N_{\rm ch}}{b_i^3-a_i^3+b_j^3-a_j^3}
\frac{(b_i+b_j)^2}{1-\nu^2}(b_ib_j)^{1/2},
\label{B}
\end{equation}
where the proportionality constant $C$ is used to calibrate the model with experiment. The total internal energy associated with pair interactions is then given by
\begin{equation}
U=\sum_{i<j=1}^{N}\, v_H(r_{ij}),
\label{U}
\end{equation}
where, for each pair in the sum, the appropriate amplitude must be taken from Eq.~(\ref{B}). Note that, according to the Hertz model, as the cavity of a hollow microgel collapses ($a$ decreases), the Young's modulus decreases, i.e., the microgel becomes softer. Thus, the presence of a cavity promotes deswelling and faceting of hollow microgels with increasing concentration.

\newpage

\section{Simulation Methods} 

Within the Open Source Physics Library~\cite{osp-sip2006}, we implemented the standard Metropolis algorithm~\cite{frenkel-smit2001,binder-heermann2010}, accepting a trial displacement and swelling ratio change ($\alpha\to \alpha'$) with probability
\begin{equation}
{\cal P}_{\rm acc} = \min\left\{\exp[-\beta(\Delta U+\Delta F)],~1\right\},
\label{Pacc}
\end{equation}
where $\Delta U$ is the change in internal energy [Eq.~(\ref{U})] associated with interparticle (Hertz) pair interactions and $\Delta F=F(\alpha')-F(\alpha)$ is the change in free energy [Eq.~(\ref{FloryF})] associated with swelling. When moving a microgel, we {\it independently} made trial changes in both the inner and outer swelling ratios. As the particles moved and changed size and structure, we updated the Hertz potential amplitudes according to Eq.~(\ref{B}). After many trial moves, the particles adopt equilibrium size and configurational distributions that minimize the free energy. 

\begin{figure}[h!]
\begin{center}
\includegraphics[width=0.25\textwidth]{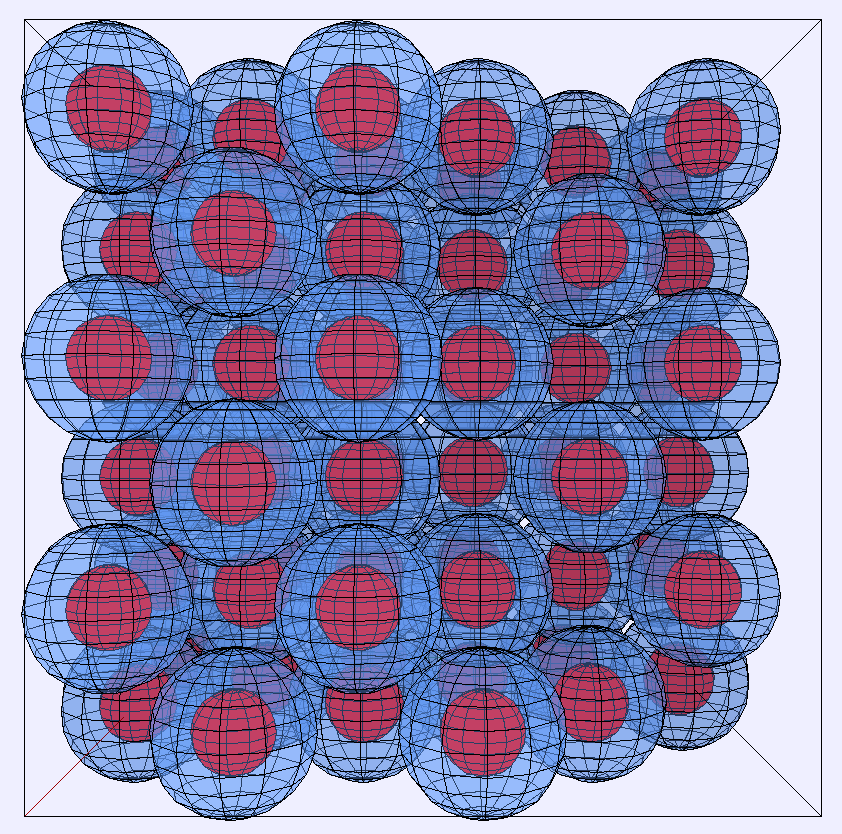}
\vspace*{-0.5cm}
\end{center}
\caption{Snapshot from a Monte Carlo simulation of a solution of soft, compressible, hollow microgels (blue spheres) with concentric cavities (red spheres).
}\label{fig:snapshot}
\end{figure}

To explore the dependence of microgel swelling and structure on concentration, we performed a series of simulations, each with $N=256$ hollow microgels. For visualization, Fig.~\ref{fig:snapshot} depicts a typical snapshot from a smaller-scale simulation of hollow microgels (blue spheres) with concentric spherical cavities (red spheres). As input, the system parameters that can be varied are the average density, collapsed particle inner and outer radii, average crosslink fraction, and Flory $\chi$ parameter. To model aqueous solutions of pNIPAM, we estimated $\chi\simeq 0.2$ from Eq.~(\ref{chi}), using fit parameters from ref.~\cite{fernandez-nieves-pre2002}. Although the synthesis yields microgels with a nominal crosslinker fraction of 5~mol\%, the actual fraction, $x$, is likely much smaller. For purposes of illustration, we chose $x=0.001$, but also examined larger ($x=0.01$) and smaller ($x=0.0001$) values. Since the collapsed radii are not known from experiments, we chose values for the inner (cavity) radius ($a_0=32$ nm or $a_0=50$ nm) and the outer radius ($b_0=63$ nm) that yield fully swollen inner and outer radii roughly consistent with experimental measurements in the dilute limit: $a=100$ nm and $b=210$ nm, respectively. For the calibration factor in Eq.~(\ref{B}), we chose $C=10$, corresponding to Young's moduli of order $10^3$ kPa.

We initialized the simulations with microgels placed either on the sites of a perfect fcc crystal lattice or in a random configuration. The latter was generated by starting the particles on a lattice and running with pair interactions turned off for several thousand steps to allow the particles to randomize.
Output data include the swollen microgel inner and outer radii, size polydispersity distributions, and the actual microgel volume fraction $\phi$. We computed $\phi$ by summing the total volume occupied by the swollen particles (cavities included) and subtracting the pair intersection volumes. The densities considered here are sufficiently low that triplet intersections may be ignored. To assess phase stability, we  computed bulk thermodynamic and structural properties, namely, the mean pair energy, osmotic pressure, radial distribution function, and static structure factor.

We computed the osmotic pressure from our simulations using the virial theorem, which relates the pressure
\begin{equation}
P=\frac{Nk_BT}{V}-\frac{1}{3V}\sum_{i<j=1}^N \la r_{ij}v_H'(r_{ij})\ra
\label{pressure}
\end{equation}
to the Hertz pair force, $-v_H'(r_{ij})$, exerted on microgel $i$ by microgel $j$, where angular brackets denote an ensemble average over configurations in the canonical ensemble.
From the particle coordinates, we computed the radial distribution function $g(r)$ by histogramming pair distances and the static structure factor via
\begin{equation}
S(q)=1+\frac{2}{N}\sum_{i<j=1}^N\la\frac{\sin(q r_{ij})}{q r_{ij}}\ra,
\label{Sq}
\end{equation}
where $q$ is the magnitude of the scattered wave vector.

\begin{figure}
\begin{center}
\includegraphics[width=0.35\textwidth]{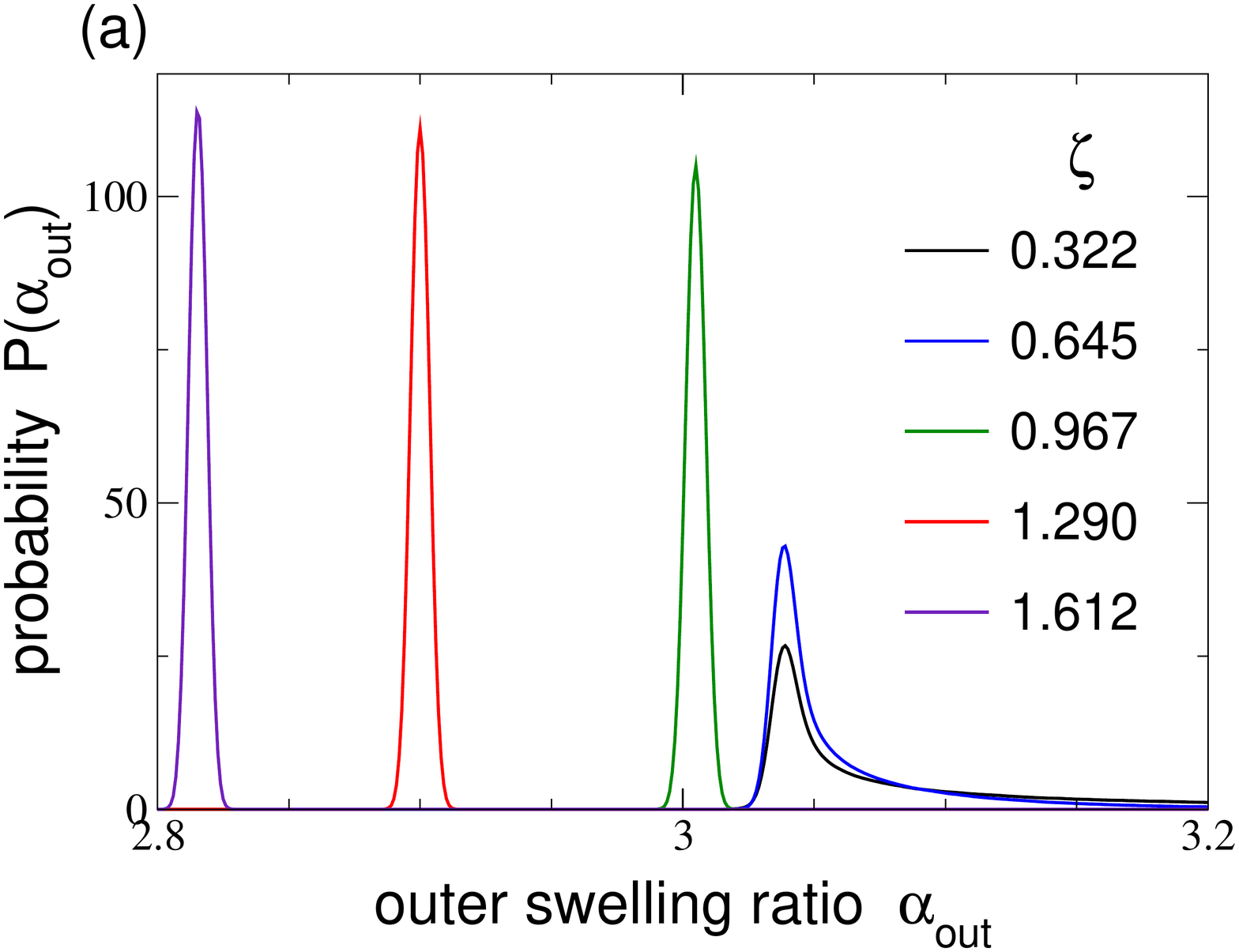}
\includegraphics[width=0.35\textwidth]{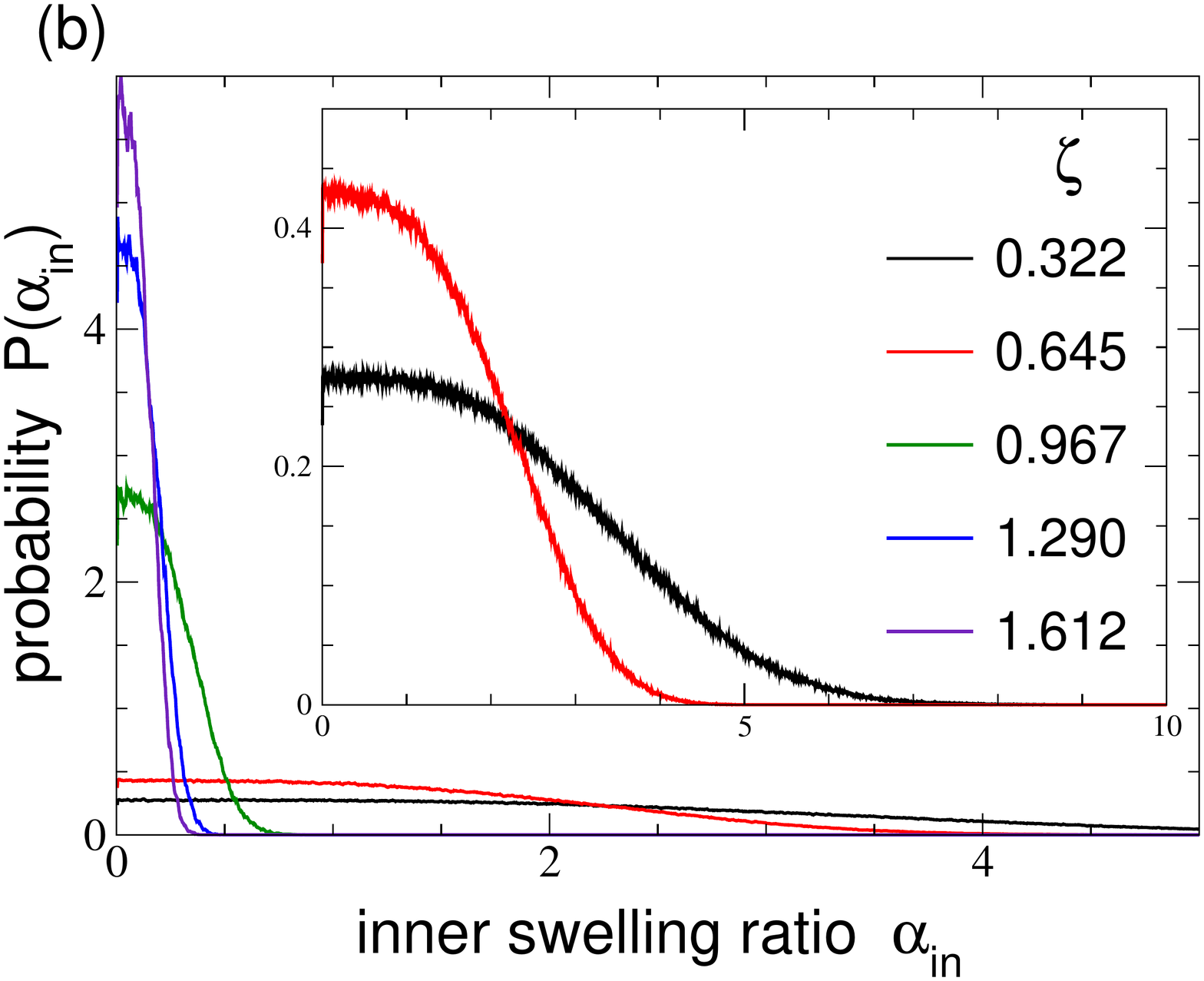}
\vspace*{-0.5cm}
\end{center}
\caption{Probability distributions of outer (a) and inner (b) swelling ratios of hollow microgels in solution with system parameters of Fig.~4, collapsed inner radius $a_0=32$ nm, crosslink fraction $x=0.001$, and generalized volume fraction $\zeta$ increasing from right to left as shown. Inset to panel (b) shows distributions for lowest $\zeta$ values on a larger scale.}
\label{fig:swelling-ratios}
\end{figure}

\section{Simulation Results}\label{simulation-results}

Figure~\ref{fig:swelling-ratios} shows our simulation results for the probability distributions of the inner and outer swelling ratios of hollow microgels, obtained by histogramming $a$ and $b$. With increasing concentration, the particles exhibit progressive deswelling and cavity collapse, consistent with Fig.~4 and with the SANS measurements.

\begin{figure}
\begin{center}
\includegraphics[width=0.35\textwidth]{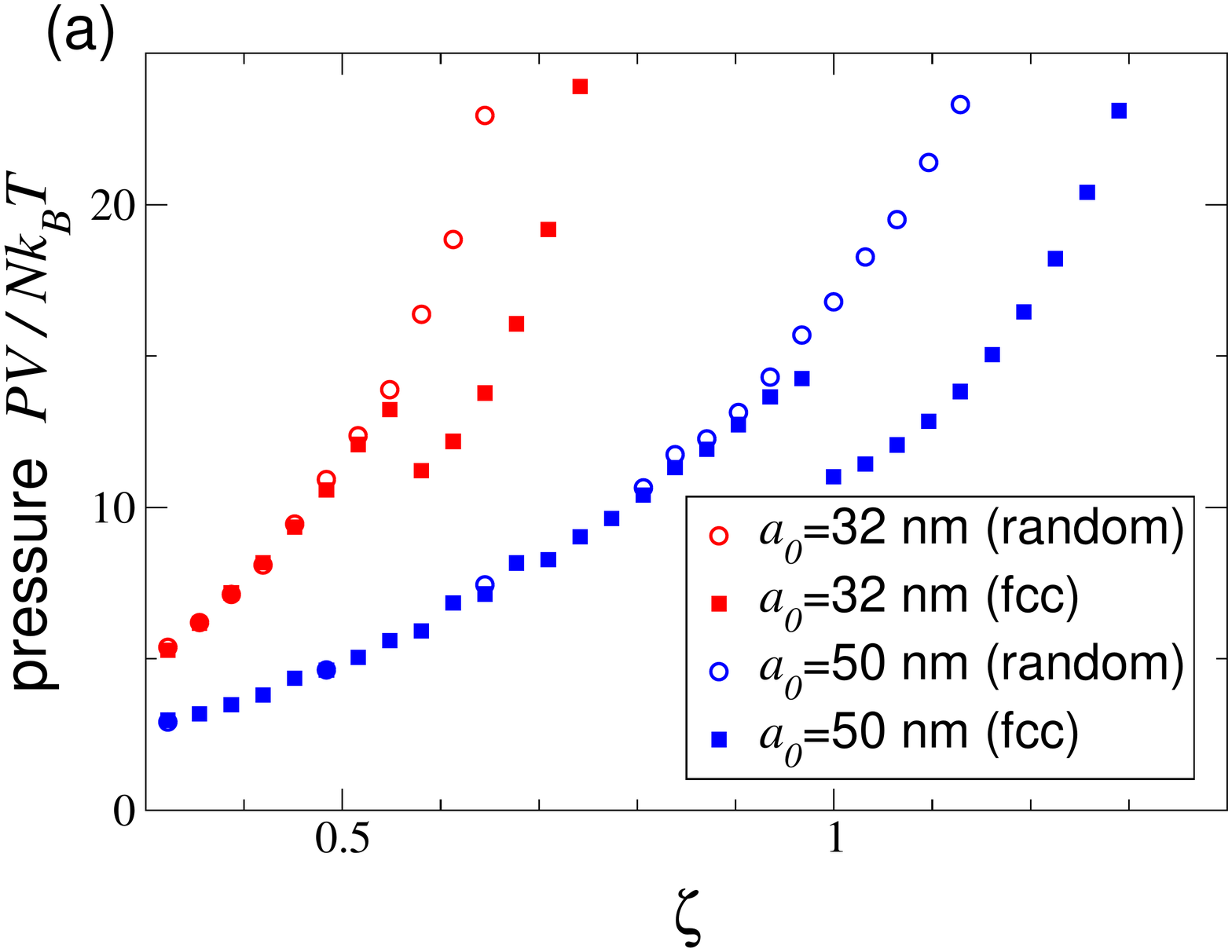}
\includegraphics[width=0.35\textwidth]{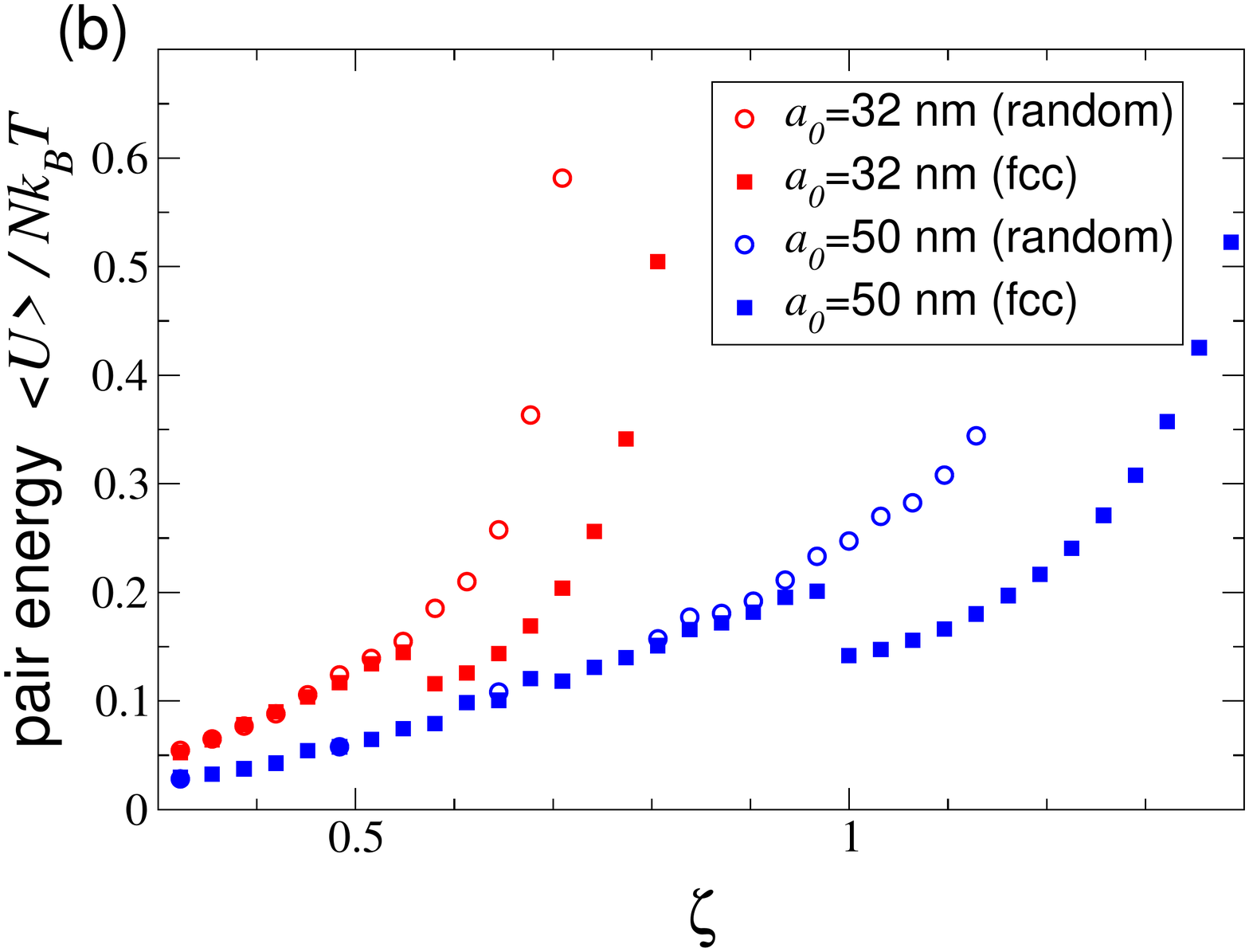}
\vspace*{-0.5cm}
\end{center}
\caption{Simulation data for bulk thermodynamic properties of hollow microgel solutions: (a) osmotic pressure $P$ and (b) mean pair interaction energy $\la U\ra$ vs.~generalized volume fraction $\zeta$ in solutions with system parameters of Fig.~4, collapsed inner radius $a_0=32$ or $50$ nm, and crosslink fraction $x=0.001$. Empty (closed) symbols represent data from runs initialized in random (fcc crystal) configurations. Statistical error bars are smaller than symbols.}
\label{fig:thermal}
\end{figure}

\begin{figure}
\begin{center}
\includegraphics[width=0.35\textwidth]{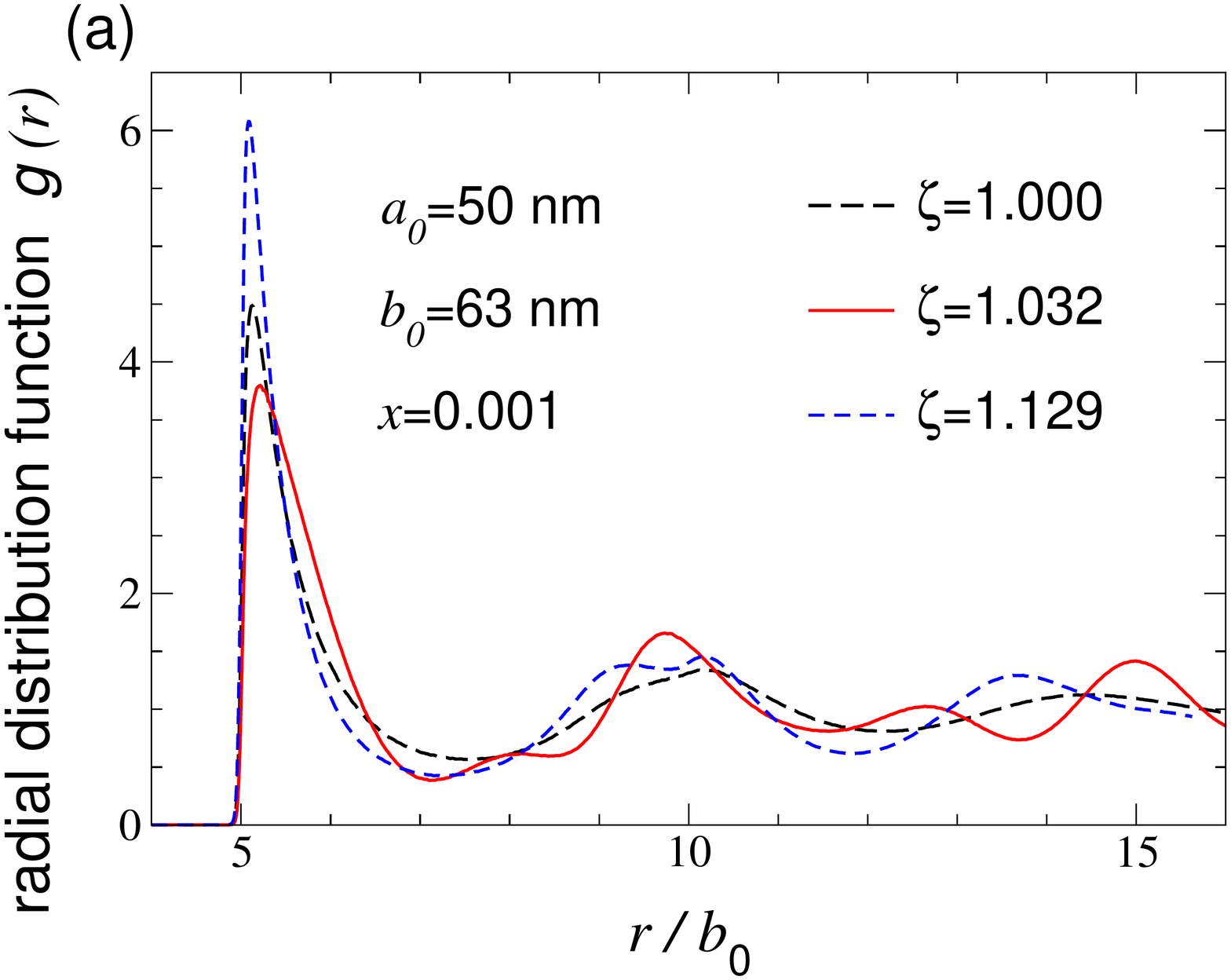}
\includegraphics[width=0.35\textwidth]{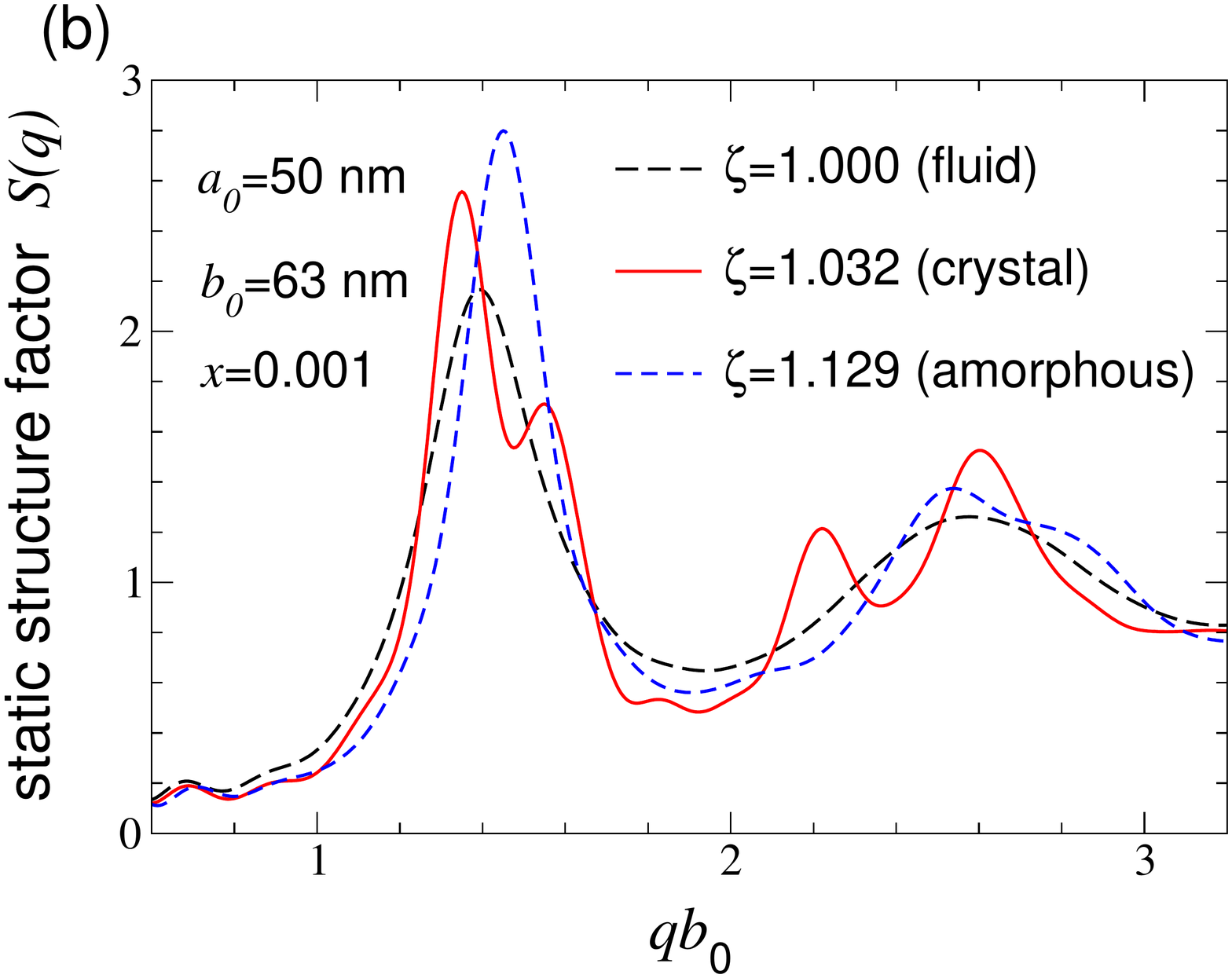}
\vspace*{-0.5cm}
\end{center}
\caption{Simulation data for bulk structural properties of hollow microgel solutions: (a) radial distribution function $g(r)$ vs.~radial distance $r$ in units of collapsed microgel outer radius $b_0$ and (b) static structure factor $S(q)$ vs.~scattered wave vector $q$ in solutions with system parameters of Fig.~4, collapsed inner radius $a_0=50$ nm, crosslink fraction $x=0.001$, and generalized volume fractions $\zeta$ from fluid (black long-dashed curve) to solid phase, which may be either an fcc crystal (red solid curve) or amorphous (blue short-dashed curve).
}
\label{fig:structure}
\end{figure}
To assess thermodynamic phase stability of hollow microgel solutions, we computed the osmotic pressure and mean pair energy vs.~concentration for systems initialized in either fcc crystal or random configurations (solid and empty symbols in Fig.~\ref{fig:thermal}, respectively). At lower $\zeta$, the two branches of the equation of state ($P$ vs.~$\zeta$) and of the mean pair energy ($\la U\ra$ vs.~$\zeta$) coincide, indicative of a fluid phase. At sufficiently high $\zeta$, such that the volume fraction $\phi\simeq 0.5$ ($\zeta\simeq 0.55$ for $a_0=32$ nm and $\zeta\simeq 1$ for $a_0=50$ nm), the fcc branch splits off and drops below the random branch. This trend strongly suggests that, at a threshold concentration, which increases with increasing cavity size, the close-packed fcc crystal becomes stable and an amorphous solid metastable relative to the fluid. Interestingly, the pressure at the fluid-solid transition has a value that is nearly independent of cavity size and is about equal to that of a hard-sphere fluid \cite{Hansen-McDonald}. While we have not computed free energies, which would require more extensive simulations for thermodynamic integrations, our results for the pressure and mean pair energy strongly support our interpretations.

To further support our conclusions regarding phase stability, we computed structural properties.
Figure~\ref{fig:structure} shows the radial distribution function and static structure factor for hollow microgel solutions at concentrations ranging from just below to just above the fluid-solid transition. At lower $\zeta$, the weak structure in $g(r)$ and $S(q)$ indicates a stable fluid phase. At higher $\zeta$, systems initialized in an fcc crystal configuration remained in that structure, as reflected by distinct peaks in $g(r)$ and by a prominent main peak in $S(q)$ and distinct secondary peaks at higher $q$. In contrast, systems initialized in a random configuration remained amorphous, evidenced by a tall main peak in $S(q)$, but broad, shallow secondary peaks. Most importantly for interpreting experiments, {\it the larger the cavity and lower the crosslink fraction, the higher the density up to which hollow microgels remain stable in the fluid phase}. Our results illustrate the sensitive dependence of particle swelling and phase stability on cavity size, and the predicted trends support the absence of crystalline solids observed in experiments.

\newpage

%At lower concentrations,
%the solution is in a disordered fluid phase, as reflected by weak structure in the
%radial distribution function $g(r)$, Fig.~\ref{fig:structure}(a), and the static factor factor $S(q)$, Fig.~\ref{fig:structure}(b). In this state, the microgels are relatively weakly confined and not significantly compressed. At higher concentrations, the solution is in an ordered solid phase (fcc crystal), as signaled by sharp peaks in $g(r)$ and $S(q)$. In this state, the microgels are strongly confined and compressed by their neighbors.

%%%%%%%%%%%%%%%%%%%%%%%%%%

%We also performed runs for different values of $\chi$ and found qualitatively similar swelling trends.

\bibliographystyle{apsrev4-1}
\bibliography{Scotti_hollowINhollow_SM}